\documentclass[a4paper,preprint,superscriptaddress,nofootinbib,longbibliography]{revtex4-1}
\usepackage[dvipdfmx]{graphicx}
\usepackage{bm}
\usepackage{amssymb}
\usepackage{amsmath}
\usepackage{color}
\usepackage{cancel}
\usepackage{comment}
\usepackage{here}
\usepackage{hyperref}
\usepackage{subfigure}
\usepackage{multirow}
\def\beq#1\eeq{\begin{equation}#1\end{equation}}
\def\bal#1\eal{\begin{align}#1\end{align}}

\begin{document} 

\title{The electric dipole moment in a model for neutrino mass, \\
dark matter and baryon asymmetry of the Universe}

\author{Kazuki Enomoto}
\email{k\_enomoto@kaist.ac.kr}
\affiliation{Department of Physics, KAIST, Daejeon 34141, Korea}

\author{Shinya Kanemura}
\email{kanemu@het.phys.osaka-u.ac.jp}
\affiliation{Department of Physics, Osaka University, Toyonaka, Osaka 560-0043, Japan}

\author{Sora Taniguchi}
\email{taniguchi@het.phys.sci.osaka-u.ac.jp}
\affiliation{Department of Physics, Osaka University, Toyonaka, Osaka 560-0043, Japan}

\preprint{OU-HET-1225}

\begin{abstract}
The electric dipole moment is examined in a three-loop neutrino mass model with dark matter originally proposed in Aoki {\it et al} [Phys. Rev. Lett. 102, 051805 (2009)]. The model contains a $CP$-violating phase in the Higgs potential which plays an important role in electroweak baryogenesis and is thus expected to explain the baryon asymmetry of the Universe simultaneously. However, such a $CP$-violating phase is severely constrained by the measurements of the electron electric dipole moment (eEDM), and a suppression mechanism for the eEDM is necessary to explain the observed baryon asymmetry while avoiding the constraint. In this paper, we examine neutrino mass, lepton-flavor-violating processes, dark matter, and the eEDM in the model. We show that the eEDM can be suppressed by destructive interference between the $CP$-violating phases in the Higgs sector and the dark sector with large $CP$-violating phases. We propose some benchmark scenarios including $O(1)$ $CP$-violating phases where tiny neutrino mass and dark matter can be explained while avoiding all current experimental and theoretical constraints. These $CP$-violating phases are expected to be large enough to generate the observed baryon asymmetry in the electroweak baryogenesis scenario.
\end{abstract}

\maketitle

\section{Introduction}

Although the standard model (SM) is consistent with various experimental results in particle physics and cosmology, we believe in the existence of physics beyond the SM because of its lack of ability to explain some phenomena such as neutrino oscillation~\cite{Super-Kamiokande:1998kpq, SNO:2001kpb}, which implies tiny neutrino masses, the existence of dark matter (DM)~\cite{Planck:2018vyg} and baryon asymmetry of the Universe (BAU)~\cite{Planck:2018vyg, Fields:2019pfx}. 
In recent decades, it has been one of the most important directions in particle physics to investigate new physics models to solve these problems. 

From the theoretical perspective, the SM is an incomplete theory, {\it i.e.}, there is no principle to determine the shape of the Higgs sector. 
Although the Higgs boson was discovered at LHC in 2012~\cite{ATLAS:2012yve}, the entire structure of the Higgs sector has not been revealed, and there remains the possibility of an extended Higgs sector. 

Given the above considerations, it would be natural to come up with a new physics model where an extended Higgs sector provides a mechanism to solve the above three problems. 
Various such mechanisms have been proposed so far. 
For example, loop-induced tiny neutrino masses via the quantum effect of additional scalar bosons~\cite{Zee:1980ai}, the scalar field DM as the weakly interacting massive particles (WIMP), and electroweak baryogenesis (EWBG) for the BAU~\cite{Kuzmin:1985mm} triggered by the strongly first-order electroweak phase transition (EWPT), which needs an extended Higgs sector~\cite{Kajantie:1996mn}.  

Models for the loop-induced neutrino masses are called radiative seesaw models~\cite{Zee:1980ai, Cheng:1980qt, Zee:1985id, Krauss:2002px, Ma:2006km, Aoki:2008av, Gustafsson:2012vj, Aoki:2022bkg, Nasri:2001ax, Gu:2007ug, Kanemura:2017haa, Enomoto:2019mzl}. 
In these models, the loop suppression factor naturally makes the size of neutrino masses small, and the mass of new particles can be at the TeV scale. 
In some radiative seesaw models, a new exact $Z_2$ symmetry is imposed to prohibit tree-level neutrino masses~\cite{Krauss:2002px, Ma:2006km, Aoki:2008av, Gustafsson:2012vj, Aoki:2022bkg, Nasri:2001ax, Gu:2007ug, Kanemura:2017haa, Enomoto:2019mzl}. 
This symmetry also plays a role in stabilizing the lightest $Z_2$-odd particle, which is a dark matter candidate. 
Thus, these models have an economical mechanism to explain the tiny neutrino mass and DM simultaneously. 

 In Ref.~\cite{Aoki:2008av}, a radiative seesaw model was proposed by Aoki, Kanemura and Seto (denoted by the AKS model in the following), where the BAU can also be explained in addition to tiny neutrino mass and DM by the Higgs sector extended by an additional Higgs doublet, a pair of charged scalar bosons and a real scalar boson. The latter two scalar bosons are dark sector particles that have an odd parity of an exact $Z_2$ symmetry.
 The extended Higgs sector can cause the strongly first-order EWPT and includes the $CP$-violating phase. 
Thus, baryon asymmetry can be generated via EWBG. 
In Ref.~\cite{Aoki:2011zg}, the Higgs potential of the AKS model and their theoretical constraints were studied in detail. 
However, in Refs.~\cite{Aoki:2008av, Aoki:2011zg}, the $CP$-violating phases were neglected for simplicity. 
Thus, the BAU was not evaluated although Ref.~\cite{Aoki:2008av} numerically showed that the strongly first-order EWPT can be achieved in the model. 

In Ref.~\cite{Aoki:2022bkg}, an extension of the AKS model was examined. 
It has a general Yukawa interaction between the SM fermions and the Higgs doublets, while those in the original AKS model are constrained by a softly broken $Z_2$ symmetry to avoid the flavor-changing neutral current (FCNC). 
By relaxing this symmetry, the model includes multiple $CP$-violating phases in not only the Higgs potential but also the Yukawa interactions, which make it possible to avoid the current severe constraint on the electron electric dipole moment (eEDM)~\cite{ACME:2018yjb, Roussy:2022cmp} by the destructive interference between these $CP$-violating phases~\cite{Kanemura:2020ibp}. 
In Ref.~\cite{Aoki:2022bkg}, the authors performed a numerical analysis of the baryon number production via EWBG in the model and found one benchmark scenario where tiny neutrino masses, DM and the BAU can be simultaneously explained while avoiding all current experimental and theoretical constraints. 
However,  Ref.~\cite{Aoki:2022bkg} assumes {\it ad hoc} flavor-alignment structure in the Yukawa interaction to restrict the FCNC~\cite{Pich:2009sp}, while the original AKS model naturally prohibits it at tree level by the softly broken $Z_2$ symmetry~\cite{Glashow:1976nt}. 
In addition, relaxing this symmetry causes too many model parameters. 

In this paper, according to the facts mentioned above, we revisit the original AKS model to investigate the $CP$ violation. 
The most significant difficulty is avoiding the constraint from the eEDM measurements unless we assume that $CP$-violating phases are very small.
We investigate the eEDM in the AKS model to see if it is possible to avoid the current severe bound while keeping a large $\mathcal{O}(1)$ $CP$-violating phase in the Higgs potential, which is necessary for successful EWBG~\cite{Turok:1990zg, Cline:1995dg, Fromme:2006cm, Cline:2011mm, Enomoto:2021dkl, Kanemura:2023juv, Aoki:2022bkg, Dorsch:2016nrg}. 
The eEDM is induced by two kinds of Feynman diagrams at the leading order. 
One is generated by the additional Higgs bosons like in the two Higgs doublet models (THDMs)~\cite{Inoue:2014nva, Modak:2018csw, Fuyuto:2019svr, Bian:2014zka, Bian:2016zba, Abe:2013qla, Cheung:2020ugr, Altmannshofer:2020shb}, which depends on a $CP$-violating phase in the Higgs potential. 
The other one is generated by a new $CP$-violating Yukawa interaction between the charged leptons and the dark sector particles, which was neglected in Ref.~\cite{Aoki:2022bkg}. 
We consider a scenario where the eEDM is suppressed by the destructive interference between the Higgs sector contribution and the dark sector contribution to avoid the current severe constraint.  
This is a different suppression mechanism from that employed in Ref.~\cite{Aoki:2022bkg}, which was proposed in Ref.~\cite{Kanemura:2020ibp} and has been studied in Ref.~\cite{Kanemura:2021atq, Enomoto:2021dkl, Kanemura:2023juv, Kanemura:2023jbz}.  
It is also different from other conventional suppression mechanisms in the THDMs~\cite{Inoue:2014nva, Modak:2018csw, Fuyuto:2019svr, Bian:2014zka, Bian:2016zba, Abe:2013qla, Cheung:2020ugr, Altmannshofer:2020shb}.

We also examine the neutrino mass, lepton-flavor-violating (LFV) processes, and DM in the model and show two benchmark scenarios where tiny neutrino masses and DM can be simultaneously explained while avoiding all current experimental and theoretical constraints.
In the two scenarios, DM is the real scalar boson, but with different masses.  
In the first scenario, the mass is about half of the 125 GeV Higgs boson, and a pair of DM mainly annihilates into the SM fermions via an almost on-shell Higgs boson. 
On the other hand, in the second scenario, DM has heavier mass, and a pair of them can annihilate into the additional Higgs bosons. 
In both scenarios, the size of the eEDM is smaller than the current limit while keeping an $\mathcal{O}(1)$ $CP$-violating phase in the Higgs potential. 

This paper is organized as follows. 
In the following section (Sec.~\ref{sec: model}), we define the model and show the lagrangian.
Theoretical and experimental constraints on the model are also shown there. 
In Sec.~\ref{sec: pheno}, we discuss some phenomenologies in the model: neutrino mass (Sec.~\ref{subsec: NuMass}), lepton flavor violation (\ref{subsec: LFV}) and DM (\ref{subsec: DM}). 
In Sec.~\ref{sec: EDM}, the eEDM in the model is investigated. 
Sec.~\ref{sec: benchmark scenarios} is assigned to discussions about numerical evaluations of the neutrino mass, the LFV processes, DM and the eEDM. 
We show two benchmark scenarios that can explain tiny neutrino masses and DM simultaneously while avoiding all the experimental and theoretical constraints.  
Some comments about the benchmark scenarios and a conclusion are presented in Sec.~\ref{sec: conclusion}.

\section{The model}
\label{sec: model}

\subsection{Lagrangian}
\label{subsec: Lagrangian}

Here, we show the lagrangian of the AKS model with $CP$-violation. 
The notation follows Ref.~\cite{Aoki:2008av}. 
The model includes new discrete symmetries; an exact $Z_2$ symmetry and a softly broken $Z_2$ symmetry, which are denoted by $Z_2$ and $\tilde{Z}_2$, respectively in the following. 
The gauge symmetry is the same as in the SM. 
The matter fields and their quantum numbers are shown in Table~\ref{table: matter fields}. 
There are three kinds of $Z_2$-odd fields: a real scalar $\eta$, a pair of charged scalars $S^\pm$, and three right-handed fermions $N_R^\alpha$ ($\alpha=1,2,3$). $N_R^\alpha$ and $\eta$ are gauge-singlet, while $S^\pm$ has a hypercharge $\pm 1$, respectively. 
$N^\alpha_R$ constitute Majorana fermions $N^\alpha \equiv N_R^\alpha + (N_R^{\alpha})^c$ with the mass
\begin{align}
\mathcal{L} = & - \frac{m_{N^\alpha}^{}}{2} \overline{(N_R^{\alpha})^c } N_R^\alpha + \mathrm{h.c.}  = - \frac{ m_{N^\alpha}^{} }{ 2 } (N^{\alpha })^\mathrm{T} C N^\alpha,  
\end{align}
where $C$ is the charge conjugate matrix. 
The SM fermions have $Z_2$-even parities, and we do not have a new $Z_2$-even fermion. 
On the other hand, $Z_2$-even scalar sector is an extension of the SM and includes two Higgs doublets $\phi_1$ and $\phi_2$, which have an even and odd parity of $\tilde{Z}_2$, respectively to avoid the FCNCs~\cite{Glashow:1976nt}. 

\begin{table}[b]
\begin{center}
\begin{tabular}{|c||c|c|c|c|c|c|c||c|c|c|} \hline
 	& $Q_L^{\prime i}$ & $u_R^{\prime i}$ & $d_R^{\prime i}$ & $L_L^{\prime i}$ & $\ell_R^{\prime i}$ & $\phi_1$ & $\phi_2$ & $N_R^\alpha$ & $S^+$ &  $\eta$ \\ \hline\hline
Spin & $1/2$ & $1/2$ & $1/2$ & $1/2$ & $1/2$ & $0$ & $0$ & $1/2$ & $0$ & $0$ \\ \hline
$SU(3)_C$ & {\bf 3} & {\bf 3} & {\bf 3} & {\bf 1} & {\bf 1} & {\bf 1} & {\bf 1} & {\bf 1} & {\bf 1} & {\bf 1} \\ \hline
$SU(2)_L$ & {\bf 2} & {\bf 1} & {\bf 1} & {\bf 2} & {\bf 1} & {\bf 2}  & {\bf 2} & {\bf 1} & {\bf 1} & {\bf 1} \\ \hline
$U(1)_Y$ & $1/6$ & $2/3$ & $-1/3$ & $-1/2$ & $-1$ & $1/2$ & $1/2$ & $0$ & $1$ & $0$ \\ \hline
$Z_2$ & \multicolumn{7}{c||}{$+$} & \multicolumn{3}{|c|}{$-$}  \\ \hline \hline
$\tilde{Z}_2$ & $+$ & $-$ & $-$ & $+$ & $+$ & $+$ & $-$ & $+$ & $+$ & $+$ \\ \hline
\end{tabular}
\caption{Matter fields in the model and their quantum numbers. 
$Z_2$ and $\tilde{Z}_2$ are exact and softly-broken $Z_2$ symmetries, respectively.}
\label{table: matter fields}
\end{center}
\end{table}

The Higgs potential $V$ is given by
\begin{align}
V =&  \ V_\mathrm{THDM} + \mu_S^2 |S^+|^2 + \frac{ \mu_\eta^2 }{ 2 } \eta^2 
+ \frac{ \lambda_S^{} }{ 4 }  |S^+|^4 
+ \frac{ \lambda_\eta^{} }{ 4! } \eta^4 
+ \frac{ \xi }{ 2 } | S^+ |^2 \eta^2
 \nonumber \\[5pt]
& + \sum_{a = 1}^2 | \phi_a^{} |^2
\Bigl( \rho_a | S^+ |^2 + \frac{ \sigma_a }{ 2 }\eta^2 \Bigr)
+ \Bigl(  2 \kappa  \tilde{\phi}_1^\dagger \phi_2 S^- \eta + \mathrm{h.c.} \Bigr). 
\end{align}
$V_\mathrm{THDM}$ is the potential in the $CP$-violating THDM;
\begin{align}
& V_\mathrm{THDM} = 
	\sum_{a=1}^2 \biggl( -\mu_a^2 |\phi_a|^2 + \frac{ \lambda_a }{ 2 } |\phi_a|^4 \biggr)
	-\bigl( \mu_{12}^2  \phi_1^\dagger \phi_2^{} + \mathrm{h.c.} \bigr)
\nonumber \\[7pt]
	& + \lambda_3 | \phi_1^{} |^2 | \phi_2 |^2 
	+ \lambda_4 | \phi_1^\dagger \phi_2 |^2 	
	+ \biggl( \frac{ \lambda_5^{} }{ 2 }  (\phi_1^\dagger \phi_2)^2 + \mathrm{h.c.} \biggr).
\end{align}
where $\tilde{\phi}_1 = i \sigma_2 \phi_1^\ast$. 
In general, $\mu_{12}^2$ and $\lambda_5$, and $\kappa$ are complex.
The rephasing of $S^\pm$ can make $\kappa$ real without loss of generality.  
On the other hand, the phase of $\mu_{12}^2$ and $\lambda_5$ remains as $CP$-violating phases as discussed below. 

The elements of the Higgs doublets are defined as 
\begin{equation}
\phi_i = \frac{ 1 }{ \sqrt{2} }
\begin{pmatrix}
\sqrt{2} w^+_i \\
v_i e^{i \xi_i}+ h_i + i a_i \\
\end{pmatrix}
\quad i=1,2, 
\end{equation}
where $v_1$ and $v_2$ are positive constant. 
In general, the vacuum expectation values (VEVs) of the doublets are complex, {\it i.e.}, the phase $\xi_i$ are not zero or $\pi$. 
However, only the relative phase $\xi_{21} = \xi_2 - \xi_1$ are physical because of the $U(1)_Y$ symmetry. 
We thus take $\xi_1 = 0$ and $\xi_2 = \xi_{21}$. 

We have three complex phases $\theta_\mu = \mathrm{arg}[\mu_{12}^2]$, $\theta_5 = \mathrm{arg}[\lambda_5]$, and $\xi_{21}$ in the Higgs potential. 
One of them can be real by rephasing $\phi_2$, and there remain two $CP$-violating phases. 
We choose $\xi_{21} = 0$ in the following. 

The stationary condition at the vacuum requires the following equations to be satisfied; 
\begin{equation}
\left\{
\begin{array}{l}
\displaystyle{-\mu_1^2 v_1 - \mu_{12}^{2R} v_2 + \frac{ \lambda_1 }{ 2 } v_1^3 + \frac{ \lambda_{345}^+ }{ 2 } v_1 v_2^2 = 0}, \\[7pt]
\displaystyle{-\mu_2^2 v_2 - \mu_{12}^{2R} v_1 + \frac{ \lambda_2 }{ 2 } v_2^3 + \frac{ \lambda_{345}^+ }{ 2 } v_1^2 v_2 = 0}, \\[7pt]
\displaystyle{-\mu_{12}^{2I} + \frac{ \lambda_5^I }{ 2 } v_1 v_2 = 0}, \\[7pt]
\end{array}
\right. 
\end{equation}
where $\lambda_{345}^+ = \lambda_3 + \lambda_4 + \lambda_5^R$, and the superscripts $R$ and $I$ represents the real and imaginary part of each quantity. 
We note that $\theta_\mu$ and $\theta_5$ are related by the third equation. 
Therefore, there is only one independent $CP$-violating phase in the Higgs sector. 
We choose $\theta_5$ as an input parameter. 
Then, $\theta_\mu$ is determined by the third equation. 
In addition, we note that dimension-two paramters $\mu_1^2$, $\mu_2^2$, and $\mu_{12}^{2I}$ are replaced by other paramters by the above equations. 
Thus, $\mu_{12}^{2R}$ is a unique dimensionful free parameter in $V_\mathrm{THDM}$.

To investigate mass eigenstates of the Higgs bosons, it is convenient to change the basis of the doublet from $\tilde{Z}_2$-parity eigenstate basis ($\phi_1$, $\phi_2$) to the Higgs basis ($\Phi_1$, $\Phi_2$), where only $\Phi_1$ obtains the real VEV~\cite{Davidson:2005cw}. 
This transformation is achieved by the rotation
\begin{equation}
\begin{pmatrix}
\Phi_1 \\
\Phi_2 \\
\end{pmatrix}
= 
\begin{pmatrix}
\cos \beta & \sin \beta \\
-\sin \beta & \cos \beta \\
\end{pmatrix}
\begin{pmatrix}
\phi_1 \\
\phi_2 \\
\end{pmatrix}, 
\end{equation}
where $\beta$ satisfies $\tan \beta = v_2/v_1$. 

$\Phi_1$ and $\Phi_2$ are given by
\begin{equation} 
\Phi_1 = \frac{ 1 }{ \sqrt{2} }
\begin{pmatrix}
\sqrt{2} G^+ \\
v + h_1^\prime + G^0 \\
\end{pmatrix}, \quad 
\Phi_2 = \frac{ 1 }{ \sqrt{2} }
\begin{pmatrix}
\sqrt{2} H^+ \\
h_2^\prime + i h_3^\prime \\
\end{pmatrix}, 
\end{equation}
where $v = \sqrt{v_1^2 + v_2^2} \simeq 246~\mathrm{GeV}$. 
Then, $G^+$ and $G^0$ are Nambu-Goldstone (NG) bosons and are absorbed into the longitudinal model of $W^\pm$ and $Z$ bosons, respectively. 
$H^\pm$ is a physical charged scalar boson, and its mass $m_{H^\pm}$ is given by
\begin{equation}
m_{H^\pm}^2 = M^2 - \frac{ v^2 }{ 2 } (\lambda_4 + \lambda_5^R), 
\end{equation}
where $M^2 = \mu_{12}^{2R}/(s_\beta c_\beta)$. 
Neutral scalar bosons $h_1^\prime$, $h_2^\prime$, and $h_3^\prime$ are not mass eigenstates, and their mass matrix is given by 
\begin{equation}
M_h^2  = v^2 
\begin{pmatrix}
\lambda_1 c_\beta^4 + \lambda_2 s_\beta^4 + 2 \lambda_{345}^+s_\beta^2 c_\beta^2 
	& \bigl\{  \lambda_2 s_\beta^2 - \lambda_1 c_\beta^2 + \lambda_{345}^+ c_{2\beta} \bigr\}s_\beta c_\beta 
		& - \lambda_5^I s_\beta c_\beta \\[10pt]
\bigl\{ \lambda_2 s_\beta^2 - \lambda_1 c_\beta^2 + \lambda_{345}^+ c_{2\beta} \bigr\}s_\beta c_\beta  
	& \displaystyle{\frac{ M^2 }{ v^2 } + (\lambda_1 + \lambda_2 - 2 \lambda_{345}^+ ) s_\beta^2 c_\beta^2} 
		& \displaystyle{- \frac{1 }{2 } \lambda_5^I c_{2\beta} } \\[10pt]
- \lambda_5^I s_\beta c_\beta 
	& \displaystyle{- \frac{1 }{2 } \lambda_5^I c_{2\beta} }
		& \displaystyle{\frac{ M^2 }{ v^2 } - \lambda_5^R} \\
\end{pmatrix}, 
\end{equation}
where $s_\beta = \sin \beta$, $c_\beta = \cos \beta$ and $c_{2\beta} = \cos 2 \beta$. 
According to Refs.~\cite{Kanemura:2015ska}, we use the following parametrization;
\begin{align}
& (M_h^2)_{11} =  \tilde{m}_h^2 s_{\beta- \tilde{\alpha}}^2 + \tilde{m}_H^2 c^2_{\beta - \tilde{\alpha}}, \\[5pt]
& (M_h^2)_{12} = (M_h^2)_{21} = (\tilde{m}_h^2 - \tilde{m}_H^2)s_{\beta - \tilde{\alpha}} c_{\beta - \tilde{\alpha}}, \\[5pt]
& (M_h^2)_{22} = \tilde{m}_h^2 c_{\beta- \tilde{\alpha}}^2 + \tilde{m}_H^2 s^2_{\beta - \tilde{\alpha}}, \\[5pt]
& (M_h^2)_{33} = \tilde{m}_A^2, 
\end{align}
where $s_{\beta - \tilde{\alpha}} = \sin(\beta - \tilde{\alpha})$ and $c_{\beta - \tilde{\alpha}} = \cos (\beta - \tilde{\alpha})$. 
We note that $\tilde{m}_h$, $\tilde{m}_H$, and $\tilde{m}_A$ are the mass eigenvalues in the $CP$-conserving limit ($\lambda_5^I = 0$), where $h_3^\prime$ is the $CP$-odd mass eigenstate, and $h_1^\prime$ and $h_2^\prime$ are mixed by the mixing angle $\beta - \tilde{\alpha}$. 

On the other hand, in the $CP$-violating case, $\lambda_5^I$ generates additional mixings, and $\tilde{m}_h$, $\tilde{m}_H$, and $\tilde{m}_A$ are no longer mass eigenvalues. 
We define the mass eigenvalues and eigenstates as follows; 
\begin{align}
& \mathrm{diag}(m_{H_1}^2, m_{H_2}^2, m_{H_3}^2) = R^\mathrm{T} M_h^2 R, \quad 
(H_1, H_2, H_3) = (h_1^\prime , h_2^\prime, h_3^\prime ) R, 
\end{align}
where the $R$ is an appropriate orthogonal matrix. 
Since it is difficult to solve the matrix $R$ analytically, 
we use $\tilde{m}_H$ and $\tilde{m}_A$ as input parameters, not $m_{H_2}$ and $m_{H_3}$. 
On the other hand, $\tilde{m}_h$ is determined so that $m_{H_1} = 125~\mathrm{GeV}$, and $H_1$ is the SM Higgs boson.
Then, the mass matrix $M_h^2$ includes five free parameters: $\tilde{m}_H$, $\tilde{m}_A$, $\tan \beta$, $s_{\beta - \tilde{\alpha}}$, and $\lambda_5^I$. 

Once the mass matrix is fixed, we can find the mass eigenvalues and the mixing matrix $R$. 
In addition, we can determine scalar couplings as follows~\cite{Kanemura:2015ska}; 
\begin{align}
& \lambda_1 v^2 =  (M_h^2)_{11} + \bigl((M_h^2)_{22} - M^2\bigr)t_\beta^2 - 2 (M_h^2)_{12} t_\beta, \\
& \lambda_2 v^2 = (M_h^2)_{11} + \bigl((M_h^2)_{22} - M^2\bigr) t_\beta^{-2} + 2 (M_h^2)_{12} t_\beta^{-1}, \\
& \lambda_3 v^2 = (M_h^2)_{11} - \bigl((M_h^2)_{22} + M^2\bigr) \cot 2\beta + 2 m_{H^\pm}^2, \\
& \lambda_4v^2 = M^2 + (M_h^2)_{33} - 2 m_{H^\pm}^2, \\
& \lambda_5^R v^2 = M^2 - (M_h^2)_{33}, 
\end{align}
where $t_\beta = \tan \beta$. 
Consequently, $V_\mathrm{THDM}$ has the following free paraemeters; 
\begin{equation}
M, \quad \tilde{m}_H, \quad \tilde{m}_A, \quad m_{H^\pm}, \quad
\tan \beta, \quad \sin (\beta - \tilde{\alpha}), \quad \theta_5. 
\end{equation}

Next, we consider the $Z_2$-odd scalar bosons. 
The masses of $S^\pm$ and $\eta$ are given by 
\begin{align}
& m_S^2 = \mu_S^2 + \frac{ v^2 }{ 2 } (\rho_1 c_\beta^2 + \rho_2 s_\beta^2), \\
& m_\eta^2 = \mu_\eta^2 + \frac{ v^2 }{ 2 } (\sigma_1 c_\beta^2 + \sigma_2 s_\beta^2). 
\end{align}
We use $m_S$ and $m_\eta$ as input parameters instead of $\mu_S$ and $\mu_\eta$. 
Thus, the remaining part of the scalar potential is determined by 
\begin{equation}
m_S, \quad m_\eta, \quad \lambda_S, \quad \lambda_\eta, \quad \xi, \quad \rho_a, \quad \sigma_a, \quad \kappa, 
\end{equation}
where $a=1,2$. 
As a result, the Higgs potential has five dimensionful parameters, two mixing angles, six dimensionless couplings and one $CP$-violating phase. 

Next, we consider the Yukawa interactions. Because of the $\tilde{Z}_2$ parity of the SM fermions shown in Table~\ref{table: matter fields}, the Yukawa interactions between the Higgs doublets and the SM fermions are given by 
\begin{align}
\mathcal{L} = & - (y_u)_{ij} \overline{Q_L^{\prime i}} \tilde{\phi}_2 u_R{\prime j} - (y_d)_{ij} \overline{Q_L^{\prime i}} \phi_2 d_R{\prime j}  - (y_\ell)_{ij} \overline{L_L^{\prime i}} \phi_1 \ell_R{\prime j} + \mathrm{h.c.}, 
\end{align}
which is the same as in the Type-X THDM~\cite{Logan:2009uf, Su:2009fz, Grossman:1994jb, Aoki:2009ha, Barger:1989fj}. 
After the bi-unitary transformations of the SM fermions, the Yukawa interactions are given by
\begin{align}
\mathcal{L} = - \sum_i \biggl\{ & \frac{ m_{u^i} }{ v } \overline{u^i} \Bigl( H_1 R_{1i} + H_2 \frac{ R_{2i} }{ t_\beta }  - H_3 \frac{ i R_{3i} }{ t_\beta }  \gamma_5 \Bigr) u^i 
	+ \frac{ m_{d^i} }{ v } \overline{d^i} \Bigl( H_1 R_{1i} +  H_2 \frac{ R_{2i} }{ t_\beta } + H_3 \frac{ iR_{3i} }{ t_\beta }  \gamma_5 \Bigr) d^i\nonumber \\
		& + \frac{ m_{\ell^i} }{ v } \overline{\ell^i} \Bigl( H_1 R_{1i} + H_2 R_{2i} t_\beta + i H_3 R_{3i} t_\beta \gamma_5 \Bigr) \ell^i \biggr\}
\nonumber \\
- \frac{ \sqrt{2} }{ v } \sum_{ij}\biggl\{ & \frac{ V_{ij} }{ t_\beta } \overline{u^i}\Bigl( m_{d^j} P_R - m_{u^i} P_L \Bigr) d^j H^+ 
	+ t_\beta m_{\ell^i} \overline{\nu^i} P_R \ell^i H^+ + \mathrm{h.c.} \biggr\}.
\end{align} 

The model has an additional Yukawa coupling among $S^\pm$, $N^\alpha_R$, and $\ell_R^i$ as follows; 
\begin{equation}
\mathcal{L} = - \sum_{i, \alpha} h_i^\alpha \overline{(N_R^\alpha)^c} \ell_R^i S^+ + \mathrm{h.c.} 
= - \sum_{i, \alpha} h_i^\alpha \overline{(N^\alpha)^c} P_R \ell^i S^+ + \mathrm{h.c.}.
\end{equation}
We note that the coupling $h_i^\alpha$ is defined for the mass eigenstates of the charged leptons (without ${}^\prime$). 
The matrix $h$ includes nine complex parameres. 
Three of them can be real by rephasing the lepton fields. 
This degree of freedom is the same as those to reduce the three unphysical phases in the (Pontecorvo-Maki-Nakagawa-Sakata) matrix $U$~\cite{ref: Pontecorvo, Maki:1962mu}. 
We have a detailed discussion about in Sec.~\ref{subsec: NuMass}. 

\subsection{Theoretical constraints}

Here, we discuss theoretical constraints on the Higgs potential, {\it i.e.}, the vacuum stability, the triviality bound, and the perturbative unitarity. 
To ensure that the local minimum is the stable vacuum of theory, a necessary condition is normally imposed such that the Higgs potential does not fall into negative infinity with a large field configuration along any direction of the scalar fields. It requires some combinations of quartic scalar couplings to be positive. 
According to Ref.~\cite{Aoki:2011zg}, we employ the following conditions;  
\begin{align}
\label{eq: VS_cond1}
& \lambda_1 > 0, \quad \lambda_2 > 0, \quad \lambda_S > 0, \quad \lambda_\eta > 0, \\
\label{eq: VS_cond2}
& \sqrt{\lambda_1 \lambda_2} + \lambda_3 + \mathrm{min}(0, \lambda_4 - |\lambda_5|) > 0, \\
\label{eq: VS_cond3}
& \sqrt{\lambda_1 \lambda_S /2 } + \rho_1 > 0, \quad 
	\sqrt{\lambda_1 \lambda_\eta/3 } + \sigma_1 > 0, 
		\sqrt{\lambda_2 \lambda_S/2 } + \rho_2 > 0, \\
\label{eq: VS_cond4}
& \sqrt{\lambda_2 \lambda_\eta /3 } > 0, \quad 
	\sqrt{\lambda_S \lambda_\eta/6} + \xi > 0, \\
\label{eq: VS_cond5}
& 2 \lambda_1 + 2 \lambda_2 + 4 \lambda_3 + 4 \rho_1 + 4 \rho_2 + \lambda_S +4 \sigma_1 + 4 \sigma_2 + \frac{ 2 }{ 3 }\lambda_\eta + 4 \xi - 16 \sqrt{2}|\kappa| > 0. 
\end{align}
In Eq.~(\ref{eq: VS_cond2}), we use $|\lambda_5|$, not $\lambda_5^R$ shown in Ref.~\cite{Aoki:2011zg}, to include the effect of the $CP$-violation in $\lambda_5$ according to discussion in the THDMs~\cite{Deshpande:1977rw, Klimenko:1984qx, Nie:1998yn, Ferreira:2004yd}. 

The triviality bound requires that all the running couplings do not infinitely blow up nor fall down below the cut-off scale $\Lambda$. 
Demanding $\Lambda$ be higher than a certain value, we find constraints on the scalar couplings. 
Ref.~\cite{Aoki:2011zg} studies the triviality bound in the AKS model by using the mass-independent renormalization group equation (RGE), where $\Lambda$ strongly depends on a scale from which the new physics effect is included in the running (the threshold effect). 

On the other hand, Ref.~\cite{Kanemura:2023wap} used the mass-dependent renormalization group equation to study the triviality bound in some new physics models with extended Higgs sectors. 
Then, the new physics effect naturally decouples in the running at low energies without setting an artificial threshold. 
They also compare the result with that in the mass-independent RGE and study how the threshold effect changes $\Lambda$. 
According to Refs.~\cite{Aoki:2011zg, Kanemura:2023wap}, we require the size of all the scalar couplings to be less than three to make $\Lambda$  higher than $10~\mathrm{TeV}$. 

Demanding the perturbative unitarity constrains the quartic scalar couplings so that the s-wave amplitude $a_0$ satisfies $|a_0| < c$ for all two-to-two-body scalar boson scattering~\cite{Lee:1977eg}, where $c = 1$~\cite{Lee:1977eg} or $1/2$~\cite{Luscher:1988gc}. 
It leads to an upper bound for the size of the scalar couplings, which implies that masses of new particles are not far from their invariant mass parameters, {\it i.e.}, $M^2$, $\mu_S^2$, and $\mu_\eta^2$~\cite{Kanemura:1993hm, Ginzburg:2005dt, Akeroyd:2000wc, Kanemura:2015ska}. 
We use the result in Ref.~\cite{Kanemura:2015ska} and requires the following conditions for the quartic couplings in $V_\mathrm{THDM}$; 
\begin{align}
\label{eq: PU_cond1}
& 3 (\lambda_1 + \lambda_2) \pm \sqrt{9 (\lambda_1 - \lambda_2)^2 + 4 ( 2 \lambda_3 + \lambda_4)^2 } < 16 \pi, \\ 
\label{eq: PU_cond2}
&  2 \lambda_3 + 2 \lambda_4 \pm 3 \lambda_5^R  < 16 \pi ,  \\
\label{eq: PU_cond3}
&\lambda_1 + \lambda_2 \pm \sqrt{(\lambda_1 - \lambda_2)^2 + 4  \lambda_4^2 } < 16 \pi, \\
\label{eq: PU_cond4}
& 2 \lambda_3 \pm 2 |\lambda_5|  < 16 \pi, \\
\label{eq: PU_cond5}
& \lambda_1 + \lambda_2 \pm \sqrt{(\lambda_1 - \lambda_2)^2 + 4 |\lambda_5|^2 }  < 16 \pi, \\
\label{eq: PU_cond6}
& \lambda_4 + \lambda_3 < 16 \pi,
\end{align}
where we employ $c = 1/2$. 
For other coupling in the Higgs potential, we simply require them to be less than three to satisfy the perturbative unitarity.

\subsection{Experimental constraints}

In this subsection, we describe experimental constraints on the new particles. 
For simplicity, we neglect the effect of the $CP$-violating couplings.
We assume that $m_{N^\alpha} = \mathcal{O}(1)~\mathrm{TeV}$, and the masses of the other particles are $\mathcal{O}(100)~\mathrm{GeV}$. 
Then, $\eta$ is the dark matter particle. 
We use the theoretical results in Refs.~\cite{Aiko:2020ksl, Arbey:2017gmh} and~\cite{Enomoto:2015wbn, Arbey:2017gmh, Haller:2018nnx} to discuss constraints on the additional Higgs bosons from the collider experiments and flavor experiments, respectively. 

$H^\pm$ has the Type-X (lepton specific) Yukawa coupling with the SM fermions \cite{Logan:2009uf, Su:2009fz, Grossman:1994jb, Aoki:2009ha, Barger:1989fj}. 
The LEP result gives the lower limit $m_{H^\pm} \gtrsim 80~\mathrm{GeV}$, which is almost independent of $\tan \beta$~\cite{ALEPH:2013htx}. 
At LHC, the main production channel for light $H^\pm$ is the top quark decay in the case that $m_{H^\pm} < m_t - m_b \simeq 170~\mathrm{GeV}$. 
The strong constraint comes from $H^\pm \to \tau \nu$ search, and it is $\tan \beta \gtrsim 10$~\cite{Arbey:2017gmh}. 
If $H^\pm$ is larger than $170~\mathrm{GeV}$, it is produced by the associated production $gg \to t b H^\pm$, and $H^\pm$ predominantly decays into $\tau \nu$ or $tb$. 
The $tb$ search gives a stronger constraint; $\tan \beta \gtrsim 2$ (1.5) for $m_{H^\pm} = 200$ (400) GeV~\cite{Aiko:2020ksl}. 

$H^\pm$ is also constrained by flavor observables~\cite{Haller:2018nnx, Enomoto:2015wbn, Arbey:2017gmh} such as $B \to X_s \gamma$~\cite{HFLAV:2019otj}, $B_s$($B_d$)  $\to \mu \mu$~\cite{ATLAS:2018cur}. 
In the current case, the strongest constraint comes from $B_d \to \mu \mu$~\cite{Haller:2018nnx, Enomoto:2015wbn, Arbey:2017gmh}, and it is $\tan \beta \gtrsim 3$ and $2$ for $m_{H^\pm} = 100$ and $400~\mathrm{GeV}$, respectively. 
Thus, the flavor experiment gives a stronger constraint than the collider experiments for $H^\pm$ heavier than $170~\mathrm{GeV}$~\cite{Arbey:2017gmh}. 

$H_2$ and $H_3$ are predominantly produced via the gluon fusion $gg \to H_{2,3}$ at LHC because we consider the type-X Yukawa interaction~\cite{Aiko:2020ksl}. 
The constraint strongly depends on their mixing with the SM Higgs boson $H_1$. 
In the case without mixing (the alignment case), they are constrained by the searches for $H_{2,3} \to \tau^+ \tau^-$ and $H_{2,3} \to t \bar{t}$. 
If they mix with $H_1$, in addition to these two processes, constraints from $H_2 \to ZZ$, $H_2 \to H_1 H_1$ and $H_3 \to Z H_1$ are important. 
All of these constraints are relaxed by considering large $\tan \beta$ because the production cross section is proportional to $\cot \beta$. 
Here, we employ $\tan \beta \gtrsim 15$ as a constraint for $H_{2,3}$ with the mass of a few hundred GeV. 

$S^\pm$ are produced by the pair production via $Z$ and $\gamma$. 
There are two decay channels $S^\pm \to N^\alpha \ell_i^\pm$ and $S^\pm \to H^\pm \eta$. 
In the benchmark scenarios shown in Sec.~\ref{sec: benchmark scenarios}, we consider $m_S = 325~\mathrm{GeV}$, $m_{H^\pm} \simeq 390~\mathrm{GeV}$, and $m_{N^\alpha} \simeq \mathcal{O}(1)~\mathrm{TeV}$. 
Thus, both the two-body decays are kinematically forbidden. 
Because of the mass hierarchy, $S^\pm \to H^{\pm \ast} \eta$ is the main decay mode unless $\kappa$ is too small. 
If $m_S - m_\eta > m_t + m_b$, $S^\pm$ predominantly decay into $t b \eta$. 
Scenario I shown in Sec.~\ref{sec: benchmark scenarios} is in this case. 
The signal of the pair production of $S^\pm$ is $t \bar{t} b \bar{b} \cancel{E}$, where $\eta$ is the missing energy. 
We expect that the constraint on this channel is weak enough at both $e^+e^-$ colliders and hadron colliders. 
If $m_S - m_\eta < m_t + m_b$, $S^\pm$ decays into $\tau \nu \eta$ unless the mass difference is too small. 
Scenario II shown in Sec.~\ref{sec: benchmark scenarios} is in this case. 
Then, the pair production of $S^\pm$ leads to the final state $\tau \bar{\tau} \cancel{E}$. 
Such a signal is constrained by the stau search in supersymmetric extensions of the SM~\cite{ATLAS:2023djh}. 
In Scenario II, we take $m_{S^\pm} = 325~\mathrm{GeV}$. 
Then, the invariant mass of the missing energy has to be larger than about $200~\mathrm{GeV}$ to avoid the costraint~\cite{ATLAS:2023djh}. 
Since $m_\eta = 250~\mathrm{GeV}$ in Scenario II, we expect that the constraint can be avoided. 

$N^\alpha$ couple only the charged leptons. 
Thus, the production at hadron colliders is suppressed. 
At $e^+ e^-$ colliders, a pair of them can be produced via the t--channel diagram mediated by $S^\pm$. 
In the following, we are interested in $N^\alpha$ with the mass $m_{N^\alpha} \simeq \mathcal{O}(1)~\mathrm{TeV}$. 
Then, $N^\alpha$ are not constrained by the collider experiments so far. 

The mass of $S^\pm$ and $N^\alpha$ and the size of the Yukawa coupling $h_i^\alpha$ are strongly constrained by the current measurements of LFV processes. 
We discuss this constraint in detail in Sec.~\ref{sec: benchmark scenarios}. 

\section{Phenomenology}
\label{sec: pheno}

In this section, we discuss the phenomenology of the model, neutrino mass, lepton flavor violating decays, and dark matter. 
Although these are already studied in previous works~\cite{Aoki:2008av, Aoki:2011zg, Aoki:2022bkg}, 
there are several updates about the parametrization of $h_i^\alpha$ and the dark matter phenomenology. 

\subsection{Neutrino mass}
\label{subsec: NuMass}

\begin{figure}[t]
\begin{center}
\includegraphics[width=0.7\textwidth]{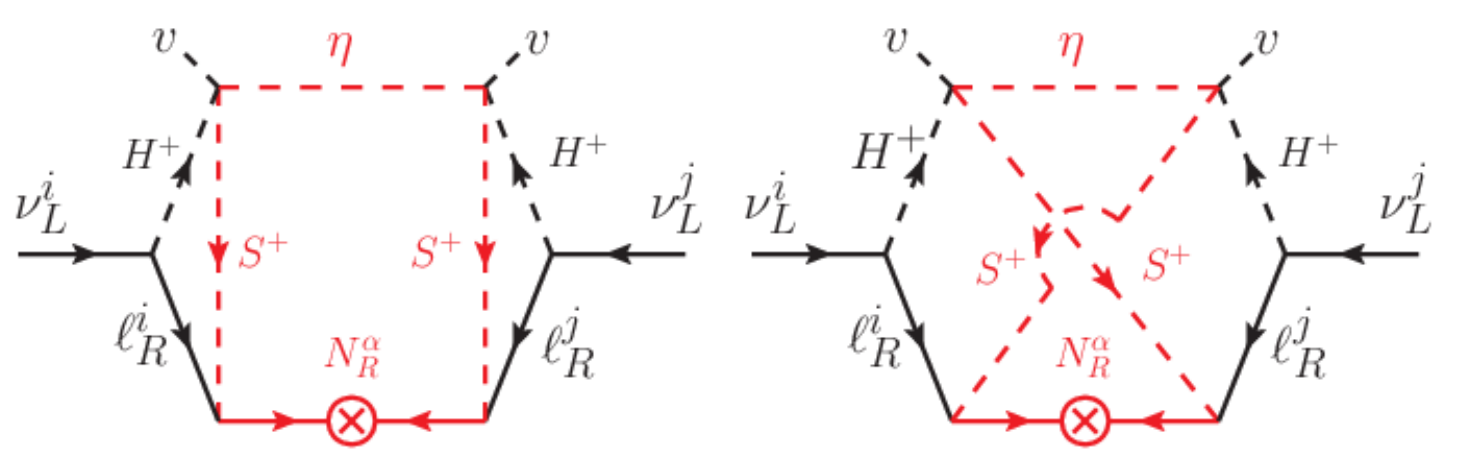}
\caption{Feynman diagram for neutrino mass}
\label{fig: neutrino mass}
\end{center}
\end{figure}

In the model, Majorana masses of neutrinos are generated by the three-loop diagram in Fig.~\ref{fig: neutrino mass}. 
The mass matrix is given by
\beq
\label{eq: neutrino mass}
(M_\nu)_{ij} = \frac{ (\kappa \tan \beta)^2 m_{\ell^i}^{} m_{\ell^j}^{} }{ (16\pi^2)^3 }
h_i^\alpha h_j^\alpha m_{N^\alpha}^{} \Bigl( F_{1\alpha} + F_{2\alpha} \Bigr),  
\eeq
where $F_{1\alpha}$ and $F_{2\alpha}$ are loop functions corresponding to the left and right diagrams in Fig.~\ref{fig: neutrino mass}, respectively. 
They are given by  
\beq
\label{eq: loop_functions_F_for_nu_mass}
F_{n \alpha} = \int_0^1 \tilde{\mathrm{d}}^4 x \int_0^\infty \mathrm{d}u \int_0^\infty \mathrm{d} v \frac{ 8 \sqrt{ u v } \tilde{F}(a_n, b_n) }{ (u + m_{H^\pm}^2 ) ( v + m_{H^\pm}^2 ) }, 
\eeq
where $n=1,2$, and 
\begin{equation}
\int \mathrm{\tilde{d}}^4x = \int_0^1 \mathrm{d}x \, \mathrm{d}y \, \mathrm{d}z \, \mathrm{d}\omega\ \delta(x+y+z+\omega -1). 
\end{equation}
The function $\tilde{F}$ and the variables $a_n$ and $b_n$ are defined as
\begin{align}
\label{eq: Ftilde}
& \tilde{F}(x, y) = \frac{ 1 }{ y^3 } \Bigl( \sqrt{x^2 - y^2 } + \frac{ x^2 }{ \sqrt{x^2 - y^2 } } - 2 x \Bigr), \\
& a_1 = (y + z ) m_S^2 + x m_{N^\alpha}^2 + \omega m_\eta^2 + z (1-z) u + y (1-y) v, \\
& a_2 = (y + z) m_S^2 + x m_{N^\alpha}^2 + \omega m_\eta^2 + (y + \omega ) (x + z) u + (x + y )( z + \omega ) v, \\
& b_1 = 2 y z \sqrt{uv}, \\
& b_2 = 2 (y z - x \omega ) \sqrt{ u v}. 
\end{align}
See Ref.~\cite{Aoki:2022bkg} for the derivation of the formulas. 

Next, we consider the parametrization of $h_i^\alpha$ to explain the neutrino oscillation data. 
It is also convenient to discuss the $CP$-violating phases in $h_i^\alpha$. 
The formula in Eq.~(\ref{eq: neutrino mass}) can be written by the following simple matrix form;
\begin{equation} 
\label{eq: simplified_neutrino_mass}
M_\nu= Y^\mathrm{T} \Lambda^{-1} Y, 
\end{equation}
where $Y$ and $\Lambda$ are $3\times3$ matrices defined as
\begin{equation}
Y_{\alpha \ell} = \frac{ |\kappa \tan \beta| }{ (16 \pi^2)^{3/2} } h^\alpha_i m_{\ell^i}, 
\quad 
\Lambda_{\alpha \beta} = \Lambda_\alpha \delta_{\alpha \beta}, 
\end{equation}
with 
\begin{equation}
\Lambda_\alpha = \bigl\{ m_{N^\alpha} (F_{1\alpha} + F_{2\alpha} ) \bigr\}^{-1}.
\end{equation} 
We note that $\Lambda_\alpha$ has a mass dimension one and is a typical new physics scale for the neutrino mass. 
$M_\nu$ is diagonalized by the PMNS matrix $U$, 
\begin{equation}
D \equiv \mathrm{diag}(m_{\nu^1}, m_{\nu^2}, m_{\nu^3}) = U^\mathrm{T} Y^\mathrm{T} \Lambda^{-1} Y U. 
\end{equation}

Since the structure of the neutrino mass matrix in Eq.~(\ref{eq: simplified_neutrino_mass}) is the same as that in the type-I seesaw mechanism~\cite{ref: Type-I seesaw}, 
we can use the Casas-Ibarra parametrization~\cite{Casas:2001sr} to study parameter regions to explain the neutrino oscillation data. 
Then, the matrix $Y$ is given by $Y = \sqrt{|\Lambda|} O \sqrt{D} U^\dagger$, where the matrix $O$ is a complex orthogonal matrix whose elements are independent of the neutrino oscillation data, $\sqrt{|\Lambda|} = \mathrm{diag}(\sqrt{|\lambda_1|}, \sqrt{|\lambda_2|}, \sqrt{|\lambda_3|})$ and $\sqrt{D} = \mathrm{diag}(\sqrt{m_{\nu^1}}, \sqrt{m_{\nu^2}}, \sqrt{m_{\nu^3}})$. 
Consequently, the Yukawa matrix $h$ has to be the following form to explain the neutrino oscillation data; 
\begin{equation}
\label{eq: CI parametrization}
h = \frac{ (16 \pi^2)^{3/2} }{ |\kappa \tan \beta| } \sqrt{|\Lambda|} O \sqrt{D} U^\dagger M_L^{-1}, 
\end{equation}
where $M_L = \mathrm{diag} (m_e, m_\mu, m_\tau)$. 

The matrix $O$ consists of three complex angles $z_1$, $z_2$, and $z_3$ as 
\begin{equation}
O = 
\begin{pmatrix}
1 & 0 & 0 \\
0 & c_{z_1} & s_{z_1} \\
0 & -s_{z_1} & c_{z_1} \\
\end{pmatrix}
\begin{pmatrix}
c_{z_2} & 0 & s_{z_2} \\
0 & 1 & 0 \\
-s_{z_2} & 0 & c_{z_2} \\
\end{pmatrix}
\begin{pmatrix}
c_{z_3} & s_{z_3} & 0 \\
-s_{z_3} & c_{z_3} & 0 \\
0 & 0 & 1 \\
\end{pmatrix}, 
\end{equation}
where $c_X = \cos X$ and $s_X = \sin X$ for any $X$. 
In this parametrization, the imaginary parts of $z_i$ are unbounded. 
Instead, we employ the following formula with six real angles; 
\begin{align}
O  = & \frac{ 1 }{ c_{\psi_1} c_{\psi_2} c_{\psi_3} }
\begin{pmatrix}
1 & 0 & 0 \\
0 & c_{\gamma_1} & s_{\gamma_1} \\
0 & -s_{\gamma_1} & c_{\gamma_1} \\
\end{pmatrix}
\begin{pmatrix}
c_{\psi_1} & 0 & 0 \\
0 & 1 & i s_{\psi_1} \\
0 & - i s_{\psi_1} & 1 \\
\end{pmatrix}
\begin{pmatrix}
c_{\gamma_2} & 0 & s_{\gamma_2} \\
0 & 1 & 0 \\
-s_{\gamma_2} & 0 & c_{\gamma_2} \\
\end{pmatrix}
\nonumber \\
& \times 
\begin{pmatrix}
1 & 0 & is_{\psi_2} \\
0 & c_{\psi_2} & 0 \\
-is_{\psi_2} & 0 & 1 \\
\end{pmatrix}
\begin{pmatrix}
c_{\gamma_3} & s_{\gamma_3} & 0 \\
-s_{\gamma_3} & c_{\gamma_3} & 0 \\
0 & 0 & 1 \\
\end{pmatrix}
\begin{pmatrix}
1 & i s_{\psi_3} & 0 \\
-i s_{\psi_3} & 1 & 0 \\
0 & 0 & c_{\psi_3} \\
\end{pmatrix}, 
\end{align}
where we used $\gamma_i = \mathrm{Re}[z_i]$, $\cosh (\mathrm{Im}[z_i]) = \cos^{-1} \psi_i$, and $\sinh (\mathrm{Im}[z_i]) = \tan \psi_i$ for $i=1$, 2, 3. 
Then, $\gamma_i$ and $\psi_i$ are bounded in the regions $[0, 2\pi)$ and $[-\pi/2, \pi/2]$, respectively. 
In the limit of $\psi_i \to 0$, the matrix $O$ is a real orthogonal matrix. 

We revisit the $CP$-violating phases in $h_i^\alpha$ by using Eq.~(\ref{eq: CI parametrization}). 
As explained in Sec.~\ref{subsec: Lagrangian}, the matrix $h$ generally includes nine complex phases, but three of them can be zero by rephasing the lepton fields. 
By using Eq.~(\ref{eq: CI parametrization}), we can understand the fact that this degree of freedom is the same as that to reduce unphysical phases from the PMNS matrix. 
The remaining $CP$-violating phases in $U_{ia}$ are the Dirac $CP$ phase $\delta$ and two Majorana $CP$ phases $\alpha_1$ and $\alpha_2$.  
As a result, there are six $CP$-violating phases in $h_i^\alpha$; $\delta$, $\alpha_1$ and $\alpha_2$ in $U_{ia}$ and $\psi_1$, $\psi_2$ and $\psi_3$ in $O_{a\alpha}$.

\subsection{Lepton flavor violation}
\label{subsec: LFV}

In the model, the lepton flavor violating decay $\ell^i \to \ell^j \gamma$ occurs at one-loop level via the Yukawa coupling $h_i^\alpha$. 
The branching ratio is given by 
\beq
\label{eq: LFVBr1}
\frac{ \mathrm{Br} ( \ell^i \to \ell^j \gamma ) }{ \mathrm{Br}(\ell^i \to \ell^j \nu_{\ell^i} \overline{\nu}_{\ell^j} ) }
= \frac{ 3 \alpha }{ 64 \pi G_F^2 }
\left| \sum_\alpha \frac{ (h^\alpha_j)^\ast h^\alpha_i f(r_\alpha) }{ m_S^2 }  \right|^2, 
\eeq
where $r_\alpha = m_{N^\alpha}^2 / m_S^2$, and 
\beq
f(x) = \frac{ 1 - 6 x + 3 x^2 + 2 x^3 - 6 x^2 \log x }{ 6 (-1 + x )^4 }. 
\eeq

Three body decays $\ell^i \to \ell^j \ell^k \overline{\ell^m}$ also occurs at one-loop level. 
There are two kinds of Feynman diagrams. 
One is the box diagram, and the other is the photon penguin diagram. 
As discussed below, we are interested in parameter regions where the size of $h_i^\alpha$ is relatively large. 
In such a case, the box diagram gives the dominant contribution. 
Thus, we neglect the contribution from the photon penguin diagram in the following. 

The branching ratio for $\ell^i \to \ell^j \ell^k \overline{\ell^m}$ is given by 
\beq
\label{eq: LFVBr2}
\frac{ \mathrm{Br}( \ell^i \to \ell^j \ell^k \overline{\ell^m} ) }
{\mathrm{Br} (\ell^i \to \ell^j \nu_{\ell^i} \overline{\nu}_{\ell^j} )}%
=
\frac{ 2 - \delta_{jk} }{ 4096 \pi^4 G_F^2 }
\bigl| A_{i,m}^{jk}  \bigr|^2, 
\eeq
where $A_{i,m}^{jk} = B_{i,m}^{jk} + B_{i,m}^{kj} + C_{i,m}^{jk} + C_{i,m}^{kj}$, and the tensors $B_{i,m}^{jk}$ and $C_{i,m}^{jk}$ are defined as
\bal
& B_{i,m}^{jk} = 
	\frac{ h^\alpha_i h^\beta_m (h^\alpha_j h^\beta_k)^\ast }{ m_S^2 }
	\left(\frac{g_2\left(r_\alpha \right) - g_2\left(r_\beta\right) }{ r_\alpha - r_\beta } \right), \\[7pt]
& C_{i,m}^{jk} = 
	\frac{ h^\alpha_i h^\alpha_m (h^\beta_j h^\beta_k)^\ast  }{ m_S^2 }
	\sqrt{r_\alpha r_\beta} \left( \frac{ g_1(r_\alpha) - g_1(r_\beta) }{ r_\alpha - r_\beta } \right), 
\eal
by using the function $g_k(x)$ ($k=1,2$) given by
\beq
g_k(x) = \frac{ 1 }{ 2^k } 
	\Bigl\{ 
		\frac{ x^k }{ (1 - x)^2 } \log x
		+ \frac{ 1 }{ 1 - x }
	\Bigr\}. 
\eeq

\subsection{dark matter}
\label{subsec: DM}

In the model, there are two dark matter candidates: the real scalar boson $\eta$ and the lightest of the Majorana fermion $N^\alpha$. 
Since $N^\alpha$ is favored to be heavier than $S^\pm$ to avoid the constraint from the lepton flavor violating decays as discussed in Sec.~\ref{sec: benchmark scenarios}, 
we consider the case that $\eta$ is the dark matter particle. 

The relic abundance of dark matter is determined by the thermally averaged cross section $\left< \sigma v \right>$, which is a function of the temperature $T$ evaluated by 
\beq
\left< \sigma v \right> = 
		\frac{ 2 x }{ K_2(x)^2 } 
		\int_1^\infty  \mathrm{d}y\  \sigma_{\text{\sf all}} vy \sqrt{y-1} K_1(2 x \sqrt{y}), 
\eeq
where $x = m_\eta / T$, $y = s/(4m_\eta^2)$ with the Mandelstam variable $s$, $\sigma_{\text{\sf all}}$ is the sum of all pair-annihilation processes of $\eta$, and $K_1(x)$ and $K_2(x)$ are the modified Bessel functions of the second kind. 
Here, we consider pair-annihilations of $\eta$ into a pair of the SM fermions ($f^i \bar{f^i}$), the weak boson's pair ($W^+W^-$ and $ZZ$), diphoton ($\gamma\gamma$) and a pair of the additional Higgs bosons ($H_i H_j$ and $H^+H^-$). The former three have already been investigated in the previous works~\cite{Aoki:2008av, Aoki:2011zg, Aoki:2022bkg}, whereas the last one is a new update in this paper. 
We show the formulas for $\sigma v$ of these annihilation channels in the following. 

The fermionic channel $\eta \eta \to f^i \bar{f^i}$, where $f=u$, $d$, $\ell$ and $i = 1,2,3$, is mediated by $H_a^{(\ast)}$ and is generated by the Yukawa interaction and the following three-point scalar interactions; 
\begin{equation}
\mathcal{L} = - \sum_{a} \frac{ v }{ 2 } \lambda_{\eta \eta a} H_a \eta \eta, 
\end{equation}
where 
\begin{equation}
\lambda_{\eta \eta a} =  (\sigma_1 c_\beta^2 + \sigma_2 s_\beta^2 ) R_{1a} - (\sigma_1 - \sigma_2 ) s_\beta c_\beta R_{2a}. 
\end{equation}
We note that there is no term proportional to $R_{3a}$ because a pair of $\eta$ does not couple to the $CP$-odd component $h_3^\prime$ at tree level. 
This is a different point from Ref.~\cite{Aoki:2022bkg} which is also a $CP$-violating extension of the original AKS model. 
The cross section is given by
\begin{align}
(\sigma v)_{f^i}
= & \frac{ N_c^f  m_{f^i}^2 }{ 4 \pi }
\Biggl\{
	 h(m_{f^i})^3
	\left| \sum_{a=1}^3 \frac{ \lambda_{\eta \eta a } Z_{f}^a}{ s - m_{H_a}^2 + i m_{H_a}^{} \Gamma_{H_a} } \right|^2
	+ h(m_{f^i}) \left| \sum_{a=1}^3 \frac{ \lambda_{\eta \eta a } X_{f}^a}{ s - m_{H_a}^2 + i m_{H_a}^{} \Gamma_{H_a} } \right|^2
\Biggr\}, 
\end{align}
where $N_c^f$ is the color factor of the fermion, $h(x) = \sqrt{1-4x^2/s}$, $\Gamma_{H_a}$ is the decay rate of $H_a$, and $Z_f^a$ and $X_f^a$ is the coupling constant of the scalar and pseudos scalar interaction between $H_a$ and the SM fermions given by 
\begin{align}
& Z_u^a = R_{1a} + R_{2a} \cot \beta, \quad X_u^a = - R_{3a} \cot \beta, \\
& Z_d^a = R_{1a} + R_{2a} \cot \beta, \quad X_d^a = R_{3a} \cot \beta, \\
& Z_\ell^a = R_{1a} + R_{2a} \tan \beta, \quad X_\ell^a = R_{3a} \tan \beta. 
\end{align}

The annihilations into the weak bosons $\eta \eta \to W^+ W^-/ZZ$ are also mediated by $H_a^{(\ast)}$ and are generated by the weak interaction and the above $\eta \eta H_a$ coupling. 
The cross section is given by
\begin{align}
(\sigma v)_{V} = & \frac{ S_V (\sigma_1 c_\beta^2 + \sigma_2 s_\beta^2)^2 }{ 16 \pi s } h(m_V^{}) \Bigl\{ (s - 2 m_V^2 )^2 + 8 m_V^4 \Bigr\} 
\biggl| \sum_{a=1}^3 \frac{ R_{1a} }{ (s - m_{H_a}^2)^2 + m_{H_a}^2 \Gamma_{H_a}^2 } \biggr|^2, 
\end{align}
where $V = W$ or $Z$, and $S_V$ is two (one) for $V = W$ ($Z$). 
We note that there is no term proportional to $R_{2a}$ or $R_{3a}$ because $\Phi_2$ does not obtain the VEV, and $h_2^\prime$ and $h_3^\prime$ do not have three-point interaction $h_a^\prime VV$. 

The diphoton annihilation is generated by two kinds of one-loop Feynman diagrams~\cite{Aoki:2008av, Aoki:2022bkg}. 
The first one includes circle and triangle diagrams constituted by a loop of $H^\pm$ or $S^\pm$ like the scalar contribution in the Higgs diphoton decay with the replacement of the external Higgs boson into two external lines $\eta \eta$.  
The second one includes box diagrams constituted by both $H^\pm$ and $S^\pm$. 
The cross section is given by 
\begin{align}
(\sigma v)_\gamma = & \frac{ \alpha^2 }{ 256 \pi^3 s } \int_{-1}^1 \mathrm{d} \cos \theta 
\Bigl\{ s^2 |A_{\gamma \gamma}|^2 + 2 s^2 m_\eta^2 \mathrm{Re}[A_{\gamma \gamma} B_{\gamma \gamma}^\ast] + |B_{\gamma \gamma}|^2 \Bigl((m_\eta^4 - u t )^2 + s^2 m_\eta^4 \Bigr) \Bigr\}, 
\end{align}
where $\theta$ is the scattering angle in the CM frame, and $t$ and $u$ are the Mandelstam variables. 
See Ref.~\cite{Aoki:2022bkg} for the definition of $A_{\gamma \gamma}$ and $B_{\gamma \gamma}$. 

Finally, we consider the annihilations into a pair of the Higgs bosons. 
The dominant contribution comes from four-point interactions $\eta \eta H_a H_b$ and $\eta \eta H^+ H^-$ unless $m_\eta$ is very close to a half of the mass of the additional Higgs bosons $H_a$, and the decay $H_a \to H_b H_c$ ($b,c \neq a$) or $H_a \to H^+ H^-$ is kinematically allowed.
Thus, we consider only the contribution from the four-point interactions. 
All such four-point interactions are generated by the coupling $\sigma_1$ and $\sigma_2$. 
It is convenient to use the interactions in the Higgs basis; 
\begin{equation}
\mathcal{L} = - \frac{ \eta^2 }{ 2 } \Bigl\{
	\sigma_1^\prime |\Phi_1|^2 + \sigma_2^\prime |\Phi_2|^2 
	+ \sigma_{12}^\prime  \bigl( \Phi_1^\dagger \Phi_2 + \mathrm{h.c.} \bigr) \Bigr\},  
\end{equation}
where 
\begin{align}
\sigma_1^\prime = \sigma_1 c_\beta^2 + \sigma_2 s_\beta^2, \quad 
\sigma_2^\prime = \sigma_1 s_\beta^2 + \sigma_2 c_\beta^2, \quad 
\sigma_{12}^\prime = (\sigma_2 - \sigma_1) s_\beta c_\beta. 
\end{align}
Then, the four-point interaction is given by
\begin{align}
\mathcal{L} =  - \frac{ \eta^2 }{ 4 } \sum_{a,b}
\Bigl\{&  \sigma_1^\prime R_{1a}R_{1b} H_a H_b 
	+ 2 \sigma_{12}^\prime R_{1a} R_{2b} H_a H_b \nonumber \\
	& + \sigma_2^\prime \Bigl(2 |H^+|^2 + (R_{2a} R_{2b} + R_{3a} R_{3b}) H_a H_b \Bigr)
\Bigr\}. 
\end{align}
In particular, in the case of the small mixing ($R_{ab} \simeq \delta_{ab}$), 
it is approximated by
\begin{equation}
\mathcal{L} \simeq  - \frac{ \eta^2 }{ 4 } 
\Bigl\{ \sigma_1^\prime R_{1a}R_{1b} H_1^2  
	+ 2 \sigma_{12}^\prime H_1 H_2 
	+ \sigma_2^\prime \Bigl(2 |H^+|^2 +H_2^2 + H_3^2 \Bigr)
\Bigr\}. 
\end{equation}
The cross section is given by
\begin{align}
(\sigma v)_H & = \frac{ \sigma_2^\prime }{ 8 \pi s } h(m_{H^\pm}^2) 
+ \sum_{a\geq b} \frac{ \lambda_{\eta \eta a b}^2 s_{ab} }{ 8 \pi s } \tilde{h}(m_{H_a}, m_{H_b}), 
\end{align}
where $\tilde{h}(x,y) = \sqrt{ 1 - 2 (x^2+y^2)/s + (x^2-y^2)^2/s^2 }$, $s_{ab} = 1/(1+\delta_{ab})$, and $\lambda_{\eta \eta a b}$ is the vertex factor of $\eta \eta H_a H_b$ given by
\begin{align}
\lambda_{\eta \eta a b} = - \frac{ 1 }{ 2 s_{ab} } 
\Bigl( & \sigma_1^\prime R_{1a}R_{1b} + 2 \sigma_{12}^\prime R_{1a} R_{2b} + \sigma_2^\prime ( R_{2a} R_{2b} + R_{3a} R_{3b} ) \Bigr). 
\end{align}

\section{Electric dipole moments}
\label{sec: EDM}

In this section, we investigate the eEDM $d_e$ in the model. 
There are two kinds of $CP$-violating phases in the model, those in the Yukawa matrix $h_i^\alpha$ and $\theta_5$ in the Higgs potential. 
They independently induce the eEDM, and they are denoted by $d_e^{h}$ and $d_e^{\theta_5}$ in the following, respectively. 

\begin{figure}[h]
\begin{center}
\subfigure[Feynman diagrams giving finite contributions]{
\includegraphics[width=0.7\textwidth]{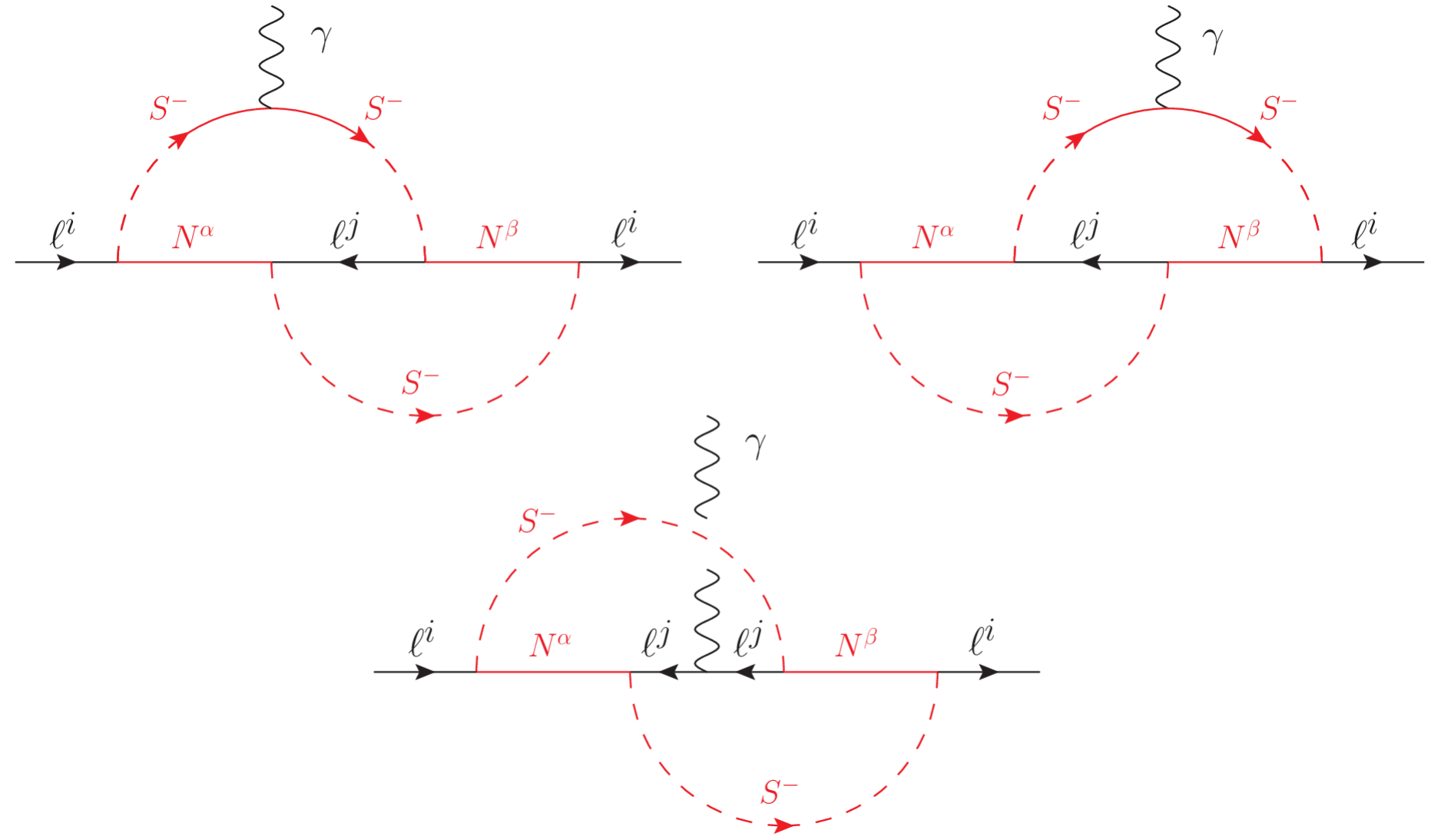}
\label{fig: EDM1}
}
\subfigure[Feynman diagrams with canceling contributions]{
\includegraphics[width=0.7\textwidth]{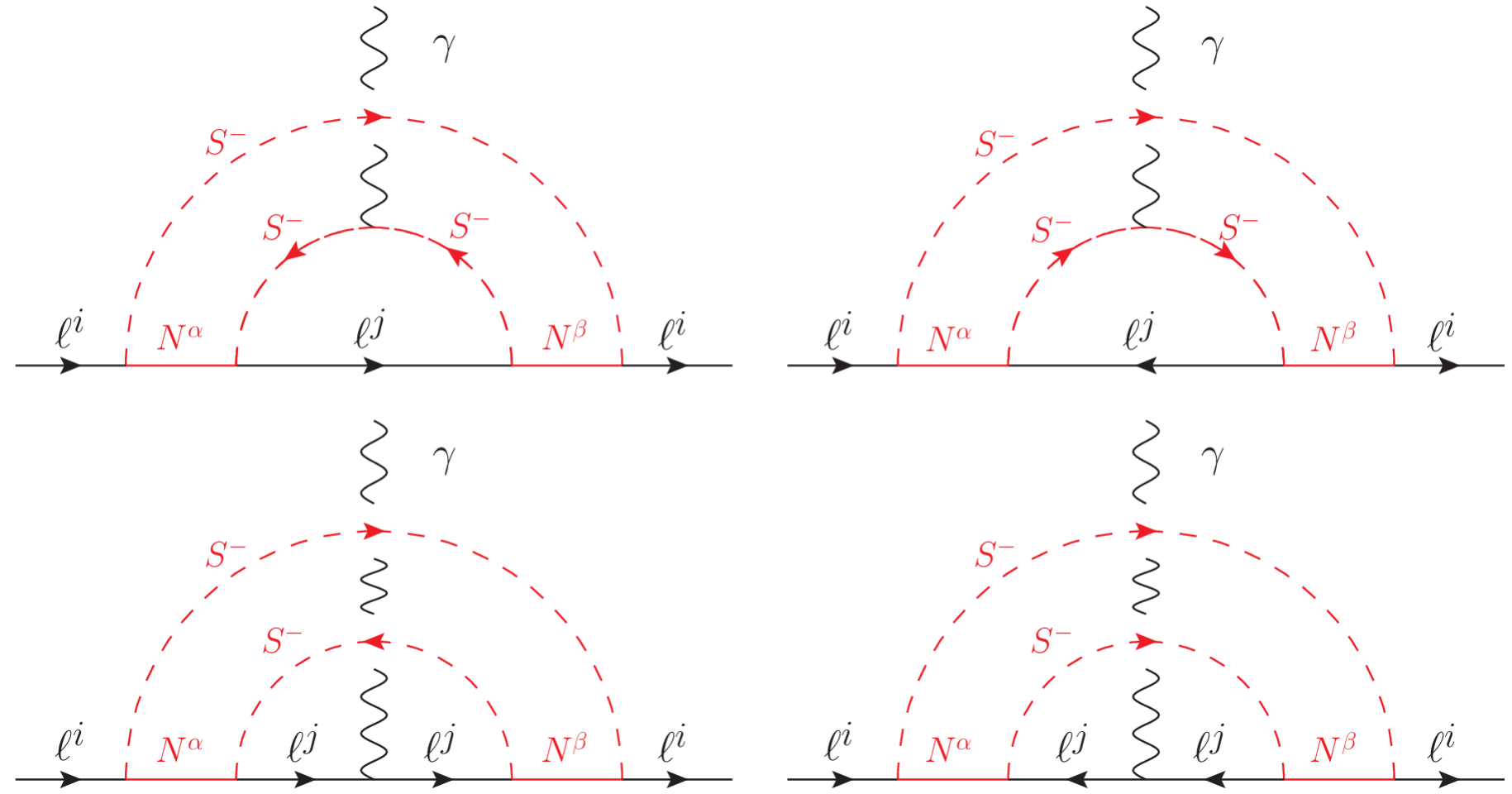}
\label{fig: EDM2}
}
\caption{Feynman diagrams for the EDM of the charged leptons via the Yukawa matrix $h$.}
\label{fig: EDM_Yukawa}
\end{center}
\end{figure}

First, we consider $d_e^h$. 
The leading contribution comes from two-loop diagrams of the quartic orders of the Yukawa matrix $h$, which are shown in Fig.~\ref{fig: EDM_Yukawa}. 
Diagrams in Fig.~\ref{fig: EDM1} give finite contributions to the EDM; on the other hand, contributions from diagrams in Fig.~\ref{fig: EDM2} are canceled although each of them is nonzero. 
The same cancellation was discussed in the scotogenic model in Ref.~\cite{Abada:2018zra}, and 
Ref.~\cite{Fujiwara:2021vam} has proved the cancellation mechanism with a more general setup by using the Ward-Takahashi identity. 

As a result, $d_e^h$ at this order is given by
\begin{equation}
\label{eq: eEDMh}
d_e^h = \frac{e m_e}{m_S^2} \left( \frac{ 1 }{ 16 \pi^2 } \right)^2 
	\sum_{\alpha, \beta, j} I_{ij}^{\alpha \beta} \sqrt{r_\alpha r_\beta}\Bigl(\mathcal{J}^1_{\alpha \beta} + \mathcal{J}^2_{\alpha \beta} \Bigr), 
\end{equation}
where $I_{ij}^{\alpha \beta} = \mathrm{Im}[ (h_i^\beta h_j^\beta)^\ast (h_i^\alpha h_j^\alpha)]$, and the function $J_{\alpha \beta}^1$ and $J_{\alpha \beta}^2$ are given by the following integrals
\begin{align}
& \mathcal{J}^1_{\alpha \beta} = \int \frac{ stz\bigl( -(1-y)(1-z) + sy(1-y) + tyz \bigr) }{ \bigl[y(1-y)s r_\beta +t x r_\alpha+ t z + uy(1-y) \bigr]^2 }, \\[10pt]
& \mathcal{J}^2_{\alpha \beta} = \int \frac{ sty(1-y) \bigl(z - (1-y)s - zt \bigr) }{ \bigl[y(1-y)s r_\beta + tx r_\alpha +  t z + u y (1-y) \bigr]^2 }, 
\end{align}
with
\begin{align}
\int =&  \int_0^1 \mathrm{d}x \mathrm{d}y \mathrm{d}z \, \delta(1-x-y-z) \int_0^1 \mathrm{d}s \mathrm{d}t \mathrm{d}u \, \delta(1-s-t-u). 
\end{align}
The loop functions $\mathcal{J}_{\alpha \beta}^1$ and $\mathcal{J}_{\alpha \beta}^2$ correspond to the sum of the upper two diagrams and the lower diagram in Fig.~\ref{fig: EDM1}, respectively. 

Here, we show some comments about comparisons with Ref.~\cite{Abada:2018zra} studying the eEDM in the scotogenic model~\cite{Ma:2006km}, which is described by Feynman diagrams with the same topology. 
We have numerically checked $\mathcal{J}_1$ and $\mathcal{J}_2$ are the same with the loop functions $I_{1+2}^M$ and $I_3^M$ defined in Ref.~\cite{Abada:2018zra}, respectively although their integral representations are different. 
The overall signs are opposite comparing Eq.~(\ref{eq: eEDMh}) with their result. 
This is due to the different chirality of the charged leptons.
In the scotogenic model, the left-handed leptons have the Yukawa interaction, while the right-handed ones do in our model. 
It induces the opposite signs of the coefficients of $\gamma_5$. 

Next, we discuss the contribution from the Higgs sector $d_e^{\theta_5}$. 
This is the same with the eEDM in the THDMs studied in Refs.~\cite{Inoue:2014nva, Modak:2018csw, Fuyuto:2019svr, Bian:2014zka, Bian:2016zba, Abe:2013qla, Cheung:2020ugr, Altmannshofer:2020shb}. 
It is known that one-loop contribution is tiny because of the mass suppression by $m_e$, and the leading contribution comes from Barr-Zee type diagrams at two-loop level~\cite{Barr:1990vd}. 
We employ the formula in Ref.~\cite{Altmannshofer:2020shb}, which includes all the two-loop diagrams and gives the gauge-invariant result.

\section{Numerical evaluations and benchmark scenarios} 
\label{sec: benchmark scenarios}

In this section, we show some benchmark scenarios and their numerical evaluations. 
Before a detailed discussion, we roughly describe our strategy to determine the benchmark scenarios. 
We first consider the constraint from the neutrino oscillation data. 
The Casas-Ibarra parametrization in Eq.~(\ref{eq: CI parametrization}) tells us the following preferred mass spectrum; 
\begin{equation}
m_{H^\pm}, \ m_S, \ m_\eta = \mathcal{O}(100)~\mathrm{GeV}, \quad 
m_{N^\alpha} = \mathcal{O}(1)~\mathrm{TeV}. 
\end{equation}
This hierarchy implies that the real scalar dark matter $\eta$ is strongly favored in our model. 

Next, we determine the structure of the Yukawa matrix $h$ with a fixed mass spectrum to avoid the current constraint on the LFV decays. We choose the values of the orthogonal matrix $O$ and unobserved parameters in the active neutrino mass matrix: the lightest neutrino mass $m_\mathrm{lightest}$ and Majorana $CP$-violating phases $\alpha_1$ and $\alpha_2$. 

$\eta$ is responsible for all the observed relic density of DM, which is made by the freeze-out mechanism. 
There are two ways to make the annihilation rate large enough to reproduce the observed value $\Omega_\mathrm{DM} h^2 \simeq 0.12$ under the current severe constraint on the direct detection of DM. 
One is the resonance effect of the on-shell Higgs boson, which is called the Higgs portal DM scenario. 
The other one uses the annihilation into the additional Higgs bosons by contact scalar interactions, which is denoted by a heavy dark matter scenario in the following. 
These make two benchmark scenarios with different values of $m_\eta$. 

The mass of the neutral additional Higgs bosons $H_2$ and $H_3$ is almost ineffective in the above discussions. 
We determine them so as to satisfy the theoretical requirements on the Higgs potential and to avoid the constraint on the $S$, $T$ and $U$ parameters. 
Other parameters in the Higgs potential are fixed to satisfy the theoretical constraints. 

Finally, we fix the $CP$-violating phase $\theta_5$ to avoid the current constraint by the destructive interference with the $CP$-violating phases in $h_i^\alpha$, which is already determined to avoid the constraint on the LFV decays. We will show that $\theta_5$ can be $\mathcal{O}(1)$, which is necessary to generate sufficient baryon asymmetry by electroweak baryogenesis.

\subsection{Neutrino mass and the size Yukawa couplings $h^\alpha_i$}
As long as $h_i^\alpha$ employs the Casas-Ibarra parametrization in Eq.~(\ref{eq: CI parametrization}), it can explain the neutrino oscillation data. 
$h_i^\alpha$ is then given by 
\begin{equation}
\label{eq: h_i_alpha_with_CI}
h_i^\alpha = \frac{ (16 \pi^2)^{3/2} }{ |\kappa \tan \beta| } 
	\sum_{a=1}^3
	\sqrt{|\Lambda_\alpha|} O_{\alpha a} \sqrt{m_{\nu^a}} U^\ast_{i a} m_{\ell^i}^{-1}. 
\end{equation}
Thus, the size of $h_i^\alpha$ can be roughly estimated by 
\begin{equation}
|h_i^\alpha| \sim 0.1 \times \frac{ (16 \pi^2)^{3/2} }{ |\kappa \tan \beta| } \frac{ \sqrt{|\Lambda_\alpha| \bar{m}_\nu } }{ m_{\ell^i} }, 
\end{equation}
where $\bar{m}_\nu$ are the typical size of the neutrino mass. 
We have assumed that the factor $O_{a\alpha} U^\ast _{ia}$ and the summation about $a$ in Eq.~(\ref{eq: h_i_alpha_with_CI}) all together make a coefficient of $\mathcal{O}(0.1)$ for simplicity.\footnote{This correctness of this assumption depends on the $CP$-violating phases $\psi_i$ in $O_{a\alpha}$. The large $CP$-violating phases ($|\psi_i| \simeq \pi/2$) make the bound stronger. As far as $|\psi_i|$ is not too large, we can use this assumption. } 
We note $|h_i^\alpha|$ is proportional to $m_{\ell^i}^{-1}$, and $N$'s strongly couple to electrons.  

To keep the perturbativity, the size of $h_i^\alpha$ should not be too large. 
The large couplings also lead to difficulty in avoiding the constraint on the LFV processes. 
We thus demand that $|h_i^\alpha|$ be less than three.  
Since $h_i^\alpha \propto \sqrt{\Lambda_\alpha}$, this requirement constraints $\Lambda_\alpha$, {\it i.e.} the mass spectrum of the new particles. 
The bound on $\Lambda_\alpha$ is roughly estimated by
\begin{equation}
\label{eq: prebound_for_Lambda}
|\Lambda_\alpha| \lesssim \frac{ m_{\ell^i}^2 |\kappa \tan \beta|^2 }{ (16\pi^2)^3 \bar{m_\nu} }. 
\end{equation}
Eq.~(\ref{eq: prebound_for_Lambda}) provides the most significant constraint in the case of $m_{\ell^i} = m_e$, which is given by
\begin{equation}
\label{eq: bound_for_Lambda}
|\Lambda_\alpha| \lesssim 1~\mathrm{TeV} \times \biggl( \frac{ |\kappa \tan \beta | }{ 40 } \biggr)^2 \biggl( \frac{ \bar{m}_\nu }{ 0.01~\mathrm{eV} } \biggr)^{-1}. 
\end{equation}

\begin{figure}[b]
\begin{center}
\includegraphics[width=0.6\textwidth]{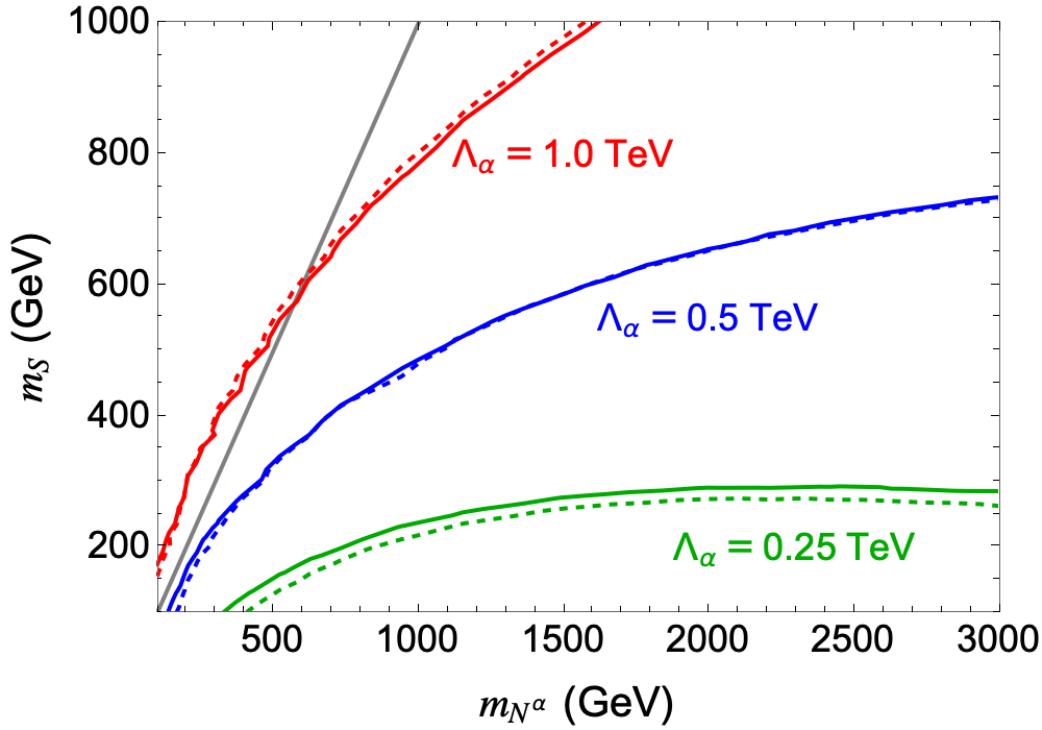}
\caption{$\Lambda_\alpha$ as a function of $m_S$ and $m_{N^\alpha}$. The red, blue and green curves correspond to $\Lambda_\alpha = 1.0~\mathrm{TeV}$, $0.5~\mathrm{TeV}$ and $0.25~\mathrm{TeV}$, respectively. The masses of the other scalar bosons are set as $m_{H^\pm} = 390~\mathrm{GeV}$ and $m_\eta = 63~\mathrm{GeV}$ ($250~\mathrm{GeV}$) on the solid (dashed) curves. The gray line shows $m_S = m_{N^\alpha}$. }
\label{fig: Lambda}
\end{center}
\end{figure}

In Fig.~\ref{fig: Lambda}, we show $\Lambda_\alpha$ as a function of $m_S$ and $m_{N^\alpha}$. 
The red, blue and green curves are contours of $\Lambda_\alpha = 1.0~\mathrm{TeV}$, $0.5~\mathrm{TeV}$ and $0.25~\mathrm{TeV}$, respectively. 
$m_{H^\pm}$ and $m_\eta$ are set to be $m_{H^\pm} = 390~\mathrm{GeV}$ and $m_\eta = 63~\mathrm{GeV}$ ($250~\mathrm{GeV}$) on the solid (dashed) curves. 
The gray line shows $m_S = m_{N^\alpha}$. 

In the case of  $\kappa \tan \beta = 40$ and $\bar{m}_\nu =0.01~\mathrm{eV}$, the upper bound on $|\Lambda_\alpha|$ is given by the red curves, and the bound is stronger for smaller $\kappa \tan \beta$. 
We can see that heavier $N^\alpha$ than $S^\pm$ are strongly preferred. This is because $\Lambda_\alpha$ is proportional to $m_S^2$ if $m_S \gg m_{N^\alpha}$, $m_{H^\pm}$, $m_\eta$, while it linearly depends on $m_{N^\alpha}$ if $N^\alpha$ is the heaviest particle. 
The hierarchy $m_{N^\alpha} > m_S$ implies that $\eta$ is the dark matter particle in the model. 

\subsection{The LFV decays}

Next, we consider the constraint from the LFV decays by using the Casas-Ibarra parametrization of $h_i^\alpha$. 
Ref.~\cite{Toma:2013zsa} performed a similar analysis in the scotogenic model~\cite{Ma:2006km}. 
For simplicity, we use an approximation such that all of $N$'s have the same mass $\bar{m}_N$. 
Even in the case that they have different masses, the result expects to give a good approximation by replacing $\bar{m}_N$ into their typical mass scale unless the mass difference is too large. 

Since $h_i^\alpha \propto m_{\ell^i}^{-1}$, the LFV decays including the light leptons $e$ and $\mu$ expect to give significant constraints. 
Thus, we show only the analysis for $\mu \to e \gamma$ and $\mu \to 3 e$. 
It is straightforward to apply the same method to the other LFV decays. 
By using Eqs.~(\ref{eq: LFVBr1}), (\ref{eq: LFVBr2}) and (\ref{eq: h_i_alpha_with_CI}), their branching ratios are evaluated as
\begin{align}
\mathrm{Br}(\mu \to e \gamma) =  & \frac{ 3 \alpha }{ 64 \pi G_F^2 m_S^4 } 
	\biggl( \frac{ (16 \pi^2 )^3 \bar{\Lambda} F }{ m_\mu m_e \kappa^2 \tan^2\beta } \biggr)^2 
	 \bigl|X_{\mu e}\bigr|^2 \nonumber \\[7pt]
	 \simeq & \, \bigl( 4\times 10^{27} \bigr) \biggl( \frac{ 1 }{ \kappa \tan \beta } \biggr)^4
		\biggl( \frac{ \bar{\Lambda}^2 F^2 }{ m_S^4 } \biggr) \bigl| X_{\mu e} \bigr|^2, \\[15pt]
\mathrm{Br}(\mu \to 3e) = & \frac{ 1 }{ 1024 \pi^4 G_F^2 m_S^4 } 
	\left( \frac{ (16\pi^2)^6 \bar{\Lambda}^2 G }{ m_\mu m_e^3 \kappa^4 \tan^4 \beta } \right)^2
	\bigl|X_{\mu e} X_{ee} - a (M_\nu)_{\mu e} (M_\nu)_{ee}^\ast \bigr|^2 \nonumber \\[7pt]
	\simeq & \bigl(10^5~\mathrm{meV}^{-4} \bigr) \biggl( \frac{ 1 }{ \kappa \tan \beta } \biggr)^8
		\biggl( \frac{ \bar{\Lambda}^4 G^2 }{ m_S^4 } \biggr)
		\bigl|X_{\mu e} X_{ee} - a (M_\nu)_{\mu e} (M_\nu)_{ee}^\ast \bigr|^2. 
\end{align}
where $\bar{\Lambda}$ is $\Lambda_\alpha$ with $m_{N^\alpha} = \bar{m}_N$, 
$F$, $G$ and $a$ is defined as 
\begin{align}
& F \equiv f(\bar{r}) = \frac{ 1 - 6 \bar{r} + 3 \bar{r}^2 + 2 \bar{r}^3 - 6 \bar{r}^2 \log \bar{r} }{ 6 (1 - \bar{r})^4 }, \\[10pt]
& G \equiv \frac{ \mathrm{d} g_2(\bar{r}) }{ \mathrm{d} \bar{r} } = \frac{ 1 }{ 4 (1 - \bar{r})^2 } \Bigl( 1 + \bar{r} + \frac{ 2 \bar{r} }{ 1 - \bar{r} } \log \bar{r} \Bigr), \\[10pt]
& a \equiv - \frac{1}{\bar{r}}\frac{ \mathrm{d} g_2(\bar{r}) }{ \mathrm{d} \bar{r} } \biggl( \frac{ \mathrm{d} g_1(\bar{r}) }{ \mathrm{d} \bar{r} } \biggr)^{-1}  = \frac{ 2 \bar{r} (2\bar{r} - 2 - (1 + \bar{r} ) \log \bar{r} ) }{ 1 - \bar{r}^2 + 2 \bar{r} \log \bar{r} }, 
\end{align}
with $\bar{r} = \bar{m}_N^2/m_S^2$, and $X_{\ell^i \ell^j}$ is given by
\begin{equation}
X_{\ell^i \ell^j} = \sum_{a,b, \alpha} \sqrt{m_{\nu^a}^{} m_{\nu^b}^{} } U_{i a }^\ast U_{j b} O_{\alpha a} O_{\alpha b}^\ast.
\end{equation}
$X_{\ell^i \ell^j}$ satisfies $X_{\ell^i \ell^j} = X_{\ell^j \ell^i}^\ast$ and has the mass dimension one. 
If the matrix $O$ is a real matrix, {\it i.e.}, $\psi_i = 0$ ($i=1,2,3$), 
$X_{\ell^i \ell^j}$ is independent of $O_{a\alpha}$ because $\sum_\alpha O_{\alpha a} O_{\alpha b}^\ast = \sum_\alpha O_{\alpha a} O_{\alpha b} = \delta_{ab}$. 
It is simplified as
\begin{equation}
X_{\ell^i \ell^j} = (U^\ast D U^\mathrm{T})_{ij} = \sum_{a} m_{\nu^a} U_{ia}^\ast U_{ja} \quad \text{if all $O_{a\alpha}$ are real.} 
\end{equation}

\begin{figure}[t]
\begin{center}
\includegraphics[width=0.7\textwidth]{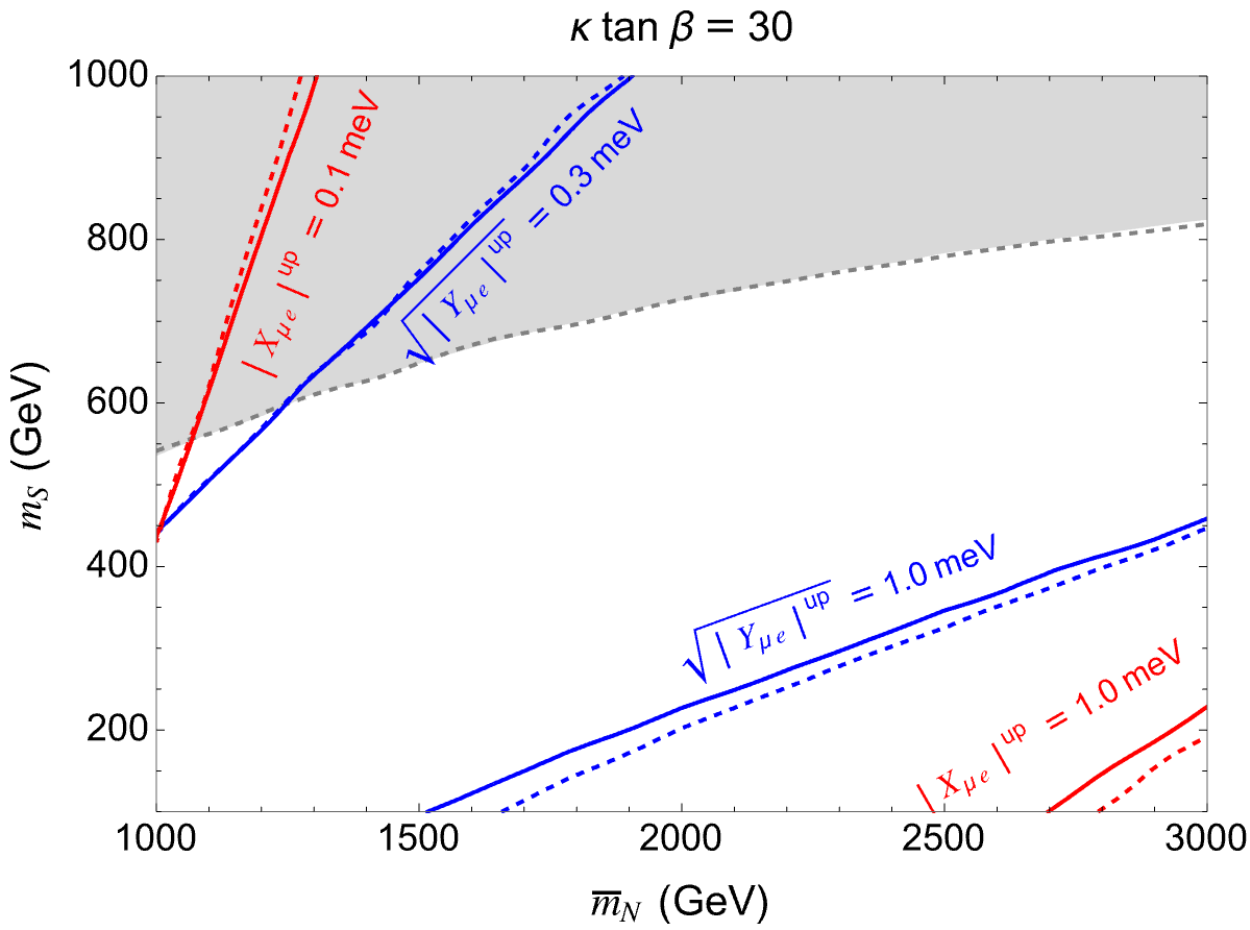}
\caption{The upper bound on $|X_{\mu e}|$ and $\sqrt{|Y_{\mu e}|}$ as functions of $\bar{m}_N$ and $m_S$ in the case of $\kappa \tan \beta = 30$. 
The red (blue) curves are contours of $|X_{\mu e}|^\mathrm{up}$ ($\sqrt{|Y_{\mu e}|^\mathrm{up}}$). The solid and dashed curves correspond to the cases of ($m_{H^\pm}$, $m_\eta$) = (390 GeV, 63 GeV) and (390 GeV, 250 GeV), respectively. The gray regions are unfavored by the bound in Eq.~(\ref{eq: bound_for_Lambda}) in the former case. The dashed gray curve is the bound in the latter case.}
\label{fig: LFV_bound}
\end{center}
\end{figure}
By using the current upper limits, $\mathrm{Br}(\mu\to e \gamma) < 3.1\times10^{-13}$~\cite{MEGII:2023ltw} and $\mathrm{Br}(\mu\to 3e) < 1.0\times10^{-12}$~\cite{SINDRUM:1987nra} at $90~\%$ confidence level, we have
\begin{align}
& \bigl|X_{\mu e} \bigr| < \bigl(0.01~\mathrm{meV} \bigr)
	\biggl( \frac{ \kappa \tan \beta }{ 40 } \biggr)^2
	\biggl| \frac{ m_S^2 / (\bar{\Lambda} F) }{ 1~\mathrm{TeV} } \biggr|, \\[5pt]
& \bigl|Y_{\mu e} \bigr| \equiv \bigl|X_{\mu e } X_{ee} - a (M_\nu)_{\mu e} (M_\nu)^\ast_{ee} \bigr|
	< (0.1~\mathrm{meV})^2
	\biggl( \frac{ \kappa \tan \beta }{ 40 } \biggr)^4
	\biggl( \frac{ m_S^2 }{ \bar{\Lambda}^2 G } \biggr). 
\end{align}
The upper bound on $|X_{\mu e}|$ and $|Y_{\mu e}|$, which are denoted as $|X_{\mu e}|^\mathrm{up}$ and $|Y_{\mu e}|^\mathrm{up}$ in the following, 
are functions of $\kappa \tan \beta$ and the masses of the new particles. 

Fig.~\ref{fig: LFV_bound} shows $|X_{\mu e}|^\mathrm{up}$ and $\sqrt{|Y_{\mu e}|^\mathrm{up}}$ in the cases of $\kappa \tan \beta = 30$.  
The red curves are contours for $|X_{\mu e}|^\mathrm{up} = 1.0~\mathrm{meV}$ and $0.1~\mathrm{meV}$, and the blue ones are those for $\sqrt{|Y_{\mu e}|^\mathrm{up}} = 1.0~\mathrm{meV}$ and $0.3~\mathrm{meV}$. 
The solid and dashed lines correspond to the cases of ($m_{H^\pm}$, $m_\eta$) = (390 GeV, 63 GeV) and (390 GeV, 250 GeV), respectively. 
The gray regions represent unfavored regions from the bound in Eq.~(\ref{eq: bound_for_Lambda}) for ($m_{H^\pm}$, $m_\eta$) = (390 GeV, 63 GeV), and the gray dashed curve is the bound for ($m_{H^\pm}$, $m_\eta$) = (390 GeV, 250 GeV). The bounds are stronger for smaller $\kappa \tan \beta$ because the Yukawa couplings $h^\alpha_i$ are proportional to $|\kappa \tan \beta |^{-1}$. 

\begin{table}[t]
\begin{center}
\caption{Input values for neutrino mass matrix~\cite{PDG}}
\label{table: neutrino oscillation data}
\begin{tabular}{|c|c|c|c|c|c|} \hline
 & $\sin^2\theta_{12}$ & $\sin^2\theta_{23}$ & $\sin^2\theta_{13}$ & $m_{\nu^2}^2-m_{\nu^1}^2$ ($\mathrm{eV^2}$) & $m_{\nu^3}^2-m_{\nu^2}^2$ ($\mathrm{eV^2}$) \\ \hline
NO & \multirow{2}{*}{$0.307$} & $0.534$ & \multirow{2}{*}{$2.20\times 10^{-2}$} &  \multirow{2}{*}{$7.53\times 10^{-5}$} & $2.437 \times 10^{-3}$ \\ \cline{1-1} \cline{3-3} \cline{6-6}
IO & 							& $0.547$ & 										  & 											& $-2.519 \times 10^{-3}$  \\ \hline
\end{tabular}
\end{center}
\end{table}
\begin{figure}[b]
\begin{center}
\subfigure[Normal Ordering]{
\includegraphics[width=0.48\textwidth]{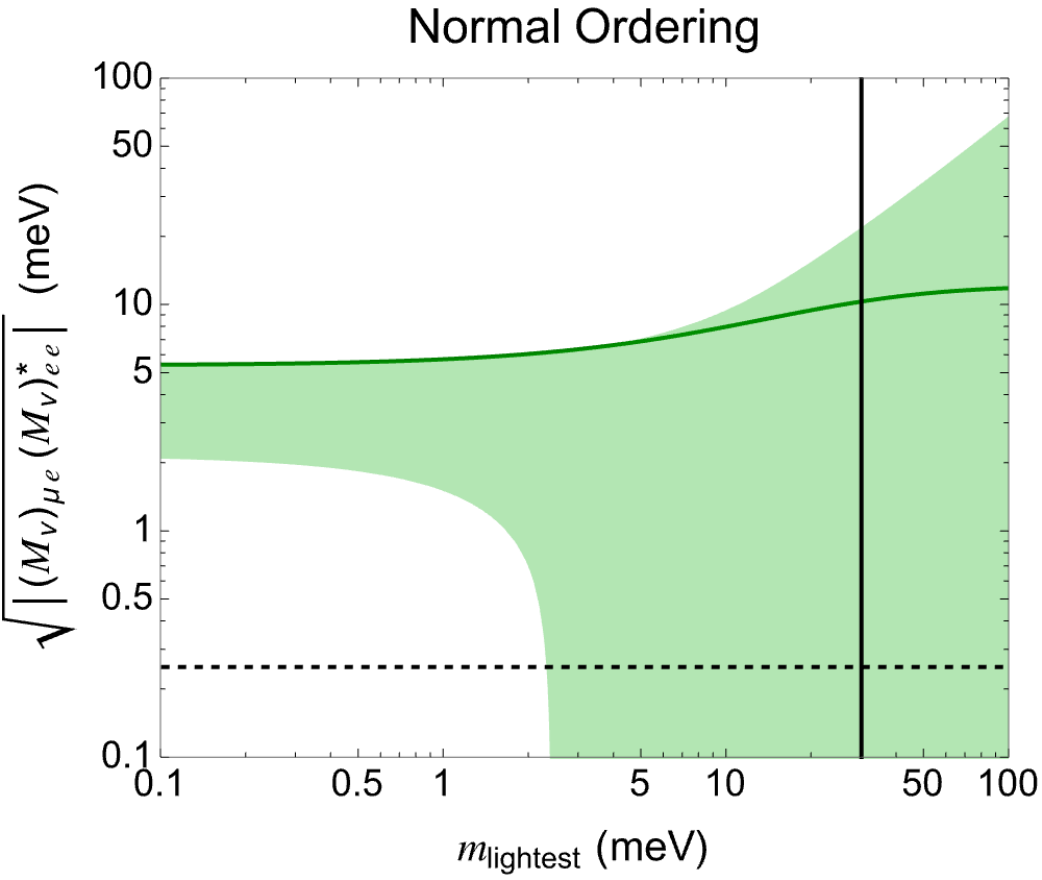}
\label{fig: MM_NO}
}
\subfigure[Inverted Ordering]{
\includegraphics[width=0.48\textwidth]{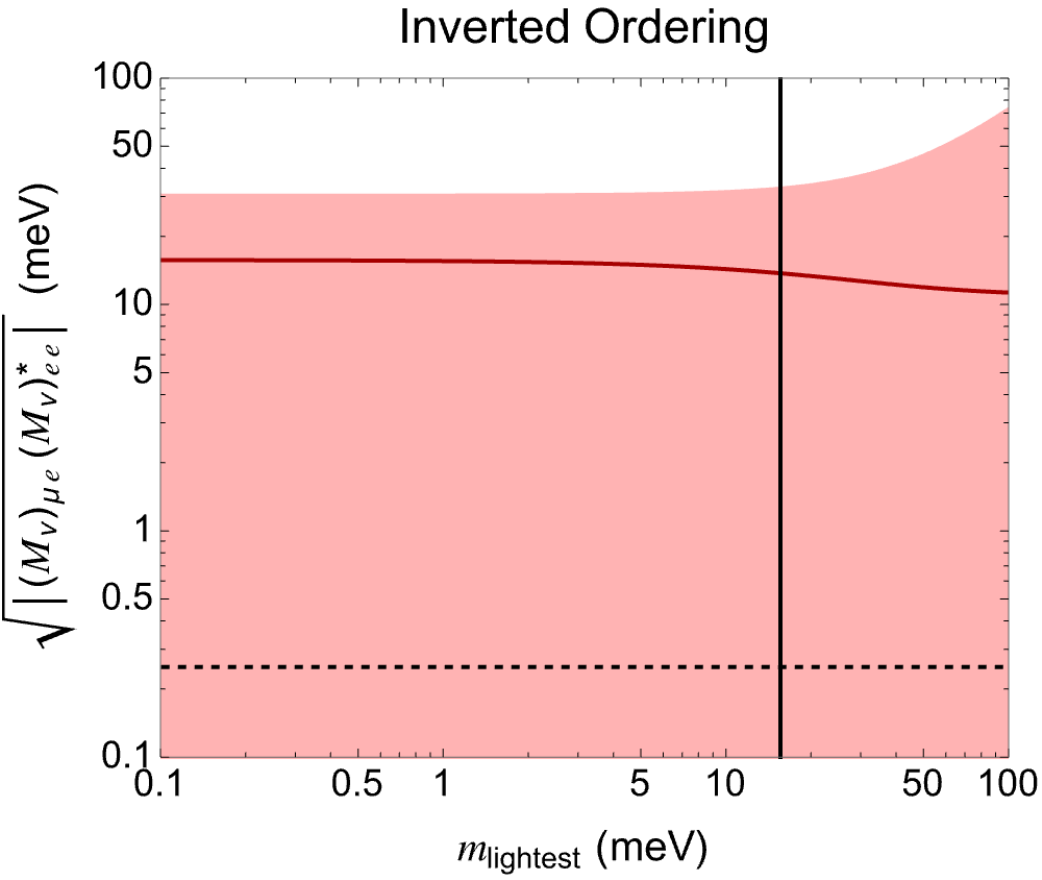}
\label{fig: MM_IO}
}
\caption{Possible values of $|(M_\nu)_{\mu e} (M_\nu)_{ee}^\ast|~\mathrm{meV^2}$ as a function of $m_\mathrm{lightest}$. The darker-colored lines represent values without $CP$-violating phases ($\delta = \alpha_1 = \alpha_2 = 0$). 
The lighter-colored regions are possible values with $CP$-violating phases in the regions $\delta \in [1.02\pi, 1.44\pi]$, $\alpha_{1,2} \in [0, 2\pi)$. }
\label{fig: MM}
\end{center}
\end{figure}

Thus, we roughly evaluate the constraint as $|X_{\mu e}| < 1.0~\mathrm{meV}$ and $\sqrt{|Y_{\mu e}|} < 1.0~\mathrm{meV}$. 
For the latter equation, we consider the sufficient condition $\sqrt{|X_{\mu e} X_{ee}|} < 1.0~\mathrm{meV}$ and $\sqrt{|a (M_\nu)_{\mu e} (M_\nu)_{ee}^\ast |} < 1.0~\mathrm{meV}$. 
In addition, in the case of $10m_S < \bar{m}_N < 100m_S$, $a$ is given by $5 < a < 16$. 
As a result, we employ the following rough conditions to avoid the LFV constraints\footnote{Even if $|(M_\nu)_{\mu e}(M_\nu)^\ast_{ee}|$ and $|X_{\mu e} X_{ee}|$ do not satisfy these bounds, it would be possible to make $|Y_{\mu e}|$ small enough to avoid the constraint from the LFV decays by a large cancellation between them. Here, we do not consider such a case.}
\begin{equation}
\label{eq: rough_LFV_constraint}
\sqrt{|(M_\nu)_{\mu e} (M_\nu)_{ee}^\ast |} < 0.25~\mathrm{meV}, \quad 
|X_{\mu e}| < 1.0~\mathrm{meV}, \quad 
\sqrt{|X_{\mu e} X_{ee}|} < 1.0~\mathrm{meV}. 
\end{equation}

\begin{figure}[t]
\begin{center}
\subfigure[Normal Ordering]{
\includegraphics[width=0.48\textwidth]{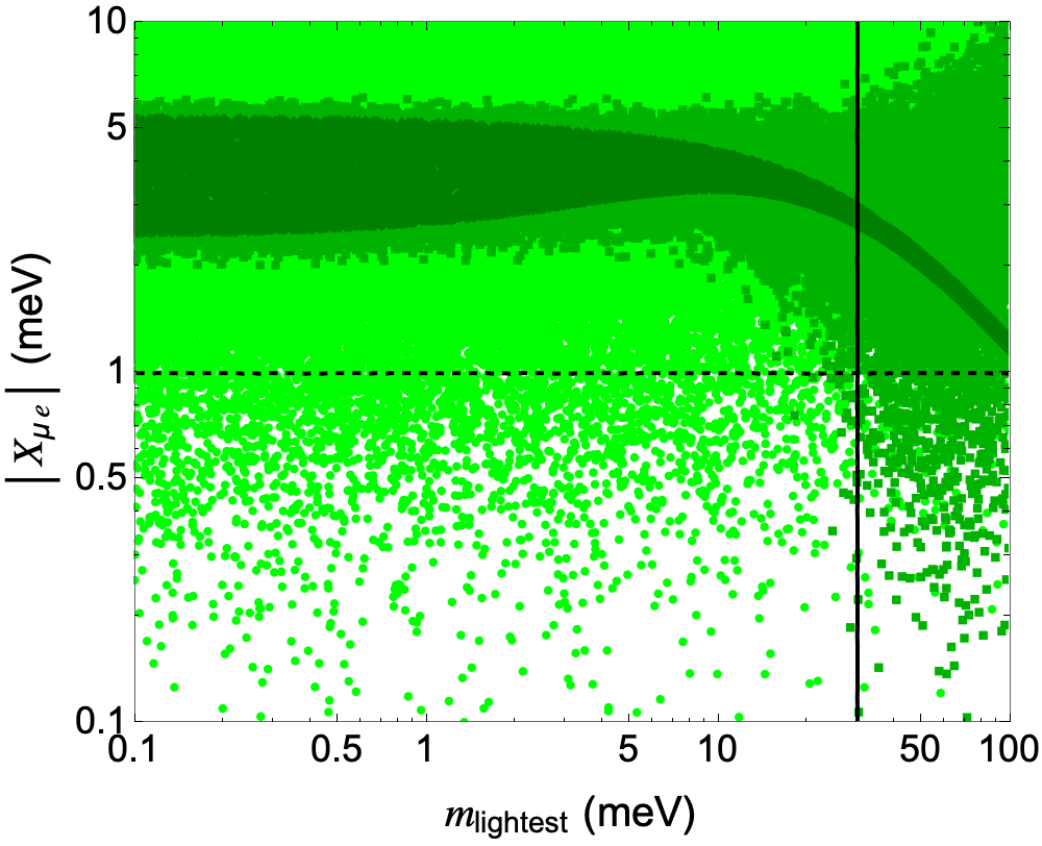}
\label{fig: XmueNO}
}
\subfigure[Inverted Ordering]{
\includegraphics[width=0.48\textwidth]{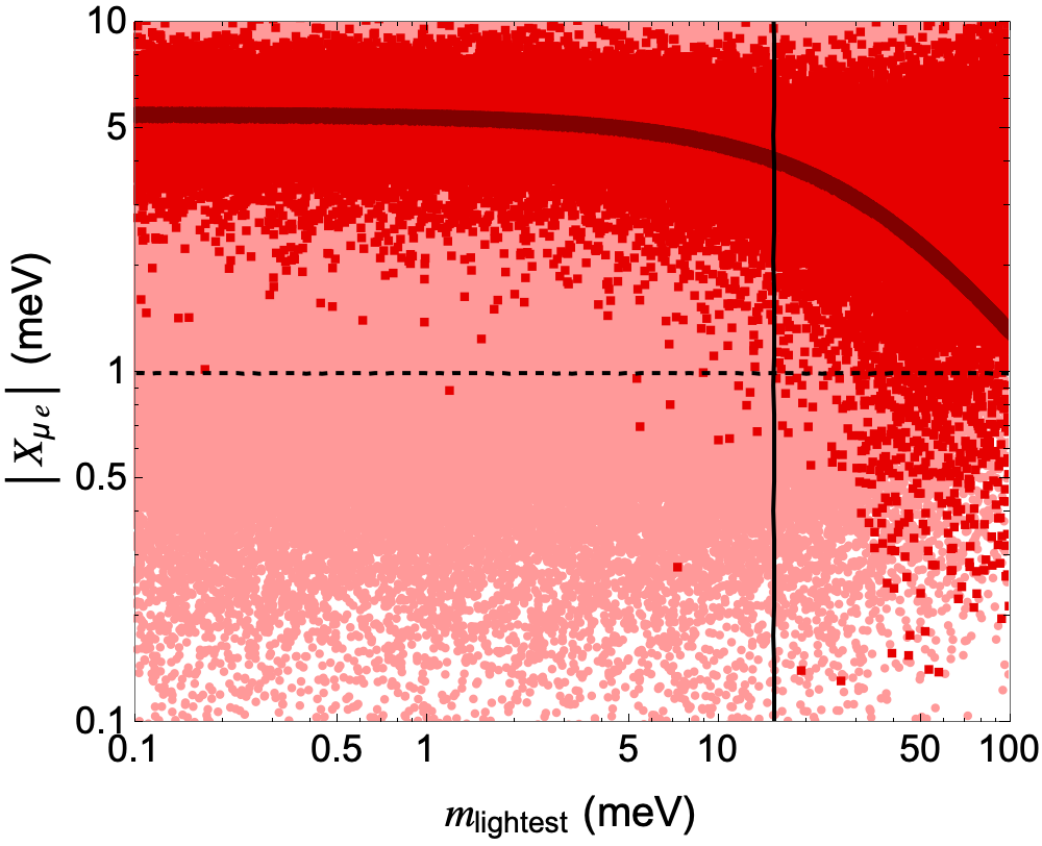}
\label{fig: XmueNO}
}
\caption{$|X_{\mu e}|~\mathrm{meV}$ for the normal ordering and inverted ordering neutrino mass data. On the darkest color points, $\psi_i$ ($i=1,2,3$) are zero. The middle dark and light color points mean that $\psi_i$ are in the range $[0,0.02]$ and $[0.02,0.2]$, respectively. The solid black line is the upper limit on $m_\mathrm{lightest}$ from the CMB data. The dashed black line represents $|X_{\mu e} | = 1~\mathrm{meV}$.}
\label{fig: Xmue}
\end{center}
\end{figure}

In Fig.~\ref{fig: MM}, we show $\sqrt{|(M_\nu)_{\mu e} (M_\nu)_{ee}^\ast |}$ as a function of the lightest neutrino mass $m_\mathrm{lightest}$ for (a) normal ordering neutrino masses (NO) and (b) Inverted ordering neutrino masses (IO). 
We used the data from Particle Data Group~\cite{PDG} in Table~\ref{table: neutrino oscillation data} to make the figure. We neglect their small uncertainties. 
The dashed black lines represent the upper limit in Eq.~(\ref{eq: rough_LFV_constraint}), and the solid black lines correspond to the cosmological bound $\sum_{i=1}^3 m_{\nu^i} < 0.12~\mathrm{eV}$~\cite{Planck:2018vyg} in each case, where heavier $m_\mathrm{lightest}$ is excluded. 
The darker-colored lines represent the value without $CP$-violating phases, {\it i.e.} $\delta = \alpha_1 = \alpha_2 = 0$.
The lighter-colored regions represent the possible values with switching on the $CP$-violating phases in the regions $\delta \in [1.02\pi, 1.44\pi]$, $\alpha_{1,2} \in [0, 2\pi)$, where $\delta$ is in the $1~\sigma$ region of the global fit data in PDG~\cite{PDG}. 
We can see that non-zero $CP$-violating phases are necessary to satisfy the condition in Eq.~(\ref{eq: rough_LFV_constraint}). 
Their interferences can make the value much smaller than the prediction in the $CP$-conserving case. 
In particular, in the inverted ordering case, we need a stronger interference effect to avoid the constraint. 

\begin{figure}[h]
\begin{center}
\subfigure[Normal Ordering]{
\includegraphics[width=0.48\textwidth]{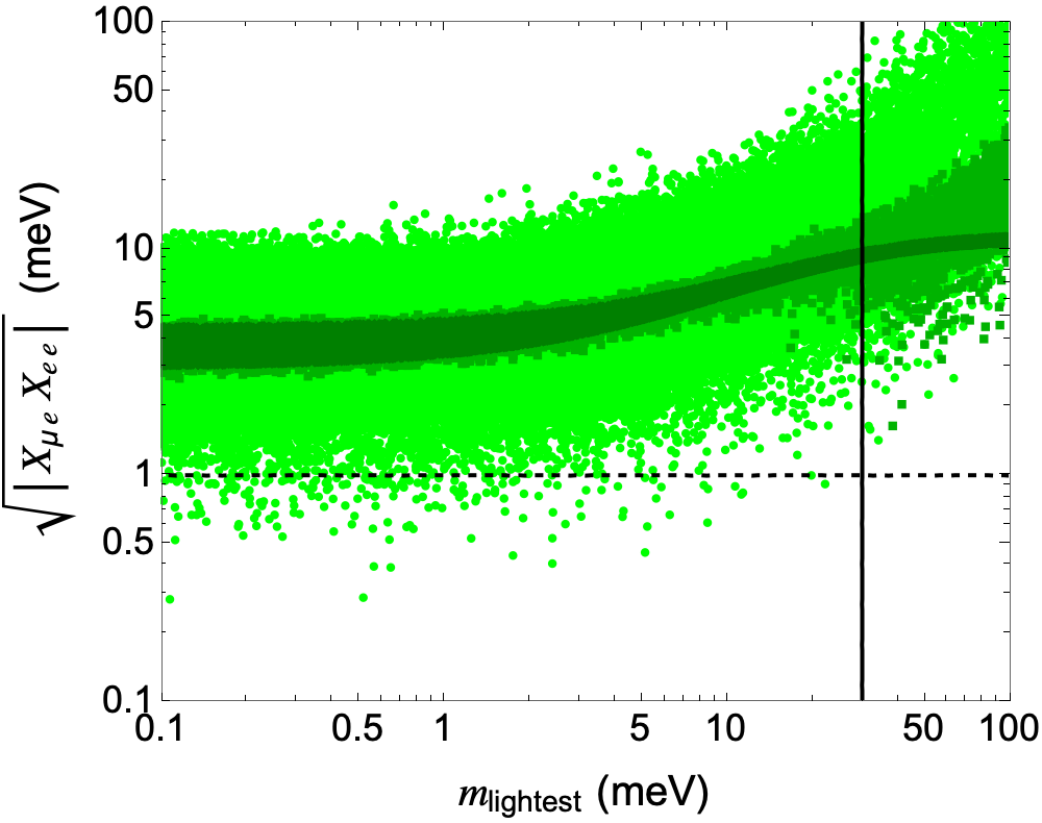}
\label{fig: XXmueNO}
}
\subfigure[Inverted Ordering]{
\includegraphics[width=0.48\textwidth]{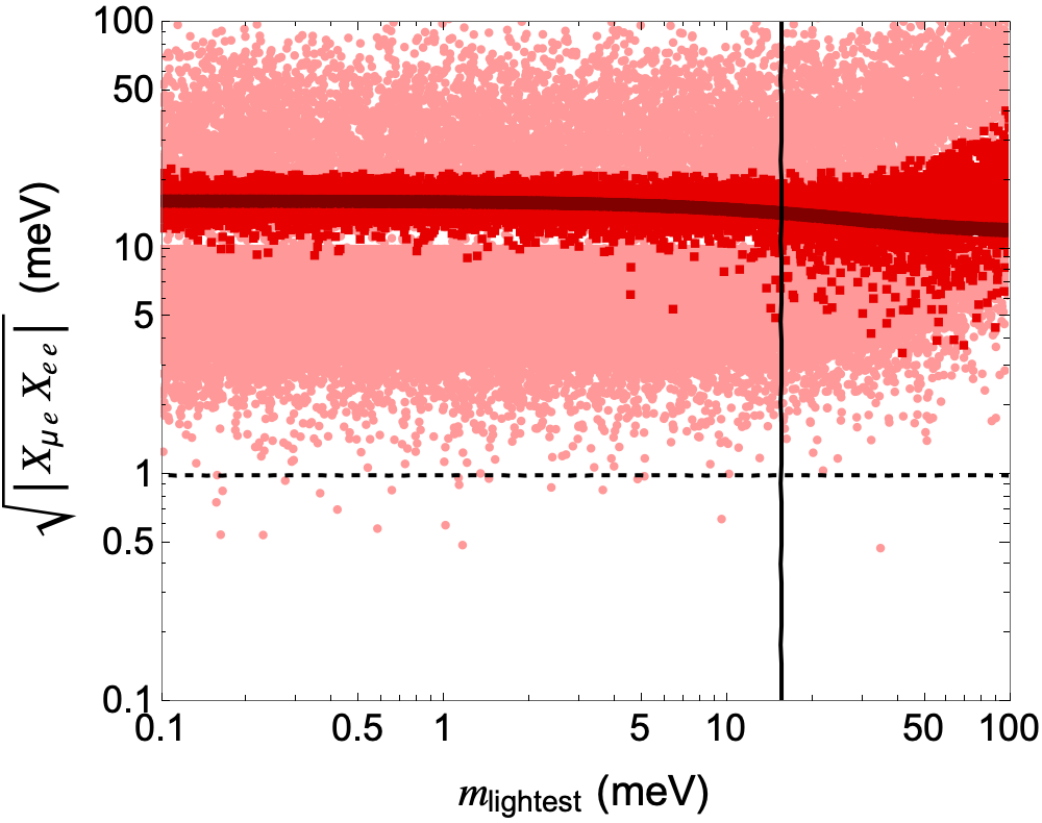}
\label{fig: XXmueIO}
}
\caption{$|X_{\mu e} X_{ee}|~\mathrm{meV}$ for the normal ordering and inverted ordering neutrino mass data. The meaning of colors and black lines is the same as Fig.~\ref{fig: Xmue}.}
\label{fig: XXmue}
\end{center}
\end{figure}

Next, we show possible values of $|X_{\mu e}|$ and $\sqrt{|X_{\mu e} X_{ee}|}$. 
These depend on the orthogonal matrix $O_{\alpha a}$ and have more parameter than $\sqrt{|(M_\nu)_{e\mu} (M_\nu)_{ee}^\ast |}$. 
Thus, we show scatter plots for each quantity, not a whole region of possible values, to reduce the cost of the calculation. We assume that the rotation angles $\gamma_i$ ($i=1$, 2, 3) are in the region $[0, 2\pi)$. 
For $CP$-violating phases $\psi_i$ ($i=1$, 2, 3), we consider three cases (i) $\psi_i = 0$, (ii) $|\psi_i| \in [0, 0.02]$ and (iii) $|\psi_i| \in [0.02, 0.2]$.
Although  $\psi_i$ can take a value in the region $[-\pi/2, \pi/2]$ in principle, we consider only these small $|\psi_i|$ regions because large $|\psi_i|$ make the Yukawa couplings $|h_i^\alpha|$ huge.
The results for $|X_{\mu e}|$ and $\sqrt{|X_{\mu e} X_{ee}|}$ are shown in Figs.~\ref{fig: Xmue} and \ref{fig: XXmue}, respectively. 
The green (red) points are data for the normal (inverted) ordering neutrino masses. 
The darker-, middle-, and lighter-colored points correspond to the cases (i), (ii) and (iii), respectively. 

We can see that in all the figures, the condition in Eq.~(\ref{eq: rough_LFV_constraint}) cannot be satisfied only with the $CP$-violating phases in the PMNS matrix, and non-zero $\psi_i$ are necessary. 
In case (ii), distributions of the points are broader; however, there are few points avoiding constraints. 
Consequently, the condition in Eq.~(\ref{eq: rough_LFV_constraint}) requires $|\psi_i| \gtrsim 0.02$. 
In case (iii) many allowed points appear in the figure for $|X_{\mu e}|$. 
Although the number of allowed points is smaller in the figure for $\sqrt{|X_{\mu e} X_{ee}|}$, we can also find some allowed points. 

Before closing the section about the constraint on the LFV decays, 
we comment on the uncertainties of the above analysis. 
As mentioned at the beginning of this section, we assumed the degenerated $m_{N^\alpha}$. If the mass differences $\Delta m_N$ are switched on, there would be deviations with the order of $\Delta m_N/ \bar{m}_N$. 
In addition, we roughly estimated the factors of the equations. 
Thus, we expect the $\mathcal{O}(100)~\%$ uncertainties for each value, {\it e.g.}, the upper values in Eq.~(\ref{eq: rough_LFV_constraint}). 
Despite this fact, it is an important observation that non-zero $CP$-violating phases are necessary not only in the PMNS matrix but also in the matrix $O_{\alpha a}$ to avoid the constraint on the LFV decays. 
This conclusion would not change even in a more detailed analysis because the rough criteria in Eq.~(\ref{eq: rough_LFV_constraint}) and the $CP$-conserving values are different by many orders of magnitude. 
Points in Figs.~\ref{fig: Xmue} and~\ref{fig: XXmue} would be a good first candidate in searching for benchmark scenarios. 

\subsection{Dark matter}

As explained above, $m_{N^\alpha}^{} > m_S^{}$ is preferred to explain the neutrino oscillation data and avoid the constraint on the LFV processes. 
Thus, $\eta$ is the dark matter particle in this model. 
The freeze-out mechanism can produce the observed relic density. 
Such a dark matter particle is strongly constrained by direct searches of the dark matter nucleus scattering~\cite{XENON:2018voc, PandaX-4T:2021bab, LZ:2022lsv}. 
We employ the latest result of the LZ collaboration~\cite{LZ:2022lsv} as a constraint. 
 
\begin{figure}[t]
\begin{center}
\includegraphics[width=0.6\textwidth]{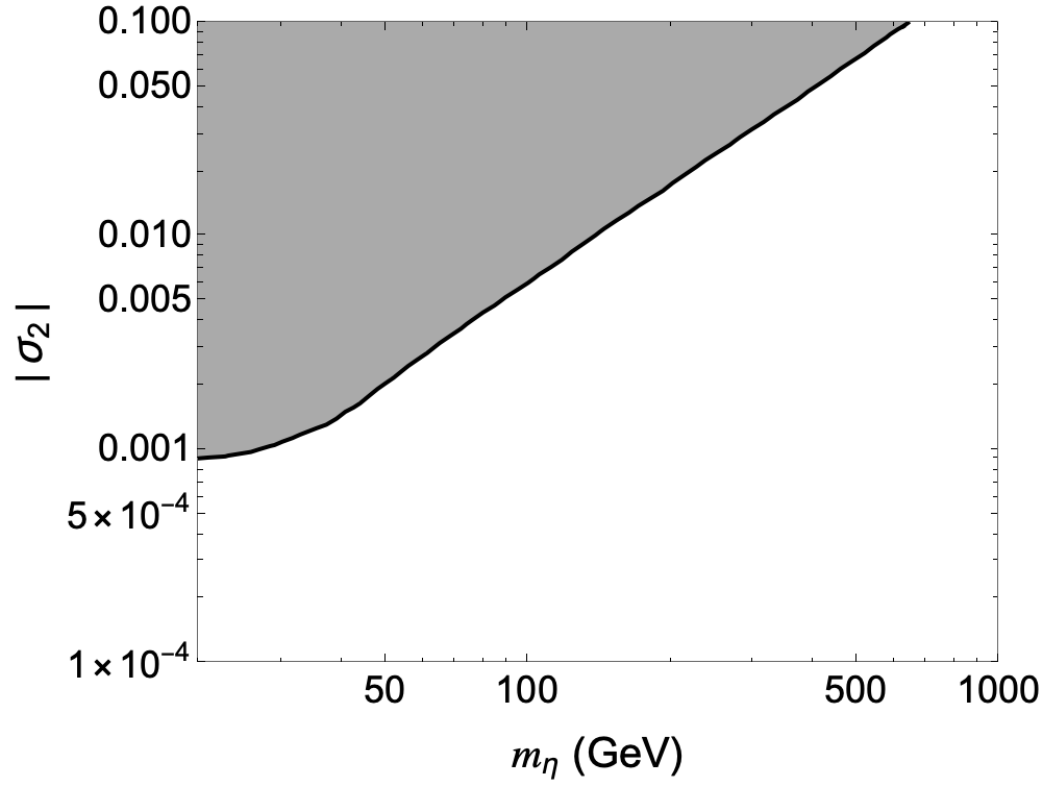}
\caption{The constraint on $m_\eta$--$|\sigma_2|$ plane from the latest constraint on the direct detection.}
\label{fig: LZconstraint}
\end{center}
\end{figure}

The spin-independent cross section for $\eta N \to \eta N$, where $N$ is a nucleon, is approximately given by 
\begin{equation}
\label{eq: sigma_SI}
\sigma_\mathrm{SI} \simeq \frac{ g^2 m_N^2 v^2 }{ 4 \pi (m_\eta + m_N)^2 m_{H_1}^4 } (\sigma_1 \cos\beta + \sigma_2 \sin \beta)^2, 
\end{equation}
where $m_N \simeq 1~\mathrm{GeV}$ is the nucleon mass, $g \simeq 10^{-3}$~\cite{Shifman:1978zn} is the coupling constant of the nucleon-Higgs interaction. 
In Eq.~(\ref{eq: sigma_SI}), we neglect the tiny mixing among the neutral Higgs bosons and the interaction between the nucleons and $H_{2,3}$, which is suppressed by $\cot\beta$. 
Since $\sigma_1 \cos \beta + \sigma_2 \sin \beta \simeq \sigma_2$ for large $\tan \beta$, $\sigma_\mathrm{SI}$ is almost independent of $\sigma_1$ and is a function of $m_\eta$ and $\sigma_2$. 
Fig.~\ref{fig: LZconstraint} shows the excluded region by the latest LZ result on the $m_\eta$--$|\sigma_2|$ plane.  

Under this constraint, there are two scenarios to produce the observed relic density: (i) Higgs portal DM scenario and (ii) Heavy DM scenario. 
In the Higgs portal DM scenario, we assume $m_\eta \simeq m_{H_1}/2 \simeq 63~\mathrm{GeV}$. Although the direct-detection constraint is severe for such a dark matter, the resonance effect of the on-shell Higgs boson can enhance the annihilation cross section $\eta \eta \to H_1 \to f \bar{f} (gg)$, and the correct relic density can be reproduced with $|\sigma_2|$ tiny enough to avoid the constraint. 

In the heavy DM scenario, we assume relatively heavy $\eta$ whose pair can be annihilated into a pair of the additional Higgs bosons $H_i H_j$ and/or $H^+H^-$. 
Neglecting small mixing among neutral Higgs bosons and large $\tan \beta$, 
the contact interactions between $\eta$ and the additional Higgs bosons are given by
\begin{equation}
\mathcal{L}_{\eta \eta H} \simeq  - \frac{ \sigma_1 }{ 2 } \Bigl\{
	 |H^\pm|^2 + \frac{ 1 }{ 2 } \bigl( H_2^2 + H_3^2 \bigr) \Bigr\}.
\end{equation}
Then, we note that the relic abundance is determined by $m_\eta$ and $\sigma_1$, not $\sigma_2$. 
Thus, the correct relic abundance can be reproduced almost independently of the direct detection constraint.\footnote{$\sigma_1$ coupling also induces three-point interaction $(\sigma_1v \sin 2 \beta) H_2 \eta^2$, which contributes to the $\eta$-$N$ scattering. However, it is negligible because the $\bar{N}NH_2$ coupling is suppressed by small $\cot \beta$ including the Yukawa interaction between $H_2$ and quarks. It also has a suppression from $\sin 2 \beta \ll 1$ in the coupling constant. }

\subsection{Benchmark scenarios} 

According to the above analysis, we propose the following two benchmark scenarios, Scenario I and Scenario II. In both scenarios, the neutrinos have the normal ordering mass hierarchy. It would be possible to find a scenario for the inverted mass hierarchy by using the above discussions although we do not show it in this paper. 
 
\subsubsection{Scenario I}

In Scenario I, we assume the following input values for $M_h^2$ in both benchmark scenarios. 
\begin{align}
& M = 270~\mathrm{GeV}, \quad \tilde{m}_A = 360~\mathrm{GeV}, \quad \tilde{m}_H^2 - M^2 = 300 ~\mathrm{GeV^2}, \nonumber \\
&  \tan \beta = 18, \quad s_{\beta - \tilde{\alpha}} = 1.0, \quad \theta_5 = - 0.315\pi \simeq -0.990. 
\end{align}
$\tilde{m}_h$ is determined to obtain the observed Higgs boson mass $H_1 = 125~\mathrm{GeV}$ and is given by $\tilde{m}_h \simeq 126~\mathrm{GeV}$. 

These input values lead to the following mass eigenvalues and the mixing matrix; 
\begin{align}
& m_{H_1} \simeq 125~\mathrm{GeV}, \quad m_{H_2} \simeq 224~\mathrm{GeV}, \quad m_{H_3} \simeq 391~\mathrm{GeV}, \nonumber \\ 
& R \simeq 
\begin{pmatrix}
0.997 & 0.0657 & -0.0307 \\
-0.0433 &  0.880 &  0.474 \\
0.0581 &  -0.471 & 0.880 \\
\end{pmatrix}
. \nonumber 
\end{align}
Since $|R_{11}-1| \simeq 0.003$, it is expected to cause $\mathcal{O}(0.1)~\%$ deviations in the Higgs couplings from the SM prediction, which is small enough to satisfy the current LHC data. 

$m_{H^\pm}$ is degenerated to $m_{H_3}$ to avoid the constraint from oblique paramters~\cite{Peskin:1990zt}, in particular the $T$ parameter;  
\begin{equation}
m_{H^\pm} = m_{H_3} \simeq 391~\mathrm{GeV}. 
\end{equation}
In the limit $m_{H^\pm} = m_{H_3}$, the custodial symmetry of the Higgs potential~\cite{Sikivie:1980hm, Haber:1992py, Pomarol:1993mu, Gerard:2007kn, Haber:2010bw, Grzadkowski:2010dj, Aiko:2020atr} is restored and the new physics contribution vanishes. We have also evaluated the $S$ and $U$ parameters by using formulas in Ref.~\cite{Branco:2011iw}. 
The result is given by 
\begin{equation}
S \simeq -0.011, \quad T \simeq 0, \quad U \simeq -1.9\times 10^{-5}. 
\end{equation}
The current result of the electroweak global fit gives $S = -0.01\pm 0.07$ and $T = 0.04\pm 0.06$ with $U = 0$~\cite{PDG}. 
Thus, the benchmark scenario satisfies this constraint. 
We note that the other new particles do not contribute to the $STU$ parameters because they are the isospin singlets. 

The original scalar quartic couplings in $V_\mathrm{THDM}$ is evaluated by using the above input values as follows; 
\begin{align}
\begin{array}{l}
\lambda_1 \simeq 1.87, \quad \lambda_2 \simeq 0.263, \quad \lambda_3 \simeq 2.91,  \\[5pt]
\lambda_4 \simeq -1.71, \quad \lambda_5^R \simeq -0.937,  \quad 
\lambda_5^I \simeq 1.43. \\
\end{array}
\end{align}
We have checked that these quartic couplings satisfy the conditions for the vacuum stability in Eqs.~(\ref{eq: VS_cond1})--(\ref{eq: VS_cond2}) and those for the perturbative unitarity in Eqs.~(\ref{eq: PU_cond1})--(\ref{eq: PU_cond6}). 
In addition, we expect that they are small enough to satisfy the triviality bound up to $10~\mathrm{TeV}$~\cite{Kanemura:2023wap}. 

Next, we consider the $Z_2$-odd particles. 
We assume the following mass spectrum; 
\begin{align}
\label{eq: Z2odd mass Scenario I}
& m_S = 325~\mathrm{GeV}, \quad m_\eta = 63~\mathrm{GeV},  \nonumber \\
& (m_{N^1}, m_{N^2}, m_{N^3}) = (2500, 3000, 3500)~\mathrm{GeV}. 
\end{align}
We note that $m_\eta \simeq m_{H_1}/2$, and the DM relic density is explained in the Higgs portal scenario as discussed below. 

The new physics scales in neutrino mass, $\Lambda_\alpha$ ($\alpha = 1$, 2, 3), are determined by $m_{H^\pm}$, $m_S$, $m_\eta$ and $m_{N^\alpha}$ as follows\footnote{We have used the Monte Carlo integration in evaluating loop integrals in $\Lambda_\alpha$ with $10^9$ sample points and have checked the uncertainties are smaller than $\mathcal{O}(0.1)~\%$.}
\begin{equation}
(\Lambda_1, \Lambda_2, \Lambda_3) \simeq (269, 271, 276)~\mathrm{GeV}.
\end{equation}
In addition, we assume the following input values for the other parameters necessary for determining $h_i^\alpha$; 
\begin{align}
& \kappa \tan \beta = 30, \quad m_{\nu^1} = 4.76~\mathrm{meV}, \quad 
\delta = 1.12\pi, \quad \alpha_1 = 1.51\pi, \quad \alpha_2 = 1.12\pi, \nonumber \\
& \gamma_1 = 0.866\pi, \quad \gamma_2 = 0.861\pi, \quad \gamma_3 = 1.56\pi, \quad \psi_1 = -0.0971, \quad \psi_2 = 0, \quad \psi_2 = 0.109\pi. 
\end{align}
Then, $h_i^\alpha$ are given by 
\begin{equation}
\label{eq: Yukawa matrix Scenario I}
\begin{pmatrix}
h^1_e & h^1_\mu & h^1_\tau \\
h^2_e & h^2_\mu & h^2_\tau \\
h^3_e & h^3_\mu & h^3_\tau \\
\end{pmatrix}
\simeq 
\begin{pmatrix}
2.85 e^{0.955\pi i} & 0.0103 e^{0.386\pi i} & 0.00242 e^{-0.0717\pi i} \\
2.31 e^{-0.438\pi i}& 0.0188 e^{-0.928\pi i} & 0.00208 e^{-0.728\pi i}\\
2.04 e^{-0.480\pi i} & 0.0543 e^{-0.0291\pi i} & 0.00221 e^{-0.0828\pi i} \\
\end{pmatrix}, 
\end{equation}
where we used the neutrino oscillation data in the normal ordering case in Table~\ref{table: neutrino oscillation data}. 
We can see the hierarchy $|h_e^\alpha| \gg |h_\mu^\alpha| > |h_\tau^\alpha|$ which is expected from the Casas-Ibarra parametrization, $h^\alpha_i \propto m_{\ell^i}^{-1}$. 

\begin{table}[b]
\begin{center}
\caption{Branching ratios for LFV decays in Scenario I}
\label{table: LFV1}
\begin{tabular}{|c|c|c|}\hline
 & Predictions & Upper bounds \\ \hline
 $\mu \to e \gamma$ & $2.95\times 10^{-14}$ & $3.1 \times 10^{-13}$~\cite{MEGII:2023ltw} \\ 
 $\tau \to e \gamma$ & $4.73\times 10^{-15}$ & $3.3 \times 10^{-8}$~\cite{BaBar:2009hkt} \\ 
  $\tau \to \mu \gamma$ & $2.49\times 10^{-18}$ & $4.4 \times 10^{-8}$~\cite{BaBar:2009hkt} \\ \hline
\end{tabular}
\hspace{0.05\textwidth}
\begin{tabular}{|c|c|c|}\hline
 & Predictions & Upper bounds \\ \hline
$\mu \to 3e$ & $4.68\times 10^{-13}$ & $1.0 \times 10^{-12}$~\cite{SINDRUM:1987nra} \\ 
$\tau \to 3e$ & $4.84\times 10^{-10}$ & $2.7 \times 10^{-8}$~\cite{Hayasaka:2010np} \\ 
$\tau \to 3\mu$ & $4.88\times 10^{-20}$ & $2.1 \times 10^{-8}$~\cite{Hayasaka:2010np} \\ \hline
\end{tabular}
\begin{tabular}{|c|c|c|}\hline
 & Predictions & Upper bounds \\ \hline
$\tau \to e \mu \overline{e}$ & $1.14 \times 10^{-16}$ & $1.8 \times 10^{-8}$~\cite{Hayasaka:2010np}  \\ 
$\tau \to \mu \mu \overline{e}$ & $5.77\times 10^{-17}$ & $1.7 \times 10^{-8}$~\cite{Hayasaka:2010np}  \\ 
$\tau \to e e \overline{\mu}$ & $1.46 \times 10^{-13}$ & $1.5\times 10^{-8}$~\cite{Hayasaka:2010np}  \\ 
$\tau \to e \mu \overline{\mu}$ & $1.14 \times 10^{-16}$ & $2.7 \times 10^{-8}$~\cite{Hayasaka:2010np}  \\ \hline
\end{tabular}
\end{center}
\end{table}
This Yukawa interaction $h_i^\alpha$ induces the LFV decays as discussed in Sec.~\ref{subsec: LFV}. 
Their branching ratios and current upper bounds are summarized in Table~\ref{table: LFV1}. The current constraints can be avoided for all processes. 

The DM relic density can be explained by the freeze-out mechanism through the annihilation $\eta \eta \to H_1 \to f \bar{f}$. 
As explained above, the annihilation cross section can be large enough due to the resonance effect of $H_1$. 
$|\sigma_2|$ determines the relic density and is assumed as 
\begin{equation}
\sigma_2 = 1.1\times 10^{-3}. 
\end{equation}
This and $m_\eta = 63~\mathrm{GeV}$ reproduce the observed relic density $\Omega_\mathrm{DM} h^2 = 0.12$~\cite{Planck:2018vyg}. 
In addition, we can see that the direct detection constraint can avoided in Fig.~\ref{fig: LZconstraint}. 

Finally, we evaluate the eEDM. 
By using Eqs.~(\ref{eq: Yukawa matrix Scenario I}) and (\ref{eq: Z2odd mass Scenario I}), the contribution from $h_i^\alpha$ is evaluated as 
\begin{equation}
d_e^h \simeq -5.82 \times 10^{-29}~\mathrm{e\ cm}. 
\end{equation}
On the other hand, the contribution from $\theta_5$ is determined by the mass spectrum of the additional Higgs bosons and the mixing matrix $O$ as follows; 
\begin{equation}
d_e^{\theta_5} \simeq 5.79 \times 10^{-29}~\mathrm{e\ cm}. 
\end{equation}
Thus, the new physics contribution is given by
\begin{equation}
d_e \simeq -3 \times 10^{-31}~\mathrm{e\ cm}. 
\end{equation}
Each contribution is larger than the current constraint $|d_e| < 4.2 \times 10^{-30}~\mathrm{e\ cm}$; however the sum of them is lower than it. 
Consequently, Scenario I avoids the current eEDM constraint although the size of the $CP$-violating phase is not small; $|\theta_5 |\simeq  0.990$. 

We have not fixed some quartic couplings in the Higgs potential because their values are irrelevant in the above discussion. 
We can choose their value to satisfy the vacuum stability conditions in Eqs.~(\ref{eq: VS_cond3})--(\ref{eq: VS_cond5}).

\subsubsection{Scenario II}

In Scenario II, almost all parameters have the same values as in Scenario I. 
We show only parameters that are different or are not explicitly shown in Scenario I. 

The most important difference is the mass of $\eta$. 
In Scenario II, we use the heavy DM scenario to explain the observed DM relic density, where the relic density is determined by the annihilation into a pair of the additional Higgs bosons. 
Thus, $\eta$ must be heavier than at least one of the additional Higgs bosons. 
We assume that 
\begin{equation}
m_\eta = 250~\mathrm{GeV}, \nonumber 
\end{equation}
and the other particles have the same mass as in Scenario I. 

$\Lambda_\alpha$ is given by 
\begin{equation}
(\Lambda_1, \Lambda_2, \Lambda_3) \simeq (277, 279, 284)~\mathrm{GeV}.
\end{equation}
The other parameters required for $h_i^\alpha$ are assumed to be 
\begin{align}
& \kappa \tan \beta = 30, \quad m_{\nu^1} = 7.17~\mathrm{meV}, \quad 
\delta = 1.22\pi, \quad \alpha_1 = 1.07\pi, \quad \alpha_2 = 0.415\pi, \nonumber \\
& \gamma_1 = 1.19\pi, \quad \gamma_2 = 1.97\pi, \quad \gamma_3 = 0.132\pi, \quad \psi_1 = -0.0852, \quad \psi_2 = 0, \quad \psi_2 = 0.0926\pi. 
\end{align}
Then the Yukawa couplings $h^\alpha_i$ are 
\begin{equation}
\label{eq: Yukawa matrix Scenario II}
\begin{pmatrix}
h^1_e & h^1_\mu & h^1_\tau \\
h^2_e & h^2_\mu & h^2_\tau \\
h^3_e & h^3_\mu & h^3_\tau \\
\end{pmatrix}
\simeq 
\begin{pmatrix}
2.66 e^{0.956\pi i} & 0.0137 e^{-0.222\pi i} & 0.00159 e^{0.854\pi i} \\
2.82 e^{0.658\pi i}& 0.0473 e^{0.755\pi i} & 0.00167 e^{-0.908\pi i}\\
2.62 e^{-0.102 \pi i} & 0.0586 e^{-0.884\pi i} & 0.00207 e^{0.971\pi i} \\
\end{pmatrix}. 
\end{equation}
The predictions for the LFV processes are shown in Table.~\ref{table: LFV2}, which are consistent with the current upper bounds. 
\begin{table}[t]
\begin{center}
\caption{Branching ratios for LFV decays in Scenario II}
\label{table: LFV2}
\begin{tabular}{|c|c|c|}\hline
 & Predictions & Upper bounds \\ \hline
 $\mu \to e \gamma$ & $5.08 \times 10^{-14}$ & $3.1 \times 10^{-13}$~\cite{MEGII:2023ltw} \\ 
 $\tau \to e \gamma$ & $1.56\times 10^{-15}$ & $3.3 \times 10^{-8}$~\cite{BaBar:2009hkt} \\ 
  $\tau \to \mu \gamma$ & $1.33\times 10^{-18}$ & $4.4 \times 10^{-8}$~\cite{BaBar:2009hkt} \\ \hline
\end{tabular}
\hspace{0.05\textwidth}
\begin{tabular}{|c|c|c|}\hline
 & Predictions & Upper bounds \\ \hline
$\mu \to 3e$ & $2.79 \times 10^{-13}$ & $1.0 \times 10^{-12}$~\cite{SINDRUM:1987nra} \\ 
$\tau \to 3e$ & $1.50\times 10^{-10}$ & $2.7 \times 10^{-8}$~\cite{Hayasaka:2010np} \\ 
$\tau \to 3\mu$ & $7.76\times 10^{-20}$ & $2.1 \times 10^{-8}$~\cite{Hayasaka:2010np} \\ \hline
\end{tabular}
\begin{tabular}{|c|c|c|}\hline
 & Predictions & Upper bounds \\ \hline
$\tau \to e \mu \overline{e}$ & $1.60 \times 10^{-16}$ & $1.8 \times 10^{-8}$~\cite{Hayasaka:2010np}  \\ 
$\tau \to \mu \mu \overline{e}$ & $2.62\times 10^{-17}$ & $1.7 \times 10^{-8}$~\cite{Hayasaka:2010np}  \\ 
$\tau \to e e \overline{\mu}$ & $6.95 \times 10^{-13}$ & $1.5\times 10^{-8}$~\cite{Hayasaka:2010np}  \\ 
$\tau \to e \mu \overline{\mu}$ & $1.60 \times 10^{-16}$ & $2.7 \times 10^{-8}$~\cite{Hayasaka:2010np}  \\ \hline
\end{tabular}
\end{center}
\end{table}

The DM relic density can be explained by the freeze-out mechanism via $\eta \eta \to H_2 H_2$, which is induced by $\sigma_1$.  
By taking
\begin{equation}
\sigma_1 = 0.0841, 
\end{equation}
we can explain the whole DM relic density by $\eta$. 
We note that this coupling is almost independent of the constraint from the DM direct detection as discussed in Sec.~\ref{subsec: DM}. 
The elastic scattering $\eta N \to \eta N$ is induced by $\sigma_2$, and the constraint can be avoided by taking the same value as in Scenario I. 

Finally, we discuss the eEDM. 
The contribution from $h_i^\alpha$ is given by 
\begin{equation}
d_e^h = -5.05 \times 10^{-29}~\mathrm{e\, cm}. 
\end{equation}
$\theta_5$ is assumed to be 
\begin{equation}
\theta_5 = -0.318 \pi \simeq -0.999.  
\end{equation}
The other input parameters for $V_\mathrm{THDM}$ are the same as in Scenario I. 
This change of $\theta_5$ leads to a slightly different mass spectrum; 
\begin{equation}
m_{H_2} \simeq 222~\mathrm{GeV}, \quad m_{H_3} = m_{H^\pm} \simeq 392~\mathrm{GeV}. 
\end{equation}
The eEDM generated by the Higgs sector is evaluated as 
\begin{equation}
d_e^{\theta_5} \simeq 4.93 \times 10^{-29}~\mathrm{e\, cm}. 
\end{equation}
Therefore, the total new contribution is given by 
\begin{equation}
d_e \simeq -1\times 10^{-30}~\mathrm{e\, cm}, 
\end{equation}
which is consistent with the current limit. 
Scenario II can also avoid the severe eEDM constraint while keeping large $|\theta_5| \simeq 0.999$.

\section{Discussions and conclusion}
\label{sec: conclusion}

In this section, we present some comments about the benchmark scenarios. 
In both benchmark scenarios, the masses of the additional Higgs bosons are a few hundred GeV, and they are expected to be tested at future HL-LHC~\cite{Aiko:2020ksl}. 
In particular, $H_2, H_3 \to \tau \tau$ channels are so effective in searching for them. 
$H_{2,3} \to Z {H}_1$ and $H_{2,3} \to H_1 H_1$ induced by the mixing are also expected to be a discovery channel in the benchmark scenarios. 
$S^\pm$ have the mass $m_S = 325~\mathrm{GeV}$, and a pair of them can be produced in high-energy hadron colliders (LHC, and HL-LHC) and future high-energy $e^+ e^-$ colliders (the upgraded ILC and FCC-ee). 
Their main decay mode is given by $S^\pm \to H^{\pm \ast} \eta \to \tau \nu \eta$. 
Since $\eta$ and $\nu$ are missing energy, the final state of $S^+ S^-$ is $\tau \bar{\tau} \cancel{E}$, which is tested at future stau searches. 
Although $N^\alpha$ have heavy masses, a pair of them can be produced at future high energy $e^+ e^-$ colliders with the beam energy $\sqrt{s} > 5~\mathrm{TeV}$. They decay into $S^\pm$ and charged leptons, which is mainly $e^\pm$ because of the hierarchy in $h_i^\alpha$. 
It would be difficult to search for $\eta$ at high-energy colliders; however, they would be tested in future DM searches. 

The LFV processes are also important to test the model. 
Because of the hierarchy in $h_i^\alpha$, the processes including $e^\pm$ are most significant. 
In particular, the predicted values for $\mu \to e \gamma$, $\mu \to 3e$, and $\tau \to 3e$ in both benchmark scenarios are close to or larger than the expected sensitivity at MEG-II~\cite{MEGII:2018kmf}, Mu3e~\cite{Blondel:2013ia, Mu3e:2020gyw} and Belle-II~\cite{Belle-II:2018jsg}, respectively. 
Thus, we expect that their signal or indication can be detected in each future experiment. 
We summarize the branching ratio in the benchmark scenarios and the expected sensitivity of these processes in Table~\ref{table: effective_LFV_channels}. 
\begin{table}[t]
\begin{center}
\caption{The branching ratio for $\mu \to e \gamma$, $\mu \to 3e$, and $\tau \to 3e$ and their expected sensitivity at future planed experiments.}
\label{table: effective_LFV_channels}
\begin{tabular}{|c|c|c|c|} \hline
& Scenario I & Scenario II & Expected sensitivity \\ \hline
$\mu \to e \gamma$ &\  $2.95 \times 10^{-14}$ \ & \ $5.08 \times 10^{-14}$ \ & $6 \times 10^{-14}$~\cite{MEGII:2018kmf} \\ \hline
$\mu \to 3e $ & \ $4.68 \times 10^{-13}$ \ & \ $2.79 \times 10^{-13}$ \ & $1 \times 10^{-16}$~\cite{Blondel:2013ia,Mu3e:2020gyw} \\ \hline
$\tau \to 3e $ & \ $4.84 \times 10^{-10}$ \ & \ $1.50 \times 10^{-10}$ \ & $4 \times 10^{-10}$~\cite{Belle-II:2018jsg} \\ \hline
\end{tabular}
\end{center}
\end{table}

In this paper, we have investigated two kinds of contributions to the eEDM at the leading order. The size of the next leading order (NLO) contribution is expected to be suppressed by $\mathcal{O}(\alpha)$. 
In the benchmark scenarios, it is roughly estimated by $d_e^\mathrm{NLO} = \mathcal{O}(10^{-31})$, which is one order smaller than the current experimental bound. 
Thus, the interference between $d_e^h$ and $d_e^{\theta_5}$ is not largely changed by the NLO contribution. 
We expect that $\theta_5$ can be $\mathcal{O}(1)$ even considering the eEDM at the NLO.
 
The future ACME-III experiment aims to improve the upper bound on the eEDM to $10^{-30}~\mathrm{e\, cm}$. 
Then, Scenario II can be tested by this improvement. 
On the other hand, the eEDM in Scenario I is smaller than the expected sensitivity. 
However, more future experiments would be able to reach the value in Scenario I. 
In this paper, we have not evaluated the neutron EDM (nEDM) or the proton EDM (pEDM) because they are suppressed by $\cot \beta$ in the quark Yukawa couplings, and the current constraints are weaker than that on the eEDM. Thus, we do not expect that they give an effective constraint on the model currently. 
However, their future improvements are important to test the $CP$-violation in the model. 
$h_i^\alpha$ do not contribute to the nEDM and the pEDM. 
Therefore, the nEDM and the pEDM can directly access to $\theta_5$, which plays an important role in electroweak baryogenesis. 

Finally, we shortly comment on EWBG in the model. 
For successful EWBG, we need to satisfy the second and third conditions of the Sakharov condition by new physics, {\it i.e.}, large enough $CP$ violation and departure from thermal equilibrium~\cite{Sakharov:1967dj}. 
$\theta_5$ is a source of $CP$ violation in the model. 
It is $\mathcal{O}(1)$ in the benchmark scenarios and is expected to be large enough to generate the observed baryon asymmetry according to the previous works~\cite{Turok:1990zg, Cline:1995dg, Fromme:2006cm, Cline:2011mm, Enomoto:2021dkl, Kanemura:2023juv, Aoki:2022bkg, Dorsch:2016nrg}. 
The departure from thermal equilibrium is achieved by the strongly first-order EWPT. 
In the model, EWPT can be strongly first order by the non-decoupling effect of the additional scalar bosons, which also induces the sizable deviation in the Higgs triple coupling~\cite{Kanemura:2002vm, Kanemura:2004ch, Kanemura:2004mg, Braathen:2019pxr, Braathen:2019zoh}. 
In our benchmark scenarios, by taking $\mu_S = 20~\mathrm{GeV}$, we have $R_{hhh} \equiv \Delta \lambda_{hhh}/\lambda_{hhh} \simeq 22~\%$, where we used the one-loop effective potential to evaluate $\Delta \lambda_{hhh}$. 
According to Refs.~\cite{Turok:1990zg, Cline:1995dg, Fromme:2006cm, Cline:2011mm, Enomoto:2021dkl, Kanemura:2023juv, Aoki:2022bkg, Dorsch:2016nrg}, this would be large enough to make EWPT strongly first-order. 
Consequently, we expect that EWBG can work in our benchmark scenarios.
Since a more detailed study needs complicated analyses using the effective potential, we will study it in our next paper to avoid making this paper too long~\cite{future}.

The non-decoupling effect in $R_{hhh}$ can be tested in the process of di-Higgs production at future high-energy colliders~\cite{Cepeda:2019klc, Fujii:2015jha, Bambade:2019fyw}. 
$R_{hhh}$ in the benchmark scenarios would be tested at the upgraded ILC with the beam energy $\sqrt{s} = 1~\mathrm{TeV}$, where the expected sensitivity is $10~\%$.  
Strongly first-order EWPT can also be probed by the gravitational waves~\cite{Grojean:2006bp, Espinosa:2010hh, Kakizaki:2015wua, Hashino:2016rvx, Hashino:2018wee} at future space-based gravitational wave observatories~\cite{LISA:2017pwj, Seto:2001qf, Corbin:2005ny}.
Furthermore, a sizable non-decoupling effect in EWPT can be tested by the observations of the primordial black holes~\cite{Hashino:2021qoq, Hashino:2022tcs}

In this paper, the electric dipole moment has been examined in a three-loop neutrino mass model with dark matter originally proposed in Ref.~\cite{Aoki:2008av}. The model contains a $CP$-violating phase in the Higgs potential which plays an important role in electroweak baryogenesis and is thus expected to explain the baryon asymmetry of the Universe simultaneously. However, such a $CP$-violating phase is severely constrained by the measurements of the electron electric dipole moment (eEDM), and a suppression mechanism for the electric dipole moment is necessary to explain the observed baryon asymmetry while avoiding the constraint. In this paper, we examine neutrino mass, lepton-flavor-violating processes, dark matter, and the eEDM in the model. We have shown that the eEDM can be suppressed by destructive interference between the $CP$-violating phases in the Higgs sector and the dark sector with large $CP$-violating phases. We have proposed some benchmark scenarios including $\mathcal{O}(1)$ $CP$-violating phases where tiny neutrino mass and dark matter can be explained while avoiding all the theoretical and experimental constraints. These $CP$-violating phases are expected to be large enough to generate the observed baryon asymmetry in the electroweak baryogenesis scenario. 

\section*{Acknowledgements}
The work of K.~E. was supported in part by the National Research Foundation of Korea (Grant No. RS-2024-00352537). The work of S. K. was supported in part by the JSPS KAKENHI Grant No. 20H00160 and No. 23K17691.


\begin{thebibliography}{99}
%\cite{Super-Kamiokande:1998kpq}
\bibitem{Super-Kamiokande:1998kpq}
Y.~Fukuda \textit{et al.} [Super-Kamiokande],
``Evidence for oscillation of atmospheric neutrinos,''
Phys. Rev. Lett. \textbf{81}, 1562-1567 (1998)
%doi:10.1103/PhysRevLett.81.1562
[arXiv:hep-ex/9807003 [hep-ex]].
%7950 citations counted in INSPIRE as of 18 Mar 2024

%\cite{SNO:2001kpb}
\bibitem{SNO:2001kpb}
Q.~R.~Ahmad \textit{et al.} [SNO],
``Measurement of the rate of $\nu_e+d \to p+p+e^-$ interactions produced by $^8$B solar neutrinos at the Sudbury Neutrino Observatory,''
Phys. Rev. Lett. \textbf{87}, 071301 (2001)
%doi:10.1103/PhysRevLett.87.071301
[arXiv:nucl-ex/0106015 [nucl-ex]]; 
%3471 citations counted in INSPIRE as of 18 Mar 2024
%
%\cite{SNO:2002tuh}
%\bibitem{SNO:2002tuh}
%Q.~R.~Ahmad \textit{et al.} [SNO],
``Direct evidence for neutrino flavor transformation from neutral current interactions in the Sudbury Neutrino Observatory,''
Phys. Rev. Lett. \textbf{89}, 011301 (2002)
%doi:10.1103/PhysRevLett.89.011301
[arXiv:nucl-ex/0204008 [nucl-ex]].
%4561 citations counted in INSPIRE as of 18 Mar 2024

%\cite{Planck:2018vyg}
\bibitem{Planck:2018vyg}
N.~Aghanim \textit{et al.} [Planck],
``Planck 2018 results. VI. Cosmological parameters,''
Astron. Astrophys. \textbf{641}, A6 (2020)
[erratum: Astron. Astrophys. \textbf{652}, C4 (2021)]
%doi:10.1051/0004-6361/201833910
[arXiv:1807.06209 [astro-ph.CO]].
%13039 citations counted in INSPIRE as of 10 Mar 2024

%\cite{Fields:2019pfx}
\bibitem{Fields:2019pfx}
B.~D.~Fields, K.~A.~Olive, T.~H.~Yeh and C.~Young,
``Big-Bang Nucleosynthesis after Planck,''
JCAP \textbf{03}, 010 (2020)
[erratum: JCAP \textbf{11}, E02 (2020)]
%doi:10.1088/1475-7516/2020/03/010
[arXiv:1912.01132 [astro-ph.CO]].
%275 citations counted in INSPIRE as of 18 Mar 2024

%\cite{ATLAS:2012yve}
\bibitem{ATLAS:2012yve}
G.~Aad \textit{et al.} [ATLAS],
``Observation of a new particle in the search for the Standard Model Higgs boson with the ATLAS detector at the LHC,''
Phys. Lett. B \textbf{716}, 1-29 (2012)
%doi:10.1016/j.physletb.2012.08.020
[arXiv:1207.7214 [hep-ex]]; 
%15105 citations counted in INSPIRE as of 18 Mar 2024
%
%\cite{CMS:2012qbp}
%\bibitem{CMS:2012qbp}
S.~Chatrchyan \textit{et al.} [CMS],
``Observation of a New Boson at a Mass of 125 GeV with the CMS Experiment at the LHC,''
Phys. Lett. B \textbf{716}, 30-61 (2012)
%doi:10.1016/j.physletb.2012.08.021
[arXiv:1207.7235 [hep-ex]].
%14641 citations counted in INSPIRE as of 18 Mar 2024

%\cite{Zee:1980ai}
\bibitem{Zee:1980ai}
A.~Zee,
``A Theory of Lepton Number Violation, Neutrino Majorana Mass, and Oscillation,''
Phys. Lett. B \textbf{93}, 389 (1980)
[erratum: Phys. Lett. B \textbf{95}, 461 (1980)]
%doi:10.1016/0370-2693(80)90349-4
%1235 citations counted in INSPIRE as of 18 Mar 2024

%\cite{Kuzmin:1985mm}
\bibitem{Kuzmin:1985mm}
V.~A.~Kuzmin, V.~A.~Rubakov and M.~E.~Shaposhnikov,
``On the Anomalous Electroweak Baryon Number Nonconservation in the Early Universe,''
Phys. Lett. B \textbf{155}, 36 (1985)
%doi:10.1016/0370-2693(85)91028-7
%3407 citations counted in INSPIRE as of 17 Mar 2024

%\cite{Kajantie:1996mn}
\bibitem{Kajantie:1996mn}
K.~Kajantie, M.~Laine, K.~Rummukainen and M.~E.~Shaposhnikov,
``Is there a~ hot electroweak phase transition at $m_H \gtrsim m_W$?,''
Phys. Rev. Lett. \textbf{77}, 2887-2890 (1996)
%doi:10.1103/PhysRevLett.77.2887
[arXiv:hep-ph/9605288 [hep-ph]]; 
%804 citations counted in INSPIRE as of 18 Mar 2024
%
%\cite{DOnofrio:2015gop}
%\bibitem{DOnofrio:2015gop}
M.~D'Onofrio and K.~Rummukainen,
``Standard model cross-over on the lattice,''
Phys. Rev. D \textbf{93}, no.2, 025003 (2016)
%doi:10.1103/PhysRevD.93.025003
[arXiv:1508.07161 [hep-ph]].
%162 citations counted in INSPIRE as of 18 Mar 2024

%\cite{Cheng:1980qt}
\bibitem{Cheng:1980qt}
T.~P.~Cheng and L.~F.~Li,
``Neutrino Masses, Mixings and Oscillations in SU(2) x U(1) Models of Electroweak Interactions,''
Phys. Rev. D \textbf{22}, 2860 (1980)
%doi:10.1103/PhysRevD.22.2860
%1292 citations counted in INSPIRE as of 18 Mar 2024

%\cite{Zee:1985id}
\bibitem{Zee:1985id}
A.~Zee,
``Quantum Numbers of Majorana Neutrino Masses,''
Nucl. Phys. B \textbf{264}, 99-110 (1986);
%doi:10.1016/0550-3213(86)90475-X
%558 citations counted in INSPIRE as of 18 Mar 2024
%
%\cite{Babu:1988ki}
%\bibitem{Babu:1988ki}
K.~S.~Babu,
``Model of 'Calculable' Majorana Neutrino Masses,''
Phys. Lett. B \textbf{203}, 132-136 (1988)
%doi:10.1016/0370-2693(88)91584-5
%814 citations counted in INSPIRE as of 18 Mar 2024

%\cite{Krauss:2002px}
\bibitem{Krauss:2002px}
L.~M.~Krauss, S.~Nasri and M.~Trodden,
``A Model for neutrino masses and dark matter,''
Phys. Rev. D \textbf{67}, 085002 (2003)
%doi:10.1103/PhysRevD.67.085002
[arXiv:hep-ph/0210389 [hep-ph]].
%378 citations counted in INSPIRE as of 18 Mar 2024

%\cite{Ma:2006km}
\bibitem{Ma:2006km}
E.~Ma,
``Verifiable radiative seesaw mechanism of neutrino mass and dark matter,''
Phys. Rev. D \textbf{73}, 077301 (2006)
%doi:10.1103/PhysRevD.73.077301
[arXiv:hep-ph/0601225 [hep-ph]].
%1469 citations counted in INSPIRE as of 10 Mar 2024

%\cite{Aoki:2008av}
\bibitem{Aoki:2008av}
M.~Aoki, S.~Kanemura and O.~Seto,
``Neutrino mass, Dark Matter and Baryon Asymmetry via TeV-Scale Physics without Fine-Tuning,''
Phys. Rev. Lett. \textbf{102}, 051805 (2009)
%doi:10.1103/PhysRevLett.102.051805
[arXiv:0807.0361 [hep-ph]]; 
%341 citations counted in INSPIRE as of 24 Sep 2023
%
%\cite{Aoki:2009vf}
%\bibitem{Aoki:2009vf}
%M.~Aoki, S.~Kanemura and O.~Seto,
``A Model of TeV Scale Physics for Neutrino Mass, Dark Matter and Baryon Asymmetry and its Phenomenology,''
Phys. Rev. D \textbf{80}, 033007 (2009)
%doi:10.1103/PhysRevD.80.033007
[arXiv:0904.3829 [hep-ph]]; 
%165 citations counted in INSPIRE as of 24 Sep 2023

%\cite{Gustafsson:2012vj}
\bibitem{Gustafsson:2012vj}
M.~Gustafsson, J.~M.~No and M.~A.~Rivera,
``Predictive Model for Radiatively Induced Neutrino Masses and Mixings with Dark Matter,''
Phys. Rev. Lett. \textbf{110}, no.21, 211802 (2013)
[erratum: Phys. Rev. Lett. \textbf{112}, no.25, 259902 (2014)]
%doi:10.1103/PhysRevLett.110.211802
[arXiv:1212.4806 [hep-ph]].
%186 citations counted in INSPIRE as of 18 Mar 2024

%\cite{Aoki:2022bkg}
\bibitem{Aoki:2022bkg}
M.~Aoki, K.~Enomoto and S.~Kanemura,
``Electroweak baryogenesis in the three-loop neutrino mass model with dark matter,''
Phys. Rev. D \textbf{107}, no.11, 115022 (2023)
%doi:10.1103/PhysRevD.107.115022
[arXiv:2212.14786 [hep-ph]].
%1 citations counted in INSPIRE as of 24 Sep 2023

%\cite{Nasri:2001ax}
\bibitem{Nasri:2001ax}
S.~Nasri and S.~Moussa,
``Model for small neutrino masses at the TeV scale,''
Mod. Phys. Lett. A \textbf{17}, 771-778 (2002)
%doi:10.1142/S0217732302007119
[arXiv:hep-ph/0106107 [hep-ph]];
%30 citations counted in INSPIRE as of 18 Mar 2024
%
%\cite{Kanemura:2011jj}
%\bibitem{Kanemura:2011jj}
S.~Kanemura, T.~Nabeshima and H.~Sugiyama,
``Neutrino Masses from Loop-Induced Dirac Yukawa Couplings,''
Phys. Lett. B \textbf{703}, 66-70 (2011)
%doi:10.1016/j.physletb.2011.07.047
[arXiv:1106.2480 [hep-ph]].
%71 citations counted in INSPIRE as of 18 Mar 2024

%\cite{Gu:2007ug}
\bibitem{Gu:2007ug}
P.~H.~Gu and U.~Sarkar,
``Radiative Neutrino Mass, Dark Matter and Leptogenesis,''
Phys. Rev. D \textbf{77}, 105031 (2008)
%doi:10.1103/PhysRevD.77.105031
[arXiv:0712.2933 [hep-ph]].
%120 citations counted in INSPIRE as of 18 Mar 2024

%\cite{Kanemura:2017haa}
\bibitem{Kanemura:2017haa}
S.~Kanemura, K.~Sakurai and H.~Sugiyama,
``Neutrino mass without lepton number violation, dark matter; and a strongly first-order phase transition,''
Phys. Rev. D \textbf{96}, no.9, 095024 (2017)
doi:10.1103/PhysRevD.96.095024
%[arXiv:1705.07040 [hep-ph]].
%6 citations counted in INSPIRE as of 18 Mar 2024

%\cite{Enomoto:2019mzl}
\bibitem{Enomoto:2019mzl}
K.~Enomoto, S.~Kanemura, K.~Sakurai and H.~Sugiyama,
``New model for radiatively generated Dirac neutrino masses and lepton flavor violating decays of the Higgs boson,''
Phys. Rev. D \textbf{100}, no.1, 015044 (2019)
%doi:10.1103/PhysRevD.100.015044
[arXiv:1904.07039 [hep-ph]].
%19 citations counted in INSPIRE as of 18 Mar 2024

%\cite{Aoki:2011zg}
\bibitem{Aoki:2011zg}
M.~Aoki, S.~Kanemura and K.~Yagyu,
``Triviality and vacuum stability bounds in the three-loop neutrino mass model,''
Phys. Rev. D \textbf{83}, 075016 (2011)
%doi:10.1103/PhysRevD.83.075016
[arXiv:1102.3412 [hep-ph]].
%71 citations counted in INSPIRE as of 24 Sep 2023

%\cite{ACME:2018yjb}
\bibitem{ACME:2018yjb}
V.~Andreev \textit{et al.} [ACME],
``Improved limit on the electric dipole moment of the electron,''
Nature \textbf{562}, no.7727, 355-360 (2018)
%doi:10.1038/s41586-018-0599-8
%699 citations counted in INSPIRE as of 18 Mar 2024

%\cite{Roussy:2022cmp}
\bibitem{Roussy:2022cmp}
T.~S.~Roussy, L.~Caldwell, T.~Wright, W.~B.~Cairncross, Y.~Shagam, K.~B.~Ng, N.~Schlossberger, S.~Y.~Park, A.~Wang and J.~Ye, \textit{et al.}
``An improved bound on the electron\textquoteright{}s electric dipole moment,''
Science \textbf{381}, no.6653, adg4084 (2023)
%doi:10.1126/science.adg4084
[arXiv:2212.11841 [physics.atom-ph]].
%110 citations counted in INSPIRE as of 18 Mar 2024

%\cite{Kanemura:2020ibp}
\bibitem{Kanemura:2020ibp}
S.~Kanemura, M.~Kubota and K.~Yagyu,
``Aligned CP-violating Higgs sector canceling the electric dipole moment,''
JHEP \textbf{08}, 026 (2020)
%doi:10.1007/JHEP08(2020)026
[arXiv:2004.03943 [hep-ph]].
%36 citations counted in INSPIRE as of 18 Mar 2024

%\cite{Pich:2009sp}
\bibitem{Pich:2009sp}
A.~Pich and P.~Tuzon,
``Yukawa Alignment in the Two-Higgs-Doublet Model,''
Phys. Rev. D \textbf{80}, 091702 (2009)
%doi:10.1103/PhysRevD.80.091702
[arXiv:0908.1554 [hep-ph]].
%379 citations counted in INSPIRE as of 18 Mar 2024

%\cite{Glashow:1976nt}
\bibitem{Glashow:1976nt}
S.~L.~Glashow and S.~Weinberg,
``Natural Conservation Laws for Neutral Currents,''
Phys. Rev. D \textbf{15} (1977), 1958
%doi:10.1103/PhysRevD.15.1958
%2088 citations counted in INSPIRE as of 28 Dec 2023

%\cite{Turok:1990zg}
\bibitem{Turok:1990zg}
N.~Turok and J.~Zadrozny,
``Electroweak baryogenesis in the two doublet model,''
Nucl. Phys. B \textbf{358}, 471-493 (1991)
%doi:10.1016/0550-3213(91)90356-3
%331 citations counted in INSPIRE as of 29 Jul 2024

%\cite{Cline:1995dg}
\bibitem{Cline:1995dg}
J.~M.~Cline, K.~Kainulainen and A.~P.~Vischer,
``Dynamics of two Higgs doublet CP violation and baryogenesis at the electroweak phase transition,''
Phys. Rev. D \textbf{54}, 2451-2472 (1996)
%doi:10.1103/PhysRevD.54.2451
[arXiv:hep-ph/9506284 [hep-ph]].
%214 citations counted in INSPIRE as of 29 Jul 2024

%\cite{Fromme:2006cm}
\bibitem{Fromme:2006cm}
L.~Fromme, S.~J.~Huber and M.~Seniuch,
``Baryogenesis in the two-Higgs doublet model,''
JHEP \textbf{11}, 038 (2006)
%doi:10.1088/1126-6708/2006/11/038
[arXiv:hep-ph/0605242 [hep-ph]].
%285 citations counted in INSPIRE as of 29 Jul 2024

%\cite{Cline:2011mm}
\bibitem{Cline:2011mm}
J.~M.~Cline, K.~Kainulainen and M.~Trott,
``Electroweak Baryogenesis in Two Higgs Doublet Models and B meson anomalies,''
JHEP \textbf{11}, 089 (2011)
%doi:10.1007/JHEP11(2011)089
[arXiv:1107.3559 [hep-ph]].
%195 citations counted in INSPIRE as of 29 Jul 2024

%\cite{Dorsch:2016nrg}
\bibitem{Dorsch:2016nrg}
G.~C.~Dorsch, S.~J.~Huber, T.~Konstandin and J.~M.~No,
``A Second Higgs Doublet in the Early Universe: Baryogenesis and Gravitational Waves,''
JCAP \textbf{05}, 052 (2017)
%doi:10.1088/1475-7516/2017/05/052
[arXiv:1611.05874 [hep-ph]].
%149 citations counted in INSPIRE as of 17 Mar 2024

%\cite{Enomoto:2021dkl}
\bibitem{Enomoto:2021dkl}
K.~Enomoto, S.~Kanemura and Y.~Mura,
``Electroweak baryogenesis in aligned two Higgs doublet models,''
JHEP \textbf{01}, 104 (2022)
%doi:10.1007/JHEP01(2022)104
[arXiv:2111.13079 [hep-ph]]; 
%24 citations counted in INSPIRE as of 18 Mar 2024
%\cite{Enomoto:2022rrl}
%\bibitem{Enomoto:2022rrl}
%K.~Enomoto, S.~Kanemura and Y.~Mura,
``New benchmark scenarios of electroweak baryogenesis in aligned two Higgs double models,''
JHEP \textbf{09}, 121 (2022)
%doi:10.1007/JHEP09(2022)121
[arXiv:2207.00060 [hep-ph]].
%17 citations counted in INSPIRE as of 18 Mar 2024

%\cite{Kanemura:2023juv}
\bibitem{Kanemura:2023juv}
S.~Kanemura and Y.~Mura,
``Electroweak baryogenesis via top-charm mixing,''
JHEP \textbf{09}, 153 (2023)
%doi:10.1007/JHEP09(2023)153
[arXiv:2303.11252 [hep-ph]].
%6 citations counted in INSPIRE as of 18 Mar 2024

%\cite{Abe:2013qla}
\bibitem{Abe:2013qla}
T.~Abe, J.~Hisano, T.~Kitahara and K.~Tobioka,
``Gauge invariant Barr-Zee type contributions to fermionic EDMs in the two-Higgs doublet models,''
JHEP \textbf{01}, 106 (2014)
[erratum: JHEP \textbf{04}, 161 (2016)]
%doi:10.1007/JHEP01(2014)106
[arXiv:1311.4704 [hep-ph]].
%109 citations counted in INSPIRE as of 18 Jan 2024

%\cite{Inoue:2014nva}
\bibitem{Inoue:2014nva}
S.~Inoue, M.~J.~Ramsey-Musolf and Y.~Zhang,
``CP-violating phenomenology of flavor conserving two Higgs doublet models,''
Phys. Rev. D \textbf{89}, no.11, 115023 (2014)
%doi:10.1103/PhysRevD.89.115023
[arXiv:1403.4257 [hep-ph]].
%149 citations counted in INSPIRE as of 19 Mar 2024

%\cite{Bian:2014zka}
\bibitem{Bian:2014zka}
L.~Bian, T.~Liu and J.~Shu,
``Cancellations Between Two-Loop Contributions to the Electron Electric Dipole Moment with a CP-Violating Higgs Sector,''
Phys. Rev. Lett. \textbf{115}, 021801 (2015)
%doi:10.1103/PhysRevLett.115.021801
[arXiv:1411.6695 [hep-ph]].
%60 citations counted in INSPIRE as of 19 Mar 2024

%\cite{Bian:2016zba}
\bibitem{Bian:2016zba}
L.~Bian and N.~Chen,
``Cancellation mechanism in the predictions of electric dipole moments,''
Phys. Rev. D \textbf{95}, no.11, 115029 (2017)
%doi:10.1103/PhysRevD.95.115029
[arXiv:1608.07975 [hep-ph]].
%24 citations counted in INSPIRE as of 19 Mar 2024

%\cite{Modak:2018csw}
\bibitem{Modak:2018csw}
T.~Modak and E.~Senaha,
``Electroweak baryogenesis via bottom transport,''
Phys. Rev. D \textbf{99}, no.11, 115022 (2019)
%doi:10.1103/PhysRevD.99.115022
[arXiv:1811.08088 [hep-ph]].
%29 citations counted in INSPIRE as of 18 Mar 2024

%\cite{Fuyuto:2019svr}
\bibitem{Fuyuto:2019svr}
K.~Fuyuto, W.~S.~Hou and E.~Senaha,
%``Cancellation mechanism for the electron electric dipole moment connected with the baryon asymmetry of the Universe,''
Phys. Rev. D \textbf{101}, no.1, 011901 (2020)
doi:10.1103/PhysRevD.101.011901
[arXiv:1910.12404 [hep-ph]].
%44 citations counted in INSPIRE as of 18 Mar 2024

%\cite{Cheung:2020ugr}
\bibitem{Cheung:2020ugr}
K.~Cheung, A.~Jueid, Y.~N.~Mao and S.~Moretti,
``Two-Higgs-doublet model with soft $CP$ violation confronting electric dipole moments and colliders,''
Phys. Rev. D \textbf{102}, no.7, 075029 (2020)
%doi:10.1103/PhysRevD.102.075029
[arXiv:2003.04178 [hep-ph]].
%36 citations counted in INSPIRE as of 18 Jan 2024

%\cite{Altmannshofer:2020shb}
\bibitem{Altmannshofer:2020shb}
W.~Altmannshofer, S.~Gori, N.~Hamer and H.~H.~Patel,
``Electron EDM in the complex two-Higgs doublet model,''
Phys. Rev. D \textbf{102}, no.11, 115042 (2020)
%doi:10.1103/PhysRevD.102.115042
[arXiv:2009.01258 [hep-ph]].
%36 citations counted in INSPIRE as of 18 Jan 2024

%\cite{Kanemura:2021atq}
\bibitem{Kanemura:2021atq}
S.~Kanemura, M.~Kubota and K.~Yagyu,
``Testing aligned CP-violating Higgs sector at future lepton colliders,''
JHEP \textbf{04}, 144 (2021)
%doi:10.1007/JHEP04(2021)144
[arXiv:2101.03702 [hep-ph]].
%12 citations counted in INSPIRE as of 18 Mar 2024

%\cite{Kanemura:2023jbz}
\bibitem{Kanemura:2023jbz}
S.~Kanemura, K.~Katayama, T.~Mondal and K.~Yagyu,
``Multiphoton signatures as a probe of CP violation in extended Higgs sectors,''
Phys. Rev. D \textbf{109}, no.3, 033002 (2024)
%doi:10.1103/PhysRevD.109.033002
[arXiv:2308.01772 [hep-ph]].
%1 citations counted in INSPIRE as of 18 Mar 2024


%\cite{Davidson:2005cw}
\bibitem{Davidson:2005cw}
S.~Davidson and H.~E.~Haber,
``Basis-independent methods for the two-Higgs-doublet model,''
Phys. Rev. D \textbf{72}, 035004 (2005)
[erratum: Phys. Rev. D \textbf{72}, 099902 (2005)]
%doi:10.1103/PhysRevD.72.099902
[arXiv:hep-ph/0504050 [hep-ph]].
%468 citations counted in INSPIRE as of 19 Mar 2024

\bibitem{Kanemura:2015ska}
S.~Kanemura and K.~Yagyu,
``Unitarity bound in the most general two Higgs doublet model,''
Phys. Lett. B \textbf{751}, 289-296 (2015)
%doi:10.1016/j.physletb.2015.10.047
[arXiv:1509.06060 [hep-ph]].
%108 citations counted in INSPIRE as of 11 Mar 2024

%\cite{Barger:1989fj}
\bibitem{Barger:1989fj}
V.~D.~Barger, J.~L.~Hewett and R.~J.~N.~Phillips,
``New Constraints on the Charged Higgs Sector in Two Higgs Doublet Models,''
Phys. Rev. D \textbf{41}, 3421-3441 (1990)
%doi:10.1103/PhysRevD.41.3421
%617 citations counted in INSPIRE as of 17 Mar 2024

%\cite{Grossman:1994jb}
\bibitem{Grossman:1994jb}
Y.~Grossman,
``Phenomenology of models with more than two Higgs doublets,''
Nucl. Phys. B \textbf{426}, 355-384 (1994)
%doi:10.1016/0550-3213(94)90316-6
[arXiv:hep-ph/9401311 [hep-ph]].
%341 citations counted in INSPIRE as of 17 Mar 2024

%\cite{Aoki:2009ha}
\bibitem{Aoki:2009ha}
M.~Aoki, S.~Kanemura, K.~Tsumura and K.~Yagyu,
``Models of Yukawa interaction in the two Higgs doublet model, and their collider phenomenology,''
Phys. Rev. D \textbf{80} (2009), 015017
%doi:10.1103/PhysRevD.80.015017
[arXiv:0902.4665 [hep-ph]].
%368 citations counted in INSPIRE as of 28 Dec 2023

%\cite{Su:2009fz}
\bibitem{Su:2009fz}
S.~Su and B.~Thomas,
``The LHC Discovery Potential of a Leptophilic Higgs,''
Phys. Rev. D \textbf{79}, 095014 (2009)
%doi:10.1103/PhysRevD.79.095014
[arXiv:0903.0667 [hep-ph]].
%86 citations counted in INSPIRE as of 17 Mar 2024

%\cite{Logan:2009uf}
\bibitem{Logan:2009uf}
H.~E.~Logan and D.~MacLennan,
``Charged Higgs phenomenology in the lepton-specific two Higgs doublet model,''
Phys. Rev. D \textbf{79}, 115022 (2009)
%doi:10.1103/PhysRevD.79.115022
[arXiv:0903.2246 [hep-ph]].
%88 citations counted in INSPIRE as of 17 Mar 2024

%\cite{Pontecorvo:1957cp}
\bibitem{ref: Pontecorvo}
B.~Pontecorvo,
``Mesonium and anti-mesonium,''
Sov. Phys. JETP \textbf{6} (1957), 429; 
%2278 citations counted in INSPIRE as of 01 Dec 2022
%
%\cite{Pontecorvo:1957qd}
%\bibitem{Pontecorvo:1957qd}
%B.~Pontecorvo,
``Inverse beta processes and nonconservation of lepton charge,''
Zh. Eksp. Teor. Fiz. \textbf{34} (1957), 247; 
%1700 citations counted in INSPIRE as of 01 Dec 2022
%
%\cite{Pontecorvo:1967fh}
%\bibitem{Pontecorvo:1967fh}
%B.~Pontecorvo,
``Neutrino Experiments and the Problem of Conservation of Leptonic Charge,''
Zh. Eksp. Teor. Fiz. \textbf{53} (1967), 1717-1725. 
%2430 citations counted in INSPIRE as of 01 Dec 2022

%\cite{Maki:1962mu}
\bibitem{Maki:1962mu}
Z.~Maki, M.~Nakagawa and S.~Sakata,
``Remarks on the unified model of elementary particles,''
Prog. Theor. Phys. \textbf{28} (1962), 870-880. 
%doi:10.1143/PTP.28.870
%4685 citations counted in INSPIRE as of 01 Dec 2022

%\cite{Deshpande:1977rw}
\bibitem{Deshpande:1977rw}
N.~G.~Deshpande and E.~Ma,
``Pattern of Symmetry Breaking with Two Higgs Doublets,''
Phys. Rev. D \textbf{18}, 2574 (1978)
%doi:10.1103/PhysRevD.18.2574
%945 citations counted in INSPIRE as of 11 Mar 2024

%\cite{Klimenko:1984qx}
\bibitem{Klimenko:1984qx}
K.~G.~Klimenko,
``On Necessary and Sufficient Conditions for Some Higgs Potentials to Be Bounded From Below,''
Theor. Math. Phys. \textbf{62}, 58-65 (1985)
%doi:10.1007/BF01034825
%119 citations counted in INSPIRE as of 11 Mar 2024

%\cite{Nie:1998yn}
\bibitem{Nie:1998yn}
S.~Nie and M.~Sher,
``Vacuum stability bounds in the two Higgs doublet model,''
Phys. Lett. B \textbf{449}, 89-92 (1999)
%doi:10.1016/S0370-2693(99)00019-2
[arXiv:hep-ph/9811234 [hep-ph]].
%194 citations counted in INSPIRE as of 11 Mar 2024

%\cite{Ferreira:2004yd}
\bibitem{Ferreira:2004yd}
P.~M.~Ferreira, R.~Santos and A.~Barroso,
``Stability of the tree-level vacuum in two Higgs doublet models against charge or CP spontaneous violation,''
Phys. Lett. B \textbf{603}, 219-229 (2004)
[erratum: Phys. Lett. B \textbf{629}, 114-114 (2005)]
%doi:10.1016/j.physletb.2004.10.022
[arXiv:hep-ph/0406231 [hep-ph]].
%196 citations counted in INSPIRE as of 11 Mar 2024

%\cite{Kanemura:2023wap}
\bibitem{Kanemura:2023wap}
S.~Kanemura and Y.~Mura,
``New application of the mass-dependent analysis for renormalization group equation to extended Higgs models,''
[arXiv:2310.15622 [hep-ph]].
%1 citations counted in INSPIRE as of 18 Jan 2024

%\cite{Lee:1977eg}
\bibitem{Lee:1977eg}
B.~W.~Lee, C.~Quigg and H.~B.~Thacker,
``Weak Interactions at Very High-Energies: The Role of the Higgs Boson Mass,''
Phys. Rev. D \textbf{16}, 1519 (1977)
%doi:10.1103/PhysRevD.16.1519
%2425 citations counted in INSPIRE as of 11 Mar 2024

%\cite{Luscher:1988gc}
\bibitem{Luscher:1988gc}
M.~Luscher and P.~Weisz,
``Is There a Strong Interaction Sector in the Standard Lattice Higgs Model?,''
Phys. Lett. B \textbf{212}, 472-478 (1988)
%doi:10.1016/0370-2693(88)91799-6
%206 citations counted in INSPIRE as of 17 Mar 2024

%\cite{Kanemura:1993hm}
\bibitem{Kanemura:1993hm}
S.~Kanemura, T.~Kubota and E.~Takasugi,
``Lee-Quigg-Thacker bounds for Higgs boson masses in a two doublet model,''
Phys. Lett. B \textbf{313}, 155-160 (1993)
%doi:10.1016/0370-2693(93)91205-2
[arXiv:hep-ph/9303263 [hep-ph]].
%442 citations counted in INSPIRE as of 11 Mar 2024
%\cite{Kanemura:2015ska}

%\cite{Akeroyd:2000wc}
\bibitem{Akeroyd:2000wc}
A.~G.~Akeroyd, A.~Arhrib and E.~M.~Naimi,
``Note on tree level unitarity in the general two Higgs doublet model,''
Phys. Lett. B \textbf{490}, 119-124 (2000)
%doi:10.1016/S0370-2693(00)00962-X
[arXiv:hep-ph/0006035 [hep-ph]].
%430 citations counted in INSPIRE as of 11 Mar 2024

%\cite{Ginzburg:2005dt}
\bibitem{Ginzburg:2005dt}
I.~F.~Ginzburg and I.~P.~Ivanov,
``Tree-level unitarity constraints in the most general 2HDM,''
Phys. Rev. D \textbf{72}, 115010 (2005)
%doi:10.1103/PhysRevD.72.115010
[arXiv:hep-ph/0508020 [hep-ph]].
%308 citations counted in INSPIRE as of 11 Mar 2024

%\cite{Aiko:2020ksl}
\bibitem{Aiko:2020ksl}
M.~Aiko, S.~Kanemura, M.~Kikuchi, K.~Mawatari, K.~Sakurai and K.~Yagyu,
``Probing extended Higgs sectors by the synergy between direct searches at the LHC and precision tests at future lepton colliders,''
Nucl. Phys. B \textbf{966}, 115375 (2021)
%doi:10.1016/j.nuclphysb.2021.115375
[arXiv:2010.15057 [hep-ph]].
%29 citations counted in INSPIRE as of 17 Mar 2024

%\cite{Arbey:2017gmh}
\bibitem{Arbey:2017gmh}
A.~Arbey, F.~Mahmoudi, O.~Stal and T.~Stefaniak,
``Status of the Charged Higgs Boson in Two Higgs Doublet Models,''
Eur. Phys. J. C \textbf{78}, no.3, 182 (2018)
%doi:10.1140/epjc/s10052-018-5651-1
[arXiv:1706.07414 [hep-ph]].
%183 citations counted in INSPIRE as of 19 Mar 2024

%\cite{Enomoto:2015wbn}
\bibitem{Enomoto:2015wbn}
T.~Enomoto and R.~Watanabe,
``Flavor constraints on the Two Higgs Doublet Models of Z$_{2}$ symmetric and aligned types,''
JHEP \textbf{05}, 002 (2016)
%doi:10.1007/JHEP05(2016)002
[arXiv:1511.05066 [hep-ph]].
%112 citations counted in INSPIRE as of 11 Mar 2024

%\cite{Haller:2018nnx}
\bibitem{Haller:2018nnx}
J.~Haller, A.~Hoecker, R.~Kogler, K.~M\"onig, T.~Peiffer and J.~Stelzer,
``Update of the global electroweak fit and constraints on two-Higgs-doublet models,''
Eur. Phys. J. C \textbf{78}, no.8, 675 (2018)
%doi:10.1140/epjc/s10052-018-6131-3
[arXiv:1803.01853 [hep-ph]].
%421 citations counted in INSPIRE as of 20 Mar 2024

%\cite{ALEPH:2013htx}
\bibitem{ALEPH:2013htx}
G.~Abbiendi \textit{et al.} [ALEPH, DELPHI, L3, OPAL and LEP],
``Search for Charged Higgs bosons: Combined Results Using LEP Data,''
Eur. Phys. J. C \textbf{73}, 2463 (2013)
%doi:10.1140/epjc/s10052-013-2463-1
[arXiv:1301.6065 [hep-ex]].
%310 citations counted in INSPIRE as of 22 Jan 2024

%\cite{HFLAV:2019otj}
\bibitem{HFLAV:2019otj}
Y.~S.~Amhis \textit{et al.} [HFLAV],
``Averages of b-hadron, c-hadron, and $\tau $-lepton properties as of 2018,''
Eur. Phys. J. C \textbf{81}, no.3, 226 (2021)
%doi:10.1140/epjc/s10052-020-8156-7
[arXiv:1909.12524 [hep-ex]].
%871 citations counted in INSPIRE as of 20 Mar 2024

%\cite{ATLAS:2018cur}
\bibitem{ATLAS:2018cur}
M.~Aaboud \textit{et al.} [ATLAS],
``Study of the rare decays of $B^0_s$ and $B^0$ mesons into muon pairs using data collected during 2015 and 2016 with the ATLAS detector,''
JHEP \textbf{04}, 098 (2019)
%doi:10.1007/JHEP04(2019)098
[arXiv:1812.03017 [hep-ex]];
%236 citations counted in INSPIRE as of 20 Mar 2024
%
%\cite{CMS:2019bbr}
%\bibitem{CMS:2019bbr}
A.~M.~Sirunyan \textit{et al.} [CMS],
``Measurement of properties of B$^0_\mathrm{s}\to\mu^+\mu^-$ decays and search for B$^0\to\mu^+\mu^-$ with the CMS experiment,''
JHEP \textbf{04}, 188 (2020)
%doi:10.1007/JHEP04(2020)188
[arXiv:1910.12127 [hep-ex]];
%132 citations counted in INSPIRE as of 20 Mar 2024
%
%\cite{LHCb:2021awg}
%\bibitem{LHCb:2021awg}
R.~Aaij \textit{et al.} [LHCb],
``Measurement of the $B^0_s\to\mu^+\mu^-$ decay properties and search for the $B^0\to\mu^+\mu^-$ and $B^0_s\to\mu^+\mu^-\gamma$ decays,''
Phys. Rev. D \textbf{105}, no.1, 012010 (2022)
%doi:10.1103/PhysRevD.105.012010
[arXiv:2108.09283 [hep-ex]].
%169 citations counted in INSPIRE as of 20 Mar 2024

%\cite{ATLAS:2023djh}
\bibitem{ATLAS:2023djh}
 [ATLAS],
``Search for electroweak SUSY production in final states with two $\tau$-leptons in $\sqrt{s} = 13$ TeV $pp$ collisions with the ATLAS detector,''
ATLAS-CONF-2023-029; 
%8 citations counted in INSPIRE as of 19 Mar 2024
%
%\cite{CMS:2022syk}
%\bibitem{CMS:2022syk}
A.~Tumasyan \textit{et al.} [CMS],
``Search for direct pair production of supersymmetric partners of $\tau$ leptons in the final state with two hadronically decaying $\tau$ leptons and missing transverse momentum in proton-proton collisions at $\sqrt{s}$ = 13 TeV,''
Phys. Rev. D \textbf{108}, no.1, 012011 (2023)
%doi:10.1103/PhysRevD.108.012011
[arXiv:2207.02254 [hep-ex]].
%32 citations counted in INSPIRE as of 20 Mar 2024

%\cite{Minkowski:1977sc}
\bibitem{ref: Type-I seesaw}
P.~Minkowski,
``$\mu \to e\gamma$ at a Rate of One Out of $10^{9}$ Muon Decays?,''
Phys. Lett. B \textbf{67}, 421-428 (1977); 
%doi:10.1016/0370-2693(77)90435-X
%4293 citations counted in INSPIRE as of 11 Jan 2022
%
%\cite{Yanagida:1979as}
%\bibitem{Yanagida:1979as}
T.~Yanagida,
``Horizontal gauge symmetry and masses of neutrinos,''
Conf. Proc. C \textbf{7902131}, 95-99 (1979); 
%KEK-79-18-95.
%2001 citations counted in INSPIRE as of 11 Jan 2022
%
%\cite{Yanagida:1980xy}
%\bibitem{Yanagida:1980xy}
%T.~Yanagida,
``Horizontal Symmetry and Masses of Neutrinos,''
Prog. Theor. Phys. \textbf{64}, 1103 (1980); 
%doi:10.1143/PTP.64.1103
%708 citations counted in INSPIRE as of 11 Jan 2022
%
%\cite{Gell-Mann:1979vob}
%\bibitem{Gell-Mann:1979vob}
M.~Gell-Mann, P.~Ramond and R.~Slansky,
``Complex Spinors and Unified Theories,''
Conf. Proc. C \textbf{790927}, 315-321 (1979)
[arXiv:1306.4669 [hep-th]]; 
%3469 citations counted in INSPIRE as of 11 Jan 2022
%
%\cite{Mohapatra:1979ia}
%\bibitem{Mohapatra:1979ia}
R.~N.~Mohapatra and G.~Senjanovic,
``Neutrino Mass and Spontaneous Parity Nonconservation,''
Phys. Rev. Lett. \textbf{44}, 912 (1980); 
%doi:10.1103/PhysRevLett.44.912
%5879 citations counted in INSPIRE as of 11 Jan 2022
%
%\cite{Schechter:1980gr}
%\bibitem{Schechter:1980gr}
J.~Schechter and J.~W.~F.~Valle,
``Neutrino Masses in SU(2) x U(1) Theories,''
Phys. Rev. D \textbf{22}, 2227 (1980). 
%doi:10.1103/PhysRevD.22.2227
%3147 citations counted in INSPIRE as of 11 Jan 2022

%\cite{Casas:2001sr}
\bibitem{Casas:2001sr}
J.~A.~Casas and A.~Ibarra,
``Oscillating neutrinos and $\mu \to e, \gamma$,''
Nucl. Phys. B \textbf{618} (2001), 171-204
%doi:10.1016/S0550-3213(01)00475-8
[arXiv:hep-ph/0103065 [hep-ph]].
%1360 citations counted in INSPIRE as of 29 Jan 2024

%\cite{Abada:2018zra}
\bibitem{Abada:2018zra}
A.~Abada and T.~Toma,
``Electric dipole moments in the minimal scotogenic model,''
JHEP \textbf{04}, 030 (2018)
[erratum: JHEP \textbf{04}, 060 (2021)]
%doi:10.1007/JHEP04(2018)030
[arXiv:1802.00007 [hep-ph]].
%23 citations counted in INSPIRE as of 18 Jan 2024

%\cite{Fujiwara:2021vam}
\bibitem{Fujiwara:2021vam}
M.~Fujiwara, J.~Hisano and T.~Toma,
``Vanishing or non-vanishing rainbow? Reduction formulas of electric dipole moment,''
JHEP \textbf{10}, 237 (2021)
%doi:10.1007/JHEP10(2021)237
[arXiv:2106.03384 [hep-ph]].
%3 citations counted in INSPIRE as of 18 Jan 2024

%\cite{Barr:1990vd}
\bibitem{Barr:1990vd}
S.~M.~Barr and A.~Zee,
``Electric Dipole Moment of the Electron and of the Neutron,''
Phys. Rev. Lett. \textbf{65}, 21-24 (1990)
[erratum: Phys. Rev. Lett. \textbf{65}, 2920 (1990)]
%doi:10.1103/PhysRevLett.65.21
%625 citations counted in INSPIRE as of 19 Mar 2024

%\cite{Toma:2013zsa}
\bibitem{Toma:2013zsa}
T.~Toma and A.~Vicente,
``Lepton Flavor Violation in the Scotogenic Model,''
JHEP \textbf{01} (2014), 160
%doi:10.1007/JHEP01(2014)160
[arXiv:1312.2840 [hep-ph]].
%204 citations counted in INSPIRE as of 29 Jan 2024

%\cite{MEGII:2023ltw}
\bibitem{MEGII:2023ltw}
K.~Afanaciev \textit{et al.} [MEG II],
``A search for $\mu^+\to e^+\gamma$ with the first dataset of the MEG II experiment,''
[arXiv:2310.12614 [hep-ex]].
%7 citations counted in INSPIRE as of 30 Jan 2024

%\cite{SINDRUM:1987nra}
\bibitem{SINDRUM:1987nra}
U.~Bellgardt \textit{et al.} [SINDRUM],
``Search for the Decay $\mu^+ \to e^+ e^+ e^-$,''
Nucl. Phys. B \textbf{299}, 1-6 (1988)
%doi:10.1016/0550-3213(88)90462-2
%1017 citations counted in INSPIRE as of 30 Jan 2024

%\cite{ParticleDataGroup:2022pth}
\bibitem{PDG}
R.~L.~Workman \textit{et al.} [Particle Data Group],
``Review of Particle Physics,''
PTEP \textbf{2022}, 083C01 (2022)
%doi:10.1093/ptep/ptac097
%2632 citations counted in INSPIRE as of 10 Mar 2024

%\cite{XENON:2018voc}
\bibitem{XENON:2018voc}
E.~Aprile \textit{et al.} [XENON],
``Dark Matter Search Results from a One Ton-Year Exposure of XENON1T,''
Phys. Rev. Lett. \textbf{121}, no.11, 111302 (2018)
%doi:10.1103/PhysRevLett.121.111302
[arXiv:1805.12562 [astro-ph.CO]].
%2171 citations counted in INSPIRE as of 10 Mar 2024

%\cite{PandaX-4T:2021bab}
\bibitem{PandaX-4T:2021bab}
Y.~Meng \textit{et al.} [PandaX-4T],
``Dark Matter Search Results from the PandaX-4T Commissioning Run,''
Phys. Rev. Lett. \textbf{127}, no.26, 261802 (2021)
%doi:10.1103/PhysRevLett.127.261802
[arXiv:2107.13438 [hep-ex]].
%382 citations counted in INSPIRE as of 10 Mar 2024

%\cite{LZ:2022lsv}
\bibitem{LZ:2022lsv}
J.~Aalbers \textit{et al.} [LZ],
``First Dark Matter Search Results from the LUX-ZEPLIN (LZ) Experiment,''
Phys. Rev. Lett. \textbf{131}, no.4, 041002 (2023)
%doi:10.1103/PhysRevLett.131.041002
[arXiv:2207.03764 [hep-ex]].
%483 citations counted in INSPIRE as of 10 Mar 2024

%\cite{Shifman:1978zn}
\bibitem{Shifman:1978zn}
M.~A.~Shifman, A.~I.~Vainshtein and V.~I.~Zakharov,
``Remarks on Higgs Boson Interactions with Nucleons,''
Phys. Lett. B \textbf{78}, 443-446 (1978)
%doi:10.1016/0370-2693(78)90481-1
%680 citations counted in INSPIRE as of 10 Mar 2024

%\cite{Peskin:1990zt}
\bibitem{Peskin:1990zt}
M.~E.~Peskin and T.~Takeuchi,
``A New constraint on a strongly interacting Higgs sector,''
Phys. Rev. Lett. \textbf{65}, 964-967 (1990); 
%doi:10.1103/PhysRevLett.65.964
%2196 citations counted in INSPIRE as of 10 Mar 2024
%\cite{Peskin:1991sw}
%\bibitem{Peskin:1991sw}
%M.~E.~Peskin and T.~Takeuchi,
``Estimation of oblique electroweak corrections,''
Phys. Rev. D \textbf{46}, 381-409 (1992)
%doi:10.1103/PhysRevD.46.381
%2692 citations counted in INSPIRE as of 10 Mar 2024

%\cite{Sikivie:1980hm}
\bibitem{Sikivie:1980hm}
P.~Sikivie, L.~Susskind, M.~B.~Voloshin and V.~I.~Zakharov,
``Isospin Breaking in Technicolor Models,''
Nucl. Phys. B \textbf{173}, 189-207 (1980)
%doi:10.1016/0550-3213(80)90214-X
%500 citations counted in INSPIRE as of 17 Mar 2024

%\cite{Haber:1992py}
\bibitem{Haber:1992py}
H.~E.~Haber and A.~Pomarol,
``Constraints from global symmetries on radiative corrections to the Higgs sector,''
Phys. Lett. B \textbf{302}, 435-441 (1993)
%doi:10.1016/0370-2693(93)90423-F
[arXiv:hep-ph/9207267 [hep-ph]].
%59 citations counted in INSPIRE as of 17 Mar 2024

%\cite{Pomarol:1993mu}
\bibitem{Pomarol:1993mu}
A.~Pomarol and R.~Vega,
``Constraints on CP violation in the Higgs sector from the rho parameter,''
Nucl. Phys. B \textbf{413}, 3-15 (1994)
%doi:10.1016/0550-3213(94)90611-4
[arXiv:hep-ph/9305272 [hep-ph]].
%120 citations counted in INSPIRE as of 17 Mar 2024

%\cite{Gerard:2007kn}
\bibitem{Gerard:2007kn}
J.~M.~Gerard and M.~Herquet,
``A Twisted custodial symmetry in the two-Higgs-doublet model,''
Phys. Rev. Lett. \textbf{98}, 251802 (2007)
%doi:10.1103/PhysRevLett.98.251802
[arXiv:hep-ph/0703051 [hep-ph]].
%145 citations counted in INSPIRE as of 17 Mar 2024

%\cite{Haber:2010bw}
\bibitem{Haber:2010bw}
H.~E.~Haber and D.~O'Neil,
``Basis-independent methods for the two-Higgs-doublet model III: The CP-conserving limit, custodial symmetry, and the oblique parameters S, T, U,''
Phys. Rev. D \textbf{83}, 055017 (2011)
%doi:10.1103/PhysRevD.83.055017
[arXiv:1011.6188 [hep-ph]].
%223 citations counted in INSPIRE as of 17 Mar 2024

%\cite{Grzadkowski:2010dj}
\bibitem{Grzadkowski:2010dj}
B.~Grzadkowski, M.~Maniatis and J.~Wudka,
``The bilinear formalism and the custodial symmetry in the two-Higgs-doublet model,''
JHEP \textbf{11}, 030 (2011)
%doi:10.1007/JHEP11(2011)030
[arXiv:1011.5228 [hep-ph]].
%57 citations counted in INSPIRE as of 17 Mar 2024

%\cite{Aiko:2020atr}
\bibitem{Aiko:2020atr}
M.~Aiko and S.~Kanemura,
``New scenario for aligned Higgs couplings originated from the twisted custodial symmetry at high energies,''
JHEP \textbf{02}, 046 (2021)
%doi:10.1007/JHEP02(2021)046
[arXiv:2009.04330 [hep-ph]].
%14 citations counted in INSPIRE as of 17 Mar 2024

%\cite{Branco:2011iw}
\bibitem{Branco:2011iw}
G.~C.~Branco, P.~M.~Ferreira, L.~Lavoura, M.~N.~Rebelo, M.~Sher and J.~P.~Silva,
``Theory and phenomenology of two-Higgs-doublet models,''
Phys. Rept. \textbf{516}, 1-102 (2012)
%doi:10.1016/j.physrep.2012.02.002
[arXiv:1106.0034 [hep-ph]].
%2756 citations counted in INSPIRE as of 10 Mar 2024

%\cite{BaBar:2009hkt}
\bibitem{BaBar:2009hkt}
B.~Aubert \textit{et al.} [BaBar],
``Searches for Lepton Flavor Violation in the Decays tau+- $\to$ e+- gamma and tau+- $\to$ mu+- gamma,''
Phys. Rev. Lett. \textbf{104}, 021802 (2010)
%doi:10.1103/PhysRevLett.104.021802
[arXiv:0908.2381 [hep-ex]].
%773 citations counted in INSPIRE as of 12 Mar 2024

%\cite{Hayasaka:2010np}
\bibitem{Hayasaka:2010np}
K.~Hayasaka, K.~Inami, Y.~Miyazaki, K.~Arinstein, V.~Aulchenko, T.~Aushev, A.~M.~Bakich, A.~Bay, K.~Belous and V.~Bhardwaj, \textit{et al.}
``Search for Lepton Flavor Violating Tau Decays into Three Leptons with 719 Million Produced Tau+Tau- Pairs,''
Phys. Lett. B \textbf{687}, 139-143 (2010)
%doi:10.1016/j.physletb.2010.03.037
[arXiv:1001.3221 [hep-ex]].
%509 citations counted in INSPIRE as of 12 Mar 2024

%\cite{MEGII:2018kmf}
\bibitem{MEGII:2018kmf}
A.~M.~Baldini \textit{et al.} [MEG II],
``The design of the MEG II experiment,''
Eur. Phys. J. C \textbf{78}, no.5, 380 (2018)
%doi:10.1140/epjc/s10052-018-5845-6
[arXiv:1801.04688 [physics.ins-det]].
%346 citations counted in INSPIRE as of 17 Mar 2024

%\cite{Blondel:2013ia}
\bibitem{Blondel:2013ia}
A.~Blondel, A.~Bravar, M.~Pohl, S.~Bachmann, N.~Berger, M.~Kiehn, A.~Schoning, D.~Wiedner, B.~Windelband and P.~Eckert, \textit{et al.}
``Research Proposal for an Experiment to Search for the Decay $\mu \to eee$,''
[arXiv:1301.6113 [physics.ins-det]].
%407 citations counted in INSPIRE as of 17 Mar 2024

%\cite{Mu3e:2020gyw}
\bibitem{Mu3e:2020gyw}
K.~Arndt \textit{et al.} [Mu3e],
``Technical design of the phase I Mu3e experiment,''
Nucl. Instrum. Meth. A \textbf{1014}, 165679 (2021)
%doi:10.1016/j.nima.2021.165679
[arXiv:2009.11690 [physics.ins-det]].
%109 citations counted in INSPIRE as of 17 Mar 2024

%\cite{Belle-II:2018jsg}
\bibitem{Belle-II:2018jsg}
E.~Kou \textit{et al.} [Belle-II],
``The Belle II Physics Book,''
PTEP \textbf{2019}, no.12, 123C01 (2019)
[erratum: PTEP \textbf{2020}, no.2, 029201 (2020)]
%doi:10.1093/ptep/ptz106
[arXiv:1808.10567 [hep-ex]].
%1346 citations counted in INSPIRE as of 17 Mar 2024

%\cite{Sakharov:1967dj}
\bibitem{Sakharov:1967dj}
A.~D.~Sakharov,
``Violation of CP Invariance, C asymmetry, and baryon asymmetry of the universe,''
Pisma Zh. Eksp. Teor. Fiz. \textbf{5}, 32-35 (1967)
%doi:10.1070/PU1991v034n05ABEH002497
%4868 citations counted in INSPIRE as of 23 Jul 2024

%\cite{Kanemura:2002vm}
\bibitem{Kanemura:2002vm}
S.~Kanemura, S.~Kiyoura, Y.~Okada, E.~Senaha and C.~P.~Yuan,
``New physics effect on the Higgs selfcoupling,''
Phys. Lett. B \textbf{558}, 157-164 (2003)
%doi:10.1016/S0370-2693(03)00268-5
[arXiv:hep-ph/0211308 [hep-ph]].
%164 citations counted in INSPIRE as of 23 Jul 2024

%\cite{Kanemura:2004ch}
\bibitem{Kanemura:2004ch}
S.~Kanemura, Y.~Okada and E.~Senaha,
``Electroweak baryogenesis and quantum corrections to the triple Higgs boson coupling,''
Phys. Lett. B \textbf{606}, 361-366 (2005)
%doi:10.1016/j.physletb.2004.12.004
[arXiv:hep-ph/0411354 [hep-ph]].
%193 citations counted in INSPIRE as of 23 Jul 2024

%\cite{Kanemura:2004mg}
\bibitem{Kanemura:2004mg}
S.~Kanemura, Y.~Okada, E.~Senaha and C.~P.~Yuan,
``Higgs coupling constants as a probe of new physics,''
Phys. Rev. D \textbf{70}, 115002 (2004)
%doi:10.1103/PhysRevD.70.115002
[arXiv:hep-ph/0408364 [hep-ph]].
%277 citations counted in INSPIRE as of 23 Jul 2024

%\cite{Braathen:2019pxr}
\bibitem{Braathen:2019pxr}
J.~Braathen and S.~Kanemura,
``On two-loop corrections to the Higgs trilinear coupling in models with extended scalar sectors,''
Phys. Lett. B \textbf{796}, 38-46 (2019)
%doi:10.1016/j.physletb.2019.07.021
[arXiv:1903.05417 [hep-ph]].
%47 citations counted in INSPIRE as of 23 Jul 2024

%\cite{Braathen:2019zoh}
\bibitem{Braathen:2019zoh}
J.~Braathen and S.~Kanemura,
``Leading two-loop corrections to the Higgs boson self-couplings in models with extended scalar sectors,''
Eur. Phys. J. C \textbf{80}, no.3, 227 (2020)
%doi:10.1140/epjc/s10052-020-7723-2
[arXiv:1911.11507 [hep-ph]].
%38 citations counted in INSPIRE as of 23 Jul 2024

\bibitem{future}
M.~Aoki, K.~Enomoto, S.~Kanemura and S.~Taniguchi, in preparation

%\cite{Cepeda:2019klc}
\bibitem{Cepeda:2019klc}
M.~Cepeda, S.~Gori, P.~Ilten, M.~Kado, F.~Riva, R.~Abdul Khalek, A.~Aboubrahim, J.~Alimena, S.~Alioli and A.~Alves, \textit{et al.}
``Report from Working Group 2: Higgs Physics at the HL-LHC and HE-LHC,''
CERN Yellow Rep. Monogr. \textbf{7}, 221-584 (2019)
%doi:10.23731/CYRM-2019-007.221
[arXiv:1902.00134 [hep-ph]].
%707 citations counted in INSPIRE as of 19 Mar 2024

%\cite{Fujii:2015jha}
\bibitem{Fujii:2015jha}
K.~Fujii, C.~Grojean, M.~E.~Peskin, T.~Barklow, Y.~Gao, S.~Kanemura, H.~D.~Kim, J.~List, M.~Nojiri and M.~Perelstein, \textit{et al.}
``Physics Case for the International Linear Collider,''
[arXiv:1506.05992 [hep-ex]].
%259 citations counted in INSPIRE as of 19 Mar 2024

%\cite{Bambade:2019fyw}
\bibitem{Bambade:2019fyw}
P.~Bambade, T.~Barklow, T.~Behnke, M.~Berggren, J.~Brau, P.~Burrows, D.~Denisov, A.~Faus-Golfe, B.~Foster and K.~Fujii, \textit{et al.}
``The International Linear Collider: A Global Project,''
[arXiv:1903.01629 [hep-ex]].
%334 citations counted in INSPIRE as of 19 Mar 2024

%\cite{Grojean:2006bp}
\bibitem{Grojean:2006bp}
C.~Grojean and G.~Servant,
``Gravitational Waves from Phase Transitions at the Electroweak Scale and Beyond,''
Phys. Rev. D \textbf{75}, 043507 (2007)
%doi:10.1103/PhysRevD.75.043507
[arXiv:hep-ph/0607107 [hep-ph]].
%402 citations counted in INSPIRE as of 23 Jul 2024

%\cite{Espinosa:2010hh}
\bibitem{Espinosa:2010hh}
J.~R.~Espinosa, T.~Konstandin, J.~M.~No and G.~Servant,
``Energy Budget of Cosmological First-order Phase Transitions,''
JCAP \textbf{06}, 028 (2010)
%doi:10.1088/1475-7516/2010/06/028
[arXiv:1004.4187 [hep-ph]].
%512 citations counted in INSPIRE as of 23 Jul 2024

%\cite{Kakizaki:2015wua}
\bibitem{Kakizaki:2015wua}
M.~Kakizaki, S.~Kanemura and T.~Matsui,
``Gravitational waves as a probe of extended scalar sectors with the first order electroweak phase transition,''
Phys. Rev. D \textbf{92}, no.11, 115007 (2015)
%doi:10.1103/PhysRevD.92.115007
[arXiv:1509.08394 [hep-ph]].
%129 citations counted in INSPIRE as of 23 Jul 2024

%\cite{Hashino:2016rvx}
\bibitem{Hashino:2016rvx}
K.~Hashino, M.~Kakizaki, S.~Kanemura and T.~Matsui,
``Synergy between measurements of gravitational waves and the triple-Higgs coupling in probing the first-order electroweak phase transition,''
Phys. Rev. D \textbf{94}, no.1, 015005 (2016)
%doi:10.1103/PhysRevD.94.015005
[arXiv:1604.02069 [hep-ph]].
%91 citations counted in INSPIRE as of 23 Jul 2024

%\cite{Hashino:2018wee}
\bibitem{Hashino:2018wee}
K.~Hashino, R.~Jinno, M.~Kakizaki, S.~Kanemura, T.~Takahashi and M.~Takimoto,
``Selecting models of first-order phase transitions using the synergy between collider and gravitational-wave experiments,''
Phys. Rev. D \textbf{99}, no.7, 075011 (2019)
%doi:10.1103/PhysRevD.99.075011
[arXiv:1809.04994 [hep-ph]].
%64 citations counted in INSPIRE as of 23 Jul 2024


%\cite{LISA:2017pwj}
\bibitem{LISA:2017pwj}
P.~Amaro-Seoane \textit{et al.} [LISA],
``Laser Interferometer Space Antenna,''
[arXiv:1702.00786 [astro-ph.IM]].
%3083 citations counted in INSPIRE as of 23 Jul 2024

%\cite{Seto:2001qf}
\bibitem{Seto:2001qf}
N.~Seto, S.~Kawamura and T.~Nakamura,
``Possibility of direct measurement of the acceleration of the universe using 0.1-Hz band laser interferometer gravitational wave antenna in space,''
Phys. Rev. Lett. \textbf{87}, 221103 (2001)
%doi:10.1103/PhysRevLett.87.221103
[arXiv:astro-ph/0108011 [astro-ph]].
%807 citations counted in INSPIRE as of 23 Jul 2024

%\cite{Corbin:2005ny}
\bibitem{Corbin:2005ny}
V.~Corbin and N.~J.~Cornish,
``Detecting the cosmic gravitational wave background with the big bang observer,''
Class. Quant. Grav. \textbf{23}, 2435-2446 (2006)
%doi:10.1088/0264-9381/23/7/014
[arXiv:gr-qc/0512039 [gr-qc]].
%394 citations counted in INSPIRE as of 23 Jul 2024

%\cite{Hashino:2021qoq}
\bibitem{Hashino:2021qoq}
K.~Hashino, S.~Kanemura and T.~Takahashi,
``Primordial black holes as a probe of strongly first-order electroweak phase transition,''
Phys. Lett. B \textbf{833}, 137261 (2022)
%doi:10.1016/j.physletb.2022.137261
[arXiv:2111.13099 [hep-ph]].
%37 citations counted in INSPIRE as of 17 Mar 2024

%\cite{Hashino:2022tcs}
\bibitem{Hashino:2022tcs}
K.~Hashino, S.~Kanemura, T.~Takahashi and M.~Tanaka,
``Probing first-order electroweak phase transition via primordial black holes in the effective field theory,''
Phys. Lett. B \textbf{838}, 137688 (2023)
%doi:10.1016/j.physletb.2023.137688
[arXiv:2211.16225 [hep-ph]].
%14 citations counted in INSPIRE as of 17 Mar 2024




\end{thebibliography}
\end{document}